\documentclass[aps, prx, superscriptaddress, 10pt]{revtex4-2}
\usepackage[margin=2.5cm]{geometry}
\usepackage[utf8]{inputenc}
\usepackage{amssymb, amsmath, amsthm,mathtools}

\usepackage{url,  hyperref}
\usepackage[dvipsnames]{xcolor}
\hypersetup{colorlinks, linkcolor={red!70!black}, citecolor={blue!70!black},
urlcolor={red!70!black} }
\usepackage{graphicx}
\usepackage{settings_custom}
\usepackage[ruled,vlined]{algorithm2e}
\usepackage{enumerate}
\usepackage{enumitem}
\usepackage[normalem]{ulem}

\newtheorem{res}{Result}
\newtheorem{prevres}[res]{Previous result}

\begin{document}

\title{Bilinear Sequence Regression: \\ A Model for Learning from Long Sequences of High-dimensional Tokens}

\author{Vittorio Erba}
\affiliation{Statistical Physics of Computation Laboratory, École Polytechnique Fédérale de Lausanne (EPFL), Switzerland}

\author{Emanuele Troiani}
\affiliation{Statistical Physics of Computation Laboratory, École Polytechnique Fédérale de Lausanne (EPFL), Switzerland}

\author{Luca Biggio}
\affiliation{Statistical Physics of Computation Laboratory, École Polytechnique Fédérale de Lausanne (EPFL), Switzerland}
\affiliation{Department of Computing Sciences, Universita Bocconi, Milan, Italy}

\author{Antoine Maillard}
\affiliation{Department of Mathematics, ETH Zürich, Switzerland}

\author{Lenka Zdeborová}
\affiliation{Statistical Physics of Computation Laboratory, École Polytechnique Fédérale de Lausanne (EPFL), Switzerland}

\begin{abstract}  
  Current progress in artificial intelligence is centered around so-called large language models that consist of neural networks processing long sequences of high-dimensional vectors called tokens. Statistical physics provides powerful tools to study the functioning of learning with neural networks and has played a recognized role in the development of modern machine learning. The statistical physics approach relies on simplified and analytically tractable models of data. However, simple tractable models for long sequences of high-dimensional tokens are largely underexplored. 
  Inspired by the crucial role models such as the single-layer teacher-student perceptron (aka generalized linear regression) played in the theory of fully connected neural networks, in this paper, we introduce and study the \textit{bilinear sequence regression} (BSR) as one of the most basic models for sequences of tokens.
  We note that modern architectures naturally subsume the BSR model due to the skip connections.
  Building on recent methodological progress, we compute the Bayes-optimal generalization error for the model in the limit of long sequences of high-dimensional tokens, and provide a message-passing algorithm that matches this performance. We quantify the improvement that optimal learning brings with respect to vectorizing the sequence of tokens and learning via simple linear regression. We also unveil surprising properties of the gradient descent algorithms in the BSR model. 
\end{abstract}

\maketitle

\section{Introduction}

\subsection{Motivation}

\paragraph{Deep learning and the statistical physics approach to understand it.}
We are witnessing unprecedented progress in artificial intelligence, largely thanks to advances in learning with large multi-layer neural networks, commonly referred to as deep learning~\cite{lecun2015deep}. 
Milestones such as the classification of images from the ImageNet dataset~\cite{deng2009imagenet,krizhevsky2012imagenet} or super-human performance in the game of Go \cite{silver2016mastering} used deep neural networks based on combinations of fully connected and convolutional layers that map rather high-dimensional vectors 
into vectors of (in general) different, but still high, dimension. While deep learning is undeniably successful in practical tasks, the underlying theoretical mechanisms behind its functioning remain covered with open questions. This led to an abundance of theoretical works aiming to explain the behaviour of deep neural networks that are observed in practice, such as the lack of overfitting in over-parameterized neural networks, the principles thanks to which gradient-based training dynamics reach configurations with good generalization performance while many other configurations of equally good training loss exist and lead to bad generalization,
or theoretically-grounded principles leading to the choice of the best-performing architecture, algorithms and hyper-parameters, for a given dataset and task. A subfield of the theory of deep learning stems from statistical physics, 
a scientific field that is particularly well suited to come up with models that are solvable in the high-dimensional limit, and to provide insights into the above questions~\cite{zdeborova2020understanding}.
This line of work was initiated 
decades ago, with e.g. \cite{hopfield1982neural,ackley1985learning,gardner1988optimal,gardner1989three,seung1992statistical}, and regained broad interest in the last decade with a number of influential works, e.g. \cite{saxe2013exact,baldassi2015subdominant,baldassi2016unreasonable,barbier2019optimal,goldt2020modeling,advani2020high,loureiro2021learning,sorscher2022beyond}. One of the instrumental models in this line of work is the \emph{teacher-student} 
model where one investigates whether a neural network can learn from data that were generated by a teacher neural network whose weights are not known to the student neural network. This teacher-student setting was introduced in \cite{gardner1989three} and used broadly, including in most of the above-cited works. 

\paragraph{Sequence modelling takes the lead.}
The landscape of research in deep learning reshaped considerably with the rise of transformer architectures~\cite{vaswani2017attention} and subsequent large language models (LLMs), sometimes also called foundation models, such as the GPT family \cite{radford2019language,mann2020language,achiam2023gpt},
leading to the well-known chatbot ChatGPT by OpenAI~\cite{achiam2023gpt,bubeck2023sparks} that took the field of AI and the whole high-tech industry by a storm. Transformers are types of deep neural networks that stand behind this progress. They are composed of combinations of fully-connected layers and, crucially, so-called attention layers. 
While
a fully-connected layer maps vectors into vectors, an attention layer maps sequences of vectors (called tokens) into sequences of vectors (tokens). In language modelling, a token would typically be associated with a word, 
and each such token is mapped (embedded) into a relatively high-dimensional (typically around a 1000-dimensional) vector. The sequence then corresponds to a text composed of words/tokens and is also long, corresponding to the number of words in a text. 
It is fair to say that the impressive performance of current LLMs was not anticipated by many. The functioning of LLMs is surrounded by even more theoretical open questions -- including the emergence of capabilities \cite{wei2022emergent}, the neural scaling laws \cite{kaplan2020scaling}, or in-context learning \cite{brown2020language}. 

In our opinion, the most fundamental theoretical question underlying transformers is: 
\vspace{0.1cm}
\begin{center}
    \fbox{{Why is it advantageous to present the data as long sequences of high-dimensional tokens?}}
\end{center}
\vspace{0.1cm} 
More specifically, why is it advantageous for network architecture to act differently 
in token-space and embedding-space?
Indeed, if one vectorized the data into a single large vector and used a fully-connected architecture, the universal approximation theorem \cite{cybenko1989approximation,hornik1989multilayer} would still imply that a generic set of functions can be represented this way. There must be a computational advantage in presenting the data as sequences of tokens that the transformer architecture exploits. This advantage may be related to the underlying structure of the data, the reduction of the number of trainable parameters or the flexibility with respect to sequence length. However, the precise reasons are not understood and the advantage with respect to learning from the vectorized data is not quantified theoretically.

One can anticipate that also in this context of learning from sequences of tokens, the statistical physics approach, based on simplified models that capture some of the intriguing properties and behaviours, will help to clarify some of the key questions surrounding LLMs, transformers, and learning with attention layers. Indeed, works in this direction started appearing in the past couple of years. From those we are aware of, several build interesting simplified models and then investigate the training of the corresponding toy-transformer architecture numerically or phenomenologically~\cite{raventos2024pretraining,cagnetta2024towards,behrens2024understanding}. Others study the propagation of a signal through a trained transformer \cite{geshkovski2023mathematical,cowsik2024geometric}. So far, only a handful of works have been able to analyze the training of a toy transformer analytically. 
Concerning works that analyze the training of a neural network for data consisting of sequences of tokens,~\cite{rende2024mapping} analyzes an attention layer learning Gaussian data generated by a model where the sequence length $L$ is large, but the token dimension $d$ is small. 
Authors of \cite{cui2024phase}
analyze minimizers of the training loss and corresponding phase transitions for a teacher-student-like data model where the token dimension $d$ is large, but the length of the sequence is small $L=O(1)$. 
Finally, the authors of~\cite{lu2024asymptotic} consider a very interesting case of in-context learning linear regression where both the token dimension $d$ and the sequence length $L$ (corresponding to the number of samples given in each context in \cite{lu2024asymptotic}) are large, but they analyze only a linear attention layer, which can be seen as a special case of ridge regression, limiting the generalizability of their approach to address a broader set of questions. In practical settings, such as the GPT architectures, both the length of the sequence $L$ and the embedding dimension $d$ are large, typically in thousands \cite{dubey2024llama}.
It is thus critical to build a theoretically-analysable model, where the thermodynamic (or high-dimensional) limit corresponds to both the length $L$ and the embedding dimension $d$ going to infinity.

Another motivation of this paper stems from the following lines. While complex non-linear neural networks perform extremely well in practice, the deep learning revolution has exposed many fundamental questions even in basic statistical methods like linear regression. Indeed, describing training procedures in non-convex optimization problems is highly non-trivial even in simple single-layer neural networks with non-linear outputs, such as in the dynamics of gradient descent for phase retrieval problems \cite{mignacco2021stochasticity,sarao2020complex}. Another example of a phenomenon that can be understood already in linear regression (or its slight variations) is double-descent \cite{belkin2019reconciling}, which has fundamentally changed our understanding of overfitting and the bias-variance trade-off, while it has been theoretically explained within the framework of linear regression \cite{belkin2019reconciling,gerace2020generalisation}. It is clear from these examples that such basic models as linear regression have been extremely useful in clarifying the properties of modern deep learning. Therefore, the question arises:
\vspace{0.1cm} 
\begin{center}
    \fbox{{{ What is the basic model, analogous to perceptron or generalized linear regression, for sequences of tokens?}}}
\end{center}
\vspace{0.1cm} 
In this paper we introduce such a model and initiate its study, thus providing a rich theoretical playground to tackle questions about learning from long sequences of high-dimensional data.  

\subsection{Definition of the bilinear sequence regression model} 

Motivated by the above questions, the present paper introduces a prototypical analytically solvable model for supervised learning from long sequences of high-dimensional tokens, which we name the Bilinear Sequence Regression (BSR) model.
We consider a supervised regression task on a dataset of $n$ input/output pairs $\caD_n = \{(X^{\mu}, y^{\mu})\}_{\mu = 1}^n$ with $X^{\mu} \in \bbR^{L \times d}$ being the inputs consisting of a sequence of length $L$ of $d$-dimensional tokens, and $y \in \bbR$ are the labels. 
We focus here on regression, as opposed to the more common next-token prediction, because several theoretical studies of neural networks focused on linear regression; we are thus able to leverage the insights gained in the linear regression literature.
 Following the statistical physics studies of the teacher-student setting to analyze learning with fully-connected neural networks, we draw each component of the input data $X_{ij}^{\mu}$ independently from a Gaussian distribution of zero mean and unit variance.
 The labels are then generated through the following (teacher) model 
 \begin{equation} \label{eq:teacher}
     y^{\mu} \sim P_{\rm out}\Big(\cdot | \frac{1}{\sqrt{dLr}} \sum_{a,i=1}^{L,d} X_{ai}^\mu \sum_{\gamma=1}^r U^*_{i \gamma} V^*_{\gamma a} \Big)  \, ,
 \end{equation} 
 where $U^* \in \bbR^{d \times r}$, $V^* \in \bbR^{r \times L}$,  and their components are taken i.i.d. from a standard Gaussian, and $P_{\rm out}$ is a probabilistic scalar output channel. The parameter $r$ will be called the \textit{width} of the model. 
 Given the dataset $\caD_n$, the task is then to learn a function $f: X \in \bbR^{L \times d} \to y \in \bbR$ and obtain a good performance on a test set. 

We consider a Bayesian setting, in which the learner knows the architecture of eq.~\eqref{eq:teacher}, i.e.\ the form of $P_{\rm out}$ and the distributions of $U^*, V^*$. In this context the main task is to recover the values of $U^*$ and $V^*$. It is well-known that the optimal performance for this task is reached by the so-called \emph{Bayes-optimal} estimator, which corresponds to the mean of the posterior distribution, as we will describe in more details in Section~\ref{sec:main_results}.

 We think of the inputs $X^{\mu}\in \bbR^{L \times d}$ as sequences of tokens and the outputs $y^{\mu}$ as labels. 
 As a concrete example, each row of $X^{\mu}$ can be thought of as a vector embedding of a word, so that $X^{\mu}$ represents a text in some language, and $y^{\mu}$ may be a sentiment score associated with the text, categorising its meaning, for example, as uplifting or depressing.
 As said above, data in the form of sequences of tokens is ubiquitous in modern machine learning, including natural language datasets (where tokens are words and sequences are phrases) or biological datasets (where tokens are amino acids and sequences are proteins), yet our understanding of the performance of learning algorithms on such data is scarce. In this sense, the model \eqref{eq:teacher} provides a benchmark dataset where the inputs $X^\mu$ are unstructured (random) and the function from $X \in \bbR^{L \times d}$ to $y \in \bbR$ is parametrized by ground-truth latent variables $U^*$ and $V^*$ in a bilinear form way which is among the simplest functional forms one can posit when inputs $X^\mu$ are sequences of tokens.

 What we perceive as a key property of a simple sequence model is that in the BSR the factor $V^*$ acts on the sequence elements while $U^*$ acts separately on the embedding dimensions of the data. This is mimicking the way transformers process data (see Section \ref{subsec:BSR_transformer} for a review). Indeed, a key aspect of the dot-product attention layer is that it transforms the representation in the embedding space in one way (via the key, query and value matrices) and the representation in the sequence space in another way (via the attention matrix). Notice also that the attention layer treats elements of the sequence in a permutationally invariant manner (unless positional embeddings are explicitly provided), and does so as well for the elements of the embedding.
All this is reproduced in our model – the permutational invariance between elements of the sequence as well as the transformation of the embedding representation (via the matrix $U^*$ in our model) being different from the transformation of the sequence representation (via the matrix $V^*$ in our model).

 In this paper, we show that \eqref{eq:teacher} is a useful toy model for supervised learning over sequential data in a similar way as the teacher-student perceptron for non-sequential data, which is widely studied  in the statistical physics literature.
 The key question, then, is how neural networks learn on such a dataset in order to be able to predict labels on previously unseen inputs. Additionally, how does the performance depend on the architecture of the network and the 
 algorithm used to learn the data?

 In the present paper we address the following questions: 
 \begin{enumerate}[label=\textbf{(Q\arabic*)},ref=(Q\arabic*)]
 \item\label{q1} What is the performance of the Bayes-optimal estimator learning from a given number of samples $n$ of data generated by the model (\ref{eq:teacher}) in the limit of $d$ and $L$ large, proportionally to each other?
 We consider the full range of possible values for the width parameter $r$, with a specific focus on the regimes where $r$ is either proportional to $d$ and $L$, or remains of order $\mathcal{O}(1)$, as $L, d \to \infty$.
 \item \label{q2} Can this Bayes-optimal performance be reached by efficient algorithms, and which ones? 
 \item \label{q3} Does the Bayes-optimal performance present sharp thresholds (phase transitions) in performance as a function of the number of samples? If yes, at which sample complexities? 
 \item \label{q4} How does the Bayes-optimal performance compare to the performance of linear regression on the vectorized data? 
 \item \label{q5} What is the performance of gradient descent minimizing a loss that uses an ansatz for the function from $X$ to $y$ that matches (\ref{eq:teacher}), and how does it depend on the parameters of the model, the learning rate and the initialization of the algorithm? Unlike in linear regression where the most natural loss is convex, the bilinear nature of the present model leads to a non-convex optimization problem with multiple minima. It is a hard endeavour to understand the behaviour of gradient descent in such cases. This question is hence addressed numerically.  
\end{enumerate}
 The present paper answers all these questions. In particular, \ref{q1} is answered analytically in Sec.~\ref{sec:MMSE_ext_rank}, 
 \ref{q2} via the GAMP-RIE algorithm in Sec.~\ref{sec:amp}, 
 \ref{q3} in the discussion of Sec.~\ref{sec:strong}, 
 \ref{q4} in Sec.~\ref{sec:regression} and Figs.~\ref{fig:varyrho}-\ref{fig:varybeta}, and 
 \ref{q5} numerically in Sec.~\ref{sec:GD}. 
 
The numerical code used to produce all presented experiments is available at \url{https://github.com/SPOC-group/bilinear-sequence-regression}.

\subsection{Bilinear sequence regression as the backbone of a transformer and a one-layer MLP mixer}
 \label{subsec:BSR_transformer}

In this section, we describe the connection between the bilinear sequence regression (BSR) model~(\ref{eq:teacher}) and modern neural network architectures that achieve state-of-the-art performance in vision and natural language processing tasks. 

The BSR model can be viewed as a simple, one-layer instance of the so-called MLP-Mixer architecture (where MLP stands for multi-layer perceptron)~\cite{tolstikhin2021mlp}. MLP-Mixers operate on token sequences by alternately applying two types of MLPs: one applied \emph{token-wise}—mixing the embedding dimensions independently for each token—and one applied {\emph{dimension-wise}—mixing information across tokens independently for each embedding coordinate. These operations are then repeated across multiple layers. 
The BSR model mirrors this structure in a minimal form: the matrix \( U \) serves as the weights of a single-layer MLP applied token-wise, while the matrix \( V \) corresponds to the weights of a single-layer MLP applied embedding-coordinate-wise. In this sense, BSR is the simplest variant of an MLP-Mixer applied to token sequences, analogous to how generalized linear regression is the simplest variant of an MLP applied to vectors.

Another way to motivate the form of the bilinear sequence regression model (\ref{eq:teacher}) is as the bare backbone of a prototypical transformer architecture \cite{vaswani2017attention} designed for a supervised regression task, i.e. where the output is a continuous scalar. Let us first describe the key components of such architectures. This part can also serve to readers not familiar with these types of neural networks, to get a concise account of their key ingredients.   

We consider the following prototypical architecture for a transformer that acts on sequences of tokens $X^\mu \in \bbR^{L \times d}$ and maps them to scalar labels $y^\mu$. 
A toy model for a transformer would typically include a first linear embedding layer, followed by an attention layer, followed by a fully connected layer with omnipresent skip connections, followed by a final linear readout. 
Written mathematically, the embedding layer implements the 
linear mapping
\begin{equation}
    f_{\rm embedding} : \bbR^{L \times d} \to \bbR^{L \times d'}
    \quad\text{s.t.}\quad
    Z_{a\gamma} \coloneqq [f_{\rm embedding}(X)]_{a\gamma} = \sum_{i=1}^d X_{ai} U_{i\gamma} \, ,
\end{equation}
with learnable weights $U \in \bbR^{d \times d'}$. 
We remark that the embedding layer usually serves to reduce the dimensionality, i.e. $d' < d$. This is because, in language data, $d$ would correspond to the size of the dictionary, which is typically much larger than the embedding dimension. 
The attention layer with a skip connection implements the mapping
\begin{equation}\label{eq:f_attention}
    f_{\rm attention} : \bbR^{L \times d'} \to \bbR^{L \times d'}
    \quad\text{s.t.}\quad
    Z'_{a\gamma} \coloneqq  [f_{\rm attention}(Z)]_{a\gamma} = Z_{a\gamma} + \sum_{b=1}^L A_{ab}(Z) \left( \sum_{j=1}^{d'} Z_{bj} (w_V)_{j \gamma} \right) \, ,
\end{equation}
where $w_V\in \bbR^{d' \times d'}$ is a learnable matrix called \textit{value}, and the dot-product attention is defined as
\begin{equation}
    A_{ab}(X) = \text{softmax}\left[ \sum_{i=1}^{d'}\left(
     \sum_{\gamma=1}^{d'} Z_{a\gamma} (w_Q)_{
\gamma i} \right)
    \left( \sum_{\gamma =1}^{d'} Z_{b\gamma} (w_K)_{\gamma i} \right)
    \right] \, ,
\end{equation}
where $w_Q,w_K \in \bbR^{d' \times d'}$ are the learnable \textit{query} and \textit{key} matrices, and the softmax function maps rows of matrices into rows of normalised probabilities as
\begin{equation}
    \text{softmax}(M_{ab}) \coloneqq \frac{e^{M_{ab}}}{{\sum_{c=1}^{L}e^{M_{ac}} }} \, .
\end{equation}
A crucial aspect of the attention layer is the fact that it acts separately in sequence and embedding space. The information along the embedding dimension is processed through the key, query and value matrices, while the attention matrix itself acts only on the sequence dimension.
The subsequent one-hidden-layer fully connected network with a skip connection across the non-linearity implements the following mapping:
\begin{equation}\label{eq:f_MLP}
    f_{\rm MLP} : \bbR^{L \times d'} \to \bbR^{L \times d'}
    \quad\text{s.t.}\quad
    Z''_{a\gamma} \coloneqq  f_{\rm MLP}(Z')_{a\gamma} = Z'_{a\gamma} + \sigma \left( \sum_{j=1}^{d'} Z'_{aj} (w_F)_{j\gamma} \right) \, ,
\end{equation}
where $w_F \in \bbR^{r \times d'}$ is a learnable matrix of weights, and MLP stands for multi-layer perceptron. 
Finally, for regression tasks the natural last mapping is a readout 
\begin{equation}
    f_{\rm readout}: \bbR^{L \times d'} \to \bbR
    \quad\text{s.t.}\quad
    y=f_{\rm readout}(Z'') = \sum_{a=1}^L \sum_{\gamma=1}^{d'} Z''_{a\gamma} V_{\gamma a},
\end{equation}
where $V \in \bbR^{d' \times L}$ is a learnable matrix.

A realization that is key to motivate the bilinear sequence regression model (\ref{eq:teacher}) is that the skip-only part of this transformer architecture
(essentially setting $\sigma = 0$ and $A = 0$ in eqs.~\eqref{eq:f_attention} and \eqref{eq:f_MLP})
reduces to
\begin{equation}
     \label{eq:student}
    y = \sum_{a,i=1}^{L,d} X_{ai} \sum_{\gamma=1}^{d'} U_{i\gamma} V_{\gamma a}\, .
\end{equation}

We are now considering how this transformer architecture aims to learn from the data produced by the BSR model (\ref{eq:teacher}).  
Notice that if we set the width of the embedding layer $d'$ to be the width of the BSR model $d'=r$, the skip-only part gives an architecture that matches the bilinear sequence regression model (\ref{eq:teacher}) (with $P_{\rm out}(y|y')=\delta(y-y')$ a noiseless output channel, which will be our main focus when applying our results). We think of the skip-only part as the bare backbone of the architecture, i.e. the transformer stripped of the attention and fully connected layers. We stress that the skip-only part thus acts as a student model that matches the architecture of the teacher (\ref{eq:teacher}).

The case where $d' \neq r$ is also of interest, particularly when $d' > r$, which corresponds to the student model~\eqref{eq:student} being overparameterized relative to the target (teacher) function~\eqref{eq:teacher}. 
A detailed analysis of this very rich setting is deferred to future work, as we focus here on the more easily analyzable Bayes-optimal case $d' = r$.
 
We also note that transformers such as the ones considered in \cite{vaswani2017attention,radford2019language} actually possess more features than the ones presented above: e.g.\ they use attention and MLP layers multiple times, they use positional encoding to represent the ordering of the sequences, and the attention layer has multiple so-called ``heads'', meaning the size of the value matrix $w_V$ is $(d'/h) \times (d'/h)$, with $h$ the number of heads and the outputs $Z'$ are concatenated together from all the heads to get back to dimension $d'$. 
The architecture presented above should be thought of as simplified. 

As a matter of fact, the above rationale is valid for any model with skip connections (not only a transformer) designed to process sequences of tokens. The BSR would be the backbone of more general sequence models with skip connections designed for regression. 

\subsection{Related works}

 As far as we know, the bilinear sequence regression model (or BSR) \eqref{eq:teacher} was not yet studied in the context of sequence modelling, neither as a model for synthetic data nor as an analytically tractable model for learning. 
 It was, however, studied in the literature under the umbrella of \textit{matrix sensing}~\cite{recht2010guaranteed,Donoho_2013,schulke16} in the context of signal processing where $U^*$ and $V^*$ represent a hidden signal that is to be recovered.

We note here that other models where the regression parameters are matrix-valued and having structure corresponding to a product of two matrices have been considered in the literature, e.g. \cite{giraud2011low,hoff2015multilinear,chen2023statistical}. None of them is exactly the same as the BSR or the matrix sensing. A more conceptual difference between such works and ours is that the models are usually proposed as ansatzes for functions that are then fitted to the data. Their focus is on regression tasks on real data,
whereas we view the BSR model as a synthetic probabilistic model to generate data that is theoretically tractable, allowing to compute the Bayes optimal performance and to propose a matching algorithm. This is in line with the statistical physics works on fully connected feed-forwards architectures such as the perceptron and its multi-layer version, for which the main contribution of statistical physics was the analysis of the optimal generalization error, in an influential line of work~\cite{gardner1989three,seung1992statistical}.

 Going back to the related work on matrix sensing where each input matrix $X^{\mu}\in \bbR^{L \times d}$ is seen as a random linear projection operator, and the task becomes to retrieve the (sometimes sparse) signals $U^*, V^*$ from the projections. The well-known line of work represented by \cite{recht2010guaranteed,Donoho_2013} proposes and analyses an algorithm based on a convex relaxation of the problem, where the nuclear norm (defined as the sum of singular values) of the matrix $S^* = U^* V^* / \sqrt{r}$ is minimized. Similarly to other convex relaxations, this algorithm, however, reaches suboptimal performance with respect to the Bayes-optimal estimator. 
 Another line of work considered the matrix sensing problem solved via gradient descent in an over-parameterized setting: their findings suggested that gradient descent with infinitesimal initialization could have implicit regularization towards the minimum nuclear norm~\cite{gunasekar2017implicit,li2020towards}.

 The Bayes-optimal performance for matrix sensing, which is an instance of model \eqref{eq:teacher}, was addressed in~\cite{schulke16} using approximate message-passing algorithms and their state evolution. Their results provide an exact characterization of the Bayes-optimal performance in the low-width limit where $L,d \to \infty$ with $L/d= \mathcal{O}(1), \, n/d = \mathcal{O}(1)$ and crucially the width remains of order $r=\mathcal{O}(1)$. The paper \cite{schulke16} also claims to provide an asymptotically exact characterization of the case where the width $r$ is extensive, i.e. $r/d = \mathcal{O}(1)$, but this claim is based on incorrect assumptions, as was later found in the closely related problem of extensive-rank (the width parameter $r$ plays the role of the rank) matrix factorization 
 in the line of work~\cite{schmidt2018statistical,maillard2022perturbative,barbier2022statistical,troiani2022optimal,semerjian2024matrix,camilli2023matrix,pourkamali2023rectangular}. The characterization of the Bayes-optimal performance for the extensive width $r/d = \mathcal{O}(1)$ thus remained open: this is the first main technical contribution of the present paper, together with the proposition of an approximate message passing algorithm that reaches, in the studied cases, the optimal performance in polynomial time.

 Another model studied in the literature that is technically related to ours would correspond to the width $r=1$, with each sample $X^{\mu}$ being a symmetric matrix and $U=V^T$. Such a model has been studied both for random labels $y$ \cite{fyodorov2019spin,kamali2023dynamical,montanari2023solving} and with labels generated by a teacher model \cite{kamali2023stochastic}. 

 On a technical level, our work builds upon the works~\cite{troiani2022optimal,troiani2024}.
 More specifically, we extend the recent analysis of~\cite{troiani2024}, that treats a model that can technically be seen as a symmetric version of the BSR, where one imposes $U=V^T$, and $X_{ai}$ is a symmetric matrix. 
 Our derivation further relies on the optimal performance in the denoising of extensive-rank non-symmetric matrices~\cite{troiani2022optimal}. 

 Regarding the dynamics of gradient descent in matrix sensing, a symmetric version where $U^* = V^*$ has been considered in~\cite{bhojanapalli2016global}. There, the authors show that all minima of the natural square loss for the problem achieve perfect reconstruction in the high-dimensional limit as long as the number of samples is of order $n \geq C \cdot dr$, where $r$ is the width, but their approach does not have access to the optimal constant $C > 0$.
 A similar model has also been studied in~\cite{ding2024flat}, where the authors show that flat minima of a natural loss generalize well if $n > C r(d+L)$, but again they do not have access to the optimal constant $C$.
 In comparison, we consider generic values of the width, in particular in Section \ref{sec:GD} the extensive one $r = \Theta(d)$, and show by combining the Bayes optimal analysis and numerical evidence on gradient descent that an averaged version of gradient descent generalizes perfectly as soon as this is information-theoretically possible, i.e. as soon as the Bayes-optimal estimator generalizes perfectly. We provide both the scaling $n = \caO(dL)$ of this threshold, as well as the constant (see Result \ref{res:strong}).
 Finally, let us mention that the dynamics of alternate minimization in a related model has been considered in \cite{okajima2024asymptotic}.

\section{Main technical results}\label{sec:main_results}

\subsection{Generalized bilinear sequence regression model}\label{sec:genmodel}

From a technical point of view, it will be advantageous to think about the BSR model (\ref{eq:teacher}) in a slightly more general way. We consider the model 
\begin{equation}\label{eq.model}
    y^{\mu} \sim P_{\rm out}( \cdot | h^\mu ) 
    \quad\text{with}\quad 
    h^{\mu} = \frac{1}{\sqrt{Ld}} \sum_{a, i=1}^{L, d} X^\mu_{ai}  S^*_{ia}
    \, ,
\end{equation} 
where $S^* \in \bbR^{d \times L}$ is a hidden or latent weight matrix, $X^{\mu} \in \bbR^{L \times d}$ for $\mu = 1, \dots, n$ are input sequences composed by $L$ tokens, each a vector in dimension $d$, and $y^{\mu} \in \bbR$ for $\mu = 1, \dots, n$ are the associated scalar labels. 

We assume the data to be Gaussian, i.e. $X^{\mu}_{ai} \sim \mathcal{N}(0,1)$ independently for each value of $(\mu, a, i)$.
We believe that this assumption can be relaxed within the context of Gaussian universality results (see e.g.\ \cite{hu2022universality, wang2023learninghierarchicalpolynomialsthreelayer})
without altering the main points of our analysis: we will pursue this generalization in future works.
The labels $y^{\mu}$ are generated through a possibly probabilistic output channel $P_{\rm out}$, conditioned on the value of the scalar pre-activations $h^{\mu}$.

The BSR model of eq.~\eqref{eq:teacher} corresponds to the specific case of factorised Gaussian prior on $S^*$, i.e. 
\begin{equation}\label{eq.prior}
    S^*_{ia} = \frac{1}{\sqrt{r}} \sum_{\gamma=1}^r U^*_{i\gamma}V^*_{\gamma a} \, ,
\end{equation}
with $U^* \in \bbR^{d \times r}$, $V^* \in \bbR^{r \times L}$ matrices of i.i.d. standard Gaussian entries. This distribution introduces non-trivial dependencies between each entry of $S^*$, coupling the token and embedding dimensions. 

Notice that the formulation of the model with structured $S^*$ in eq.~\eqref{eq.model} is strictly more general than the BSR, and includes the factorized form of $S^\star$ in eq.~\eqref{eq:teacher} as a special case. In particular, all our results on Bayes optimal learning of Sections~\ref{sec:main_results} and~\ref{sec:cons} apply directly to the BSR~\eqref{eq:teacher}, and take into account fully its factorized structure (through the prior on $S^*$).

We will consider the model \eqref{eq.model} in the high-dimensional setting, where $L, d \to \infty$ with fixed ratio $L = \Theta(d)$. In particular, we define $$\beta \coloneqq \frac{\max(L,d)}{ \min(L,d)} \geq 1,$$
which remains finite as $L,d \to \infty$. $\beta$ measures the aspect ratio of the matrices $X^\mu$ and $S^*$, irrespective of which among $L$ and $d$ is bigger. In general, the scaling for the number of samples $n$ in the high-dimensional limit, i.e. the number of samples needed to at least partially retrieve the signal $S^*$, depends on the choice of its distribution (and it usually scales with the total amount of unknowns included in $S^*$). 

The main novel results of this paper relate to the so-called \textit{extensive-width} limit where the width $r$ is also proportional to the dimensionality. We thus define a width-related parameter $$\rho\coloneqq  \frac{r}{\min(d,L)}$$ that will remain finite in the high-dimensional limit of the model. 
 Note that the rank of the matrix $S^*$ is constrained to be at most $r$, and generically, when the width $r$ is extensive, the rank of the matrix $S^*$ is also extensive. We will thus call the corresponding limit the extensive-width or extensive-rank case.
 For $\rho \to \infty$, we expect by the central limit theorem that the distribution of $S^*$ approaches the one of a matrix with i.i.d. standard Gaussian entries. 
 In the regime where $r$ scales linearly with $d, L$, we will see that the correct sample scale to observe non-trivial retrieval of the ground-truth signal is $n = \mathcal{O}(dL)$. We define $\alpha$ as $$\alpha \coloneqq \frac{n}{dL}$$ with $\alpha > 0$ finite in the high-dimensional limit. The low-width limit where $\rho$ is small down to the width $r = \mathcal{O}(1)$ is considered for comparison in section \ref{sec:low_rank}, building on the result of \cite{schulke16}.

We stress that our main technical results apply to a much wider class of distributions for $S^*$ (priors), more specifically rotationally invariant distributions $P_0$, as long they admit a well-defined limiting spectral density in the high-dimensional limit.
Without loss of generality we assume that $P_0$ is normalised as
\begin{equation}\label{eq:Qstar}
     Q_* \coloneqq \lim_{d, L\to \infty} \EE_{S \sim P_0} \frac{1}{dL} \sum_{a, i=1}^{L, d} (S^*_{ia})^2 = 1 \, .
\end{equation}
In informal terms, this ensures that the entries of $S^* \sim P_0$ are on average of order $\mathcal{O}(1)$ in the high-dimensional limit.
Rotational invariance means that $P_0(S) = P_0(O_1 S O_2)$ for any pair of rotation matrices $O_1, O_2$ in dimensions respectively $d, L$, while having a well defined limiting (symmetrised) spectral density means that 
\begin{equation}
    \lim_{d,L \to \infty} \frac{1}{2 \min(d, L)} \sum_{i = 1}^{\min(d,L)} \left[ \delta(x - \sigma_i(S)) + \delta(x + \sigma_i(S)) \right]= \mu_S(x) \, ,
\end{equation}
for some density $\mu_S(x)$, where $\sigma_i(S)$ are the singular values of $S/\sqrt[4]{dL}$ (the normalisation ensures that $\sigma_i(S)$ remains of order $\mathcal{O}(1)$ in the high-dimensional limit).

\subsection{Bayes-optimal estimation} 

We are interested in predicting the information-theoretical limits for retrieving the hidden parameters $S^*$ from a typical dataset $\caD_n = \{ (X^{\mu}, y^{\mu} \}_{\mu=1}^n$. 
To achieve this goal, we will study the Bayes Optimal (BO) estimator and its performance.
We are interested in two performance metrics.
The first one is
the averaged \textit{test error} (also called generalization error), which is the error obtained when predicting the label on a newly-sampled $(\uX, \uy)$ data pair, and is defined as
\begin{equation}\label{eq:gen}
    E_{\rm gen}(\hS) \coloneqq \EE_{(\uX, \uy)} \EE_{S^*} \EE_{\caD | S^*} \left(
    \uy - P_{\rm out}\left( \frac{1}{\sqrt{Ld}} \sum_{ai} \uX_{ai} \hS(\caD)^\top_{ai}\right) \right)^2 \, ,
\end{equation}
where $\hS$ is a generic estimator (a function mapping the dataset $\caD$ to a candidate set of weights $\hS(\caD)$) and by $P_{\rm out}(x)$ we mean the (possibly random) output of the output channel conditioned on $x$.
Our second metric is
the averaged \textit{estimation error}, which is the discrepancy between the estimated weights and the true weights, and is defined as 
\begin{equation}\label{eq:est}
    E_{\rm est}(\hS) \coloneqq \frac{1}{dL} \EE_{S^*} \EE_{\caD | S^*}|| S^* - \hS(\caD) ||^2 \, .
\end{equation}
The BO estimator w.r.t. either one of the two performance metrics is defined as the estimator function $\hS: \caD \to \hS(\caD)$ minimising the respective metric.
It is a very classical result that
the BO estimator w.r.t. the test error is given by
\begin{equation}
    \hS_{\rm BO, gen}(\caD) 
    = \EE_{S \sim P(S | \caD)} \EE_{(\uX, \uy) | S} \, [\uy \uX]
    \, ,
\end{equation}
where $(\uX, \uy)$ is  a newly sampled input-output pair conditioned on a set of weights $S$, and $P(S|\caD)$ is the posterior distribution, i.e. the probability that a candidate signal $S$ has been used to generate the dataset $\caD$, which can be expressed (through Bayes' theorem) using the prior distribution and the output channel:
\begin{equation}
    P(S|\caD) \propto  P_0(S) \prod_\mu 
    P_{\rm out}\left( y^{\mu} \bigg| \frac{1}{\sqrt{Ld}} \sum_{ai} X_{ai} S_{ai}^\top \right)
    \, .
\end{equation}
The BO estimator w.r.t. the estimation error instead equals the posterior mean, i.e.
\begin{equation}
    \hS_{\rm BO, est}(\caD) 
    = \EE_{S \sim P(S | \caD)} [S]
    \, .
\end{equation}
Both these expressions are standard (see for example \cite{cover1991information}) and can be recovered by taking the derivative w.r.t. the estimator $\hS$ in the errors' definitions and setting it equal to zero.

We remark that for generic $P_{\rm out}$ the two BO estimators may differ. However, in the case of Gaussian inputs $X$ and Gaussian label noise (i.e. Gaussian output channel $P_{\rm out}(\cdot | h) = N(h, \Delta)$, which includes noiseless observations), one can show that
\begin{equation}
    E_{\rm gen}(\hS) = E_{\rm est}(\hS) + \Delta \, ,
\end{equation}
meaning that the estimation and test BO estimators coincide, and that the respective values of the errors differ only by a constant additive shift quantifying the amount of label noise. 
In the following we will focus on this special case,
and for this reason we here restrict our analysis to the BO estimator w.r.t. the estimation error. 
Finally, we remark that the BO estimation error is also called Minimal Mean Square Error (MMSE) in the literature.

We recall that the estimation error of any estimator $\hS$ is given by
\begin{equation}
    \begin{split}
        E_{\rm est}(\hS) 
        &=
        \frac{1}{dL} \EE_{S^*} \EE_{\caD | S^*}|| S^* - \hS(\caD) ||^2
        = 1 + q(\hS)  - 2m(\hS) 
    \end{split}
\end{equation}
where we defined (using conventions originating in statistical physics~\cite{zdeborova2016statistical}) the average ``magnetisation'' $m(\hS)$ and ``overlap'' $q(\hS)$ of the estimator $\hS$ as 
\begin{equation}
    \begin{split}
        m(\hS) \coloneqq \frac{1}{dL} \EE_{S^*} \EE_{\caD | S^*} \Tr((S^*)^T \hS(\caD))
        \mathand
        q(\hS) \coloneqq \frac{1}{dL} \EE_{S^*} \EE_{\caD | S^*} \Tr(\hS^T(\caD) \hS(\caD)) \, ,
    \end{split}
\end{equation}
and used that the prior is normalised to have self-overlap $Q_* = 1$, see eq.~\eqref{eq:Qstar}.
For the BO estimator, Nishimori's identities~\cite{zdeborova2016statistical} imply $m_{\rm BO}=q_{\rm BO}$, from which we obtain
\begin{equation}
    \MMSE = E_{\rm est}(\hat{S}_{\rm BO}) =  1 - q_{\rm BO} \, ,
\end{equation}
and moreover $q_{\rm BO}$ reduces to the overlap between two independent samples of the posterior distribution $P(S | \caD)$, i.e. 
\begin{equation}
        q_{\rm BO} = \frac{1}{dL} \EE_{\caD} \EE_{S_1, S_2 \sim P(S|\caD)} \Tr(S_1^T S_2) \, .
\end{equation}
We stress that in the case of a generic non-Gaussian $P_{\rm out}$, the BO test error can differ from the BO estimation error. Yet, it can still be computed as a function of the same order parameter $q_{\rm BO}$, which is the quantity for which we provide a precise asymptotic analytical treatment in the following sections. 

Finally, we notice that the estimation error for the Gibbs sampler of the posterior, i.e.\ the expected estimation error of a uniform sample of the posterior, satisfies
\begin{equation}\label{eq-gibbs-error}
    \begin{split}
        \EE_{S^*} \EE_{\caD | S^*} \EE_{S \sim P(S|\caD)}
    \frac{1}{dL} || S^* - S ||^2
    &=
    1 - 2 
     \frac{1}{dL} \EE_{S^*} \EE_{\caD | S^*} \EE_{S \sim P(S|\caD)}\Tr((S^*)^T S)
      \\&\quad
      + \frac{1}{dL} \EE_{S^*} \EE_{\caD | S^*} \EE_{S \sim P(S|\caD)}\Tr(S^T S) 
      \\
      &=
    1 - 2 m_{\rm BO}
      + \frac{1}{dL} \EE_{S^*} \EE_{\caD | S^*} \EE_{S \sim P(S|\caD)}\Tr(S^T S) 
      \\
      &= 2(1-q_{\rm BO}) \, ,
    \end{split}
\end{equation}
where in the last step we use 
Nishimori's identities, stating that a sample from the posterior is statistically equivalent to the ground-truth.
Notice the crucial difference between the BO estimator and a sample of the posterior: in the overlap term $q(S)$, the BO estimator uses the overlap between two independent samples of the posterior, while the Gibbs sampler uses the self-overlap of a single sample of the posterior. This implies that the estimation error of the Gibbs sampler, on average, is twice the BO estimation error.

\subsection{Optimal error for extensive-width BSR}
\label{sec:MMSE_ext_rank}

In this section, we present our novel results concerning the asymptotic characterisation of the BO estimator and of the associated optimal estimation error in the high-dimensional limit for all output channels. These results provide the answer to \ref{q1} posed in the introduction. We start with the general case of arbitrary rotationally-invariant priors on $S$, and follow up with the results for the extensive-width BSR model as a consequence.

\begin{res}[MMSE for generalized BSR]\label{res1}
    Consider any rotationally invariant prior $P_0(S)$ such that the empirical symmetrised singular value density of $\uS = S / \sqrt[4]{dL}$ with $S \sim P_0$ converges to a well defined probability distribution for $d \to \infty$ with $\beta = \frac{\max(L,d)}{\min(L,d)} \geq 1$ finite.
    Call $\hmu_{\uY}$ the symmetrised singular value density of $\uY = \uS + \sqrt{\delta} \uZ$ where $\uZ$ is a matrix of i.i.d. Gaussian entries $N(0, 1/ \sqrt[4]{dL})$ and $\delta > 0$.
    Define the sample ratio as $\alpha = n / (dL)$.

    Then $\lim_{d \to \infty} \MMSE = 1 - q$
    where $(q, \hq) \in \bbR^2$ are a solution to the non-linear system of equations
    \begin{equation}\label{eqres1}
        \begin{split}
            q &= 1 - \frac{1}{\hq} 
            + \frac{2}{\beta^{3/2} \hq^{2}}
            \dashint dx \, \hmu_{\uY}(x) \left[
                \frac{(\beta-1)^2}{2x^2} 
                + \frac{2\pi^2}{3} \hmu_{\uY}(x) ^2
                \right] \, , \\ 
            \hq &= 
            \frac{\alpha}{q} \int Dz \, dy \, \frac{\left(\del_z I_{\rm out}(z, y; q)\right)^2}{I_{\rm out}(z, y; q)}  
            \, ,
        \end{split}
    \end{equation}
    where
    \begin{equation}
        I_{\rm out}(z, y; q) \coloneqq \int \frac{dh d\hh}{2\pi}  
        P_{\rm out}\Big(y| h \Big)
        \exp\Big( 
            - \frac{1 - q}{2} \hh^2
            + (\sqrt{q} z + h) i \hh
        \Big) \, .
    \end{equation}
    The integral over $y$ is intended over the image of $P_{\rm out}$. The dashed integral is regularized as specified in \cite[Appendix A]{troiani2022optimal}. In the case where eq.~\eqref{eqres1} admits multiple solutions, one should pick the one maximising an associated free entropy, whose expression we provide in Appendix \ref{app:replicas}, see eq.~\eqref{eq:freeentropy}. 
    Notice that for Gaussian output channels $P_{\rm out}( \cdot | h) = N(h, \Delta)$, the equation for $\hq$ simplifies to
    \begin{equation}
        \hq = \frac{\alpha }{\Delta +1 - q} \, .
    \end{equation}
\end{res}

Result \ref{res1} can be obtained by performing a non-rigorous (hence the phrasing ``Result'' rather than theorem) but exact computation based on replica theory, detailed in Appendix~\ref{app:replicas}. One writes the partition function associated to the posterior distribution, computes it through replica theory and obtains a saddle-point characterisation for the overlap order parameter, i.e. eq.~\eqref{eqres1}. 
The main technical novelty in Result \ref{res1} 
is that for rotationally-invariant priors with extensive rank, a standard factorisation passage of the computation fails (contrary to what has been claimed by~\cite{schulke16}, where the authors' analysis is not correct in the extensive-rank regime, see the references cited in the introduction on the matter). 
However, we can now overcome this difficulty by adapting recent results for Bayes optimal extensive-rank matrix denoising \cite{maillard2022perturbative, troiani2022optimal}.

The main difficulty in solving \eqref{eqres1} is the computation of $\hmu_{\uY}(x)$, which for generic rotationally-invariant priors is \textit{a priori} non-trivial. For the factorised Gaussian prior on $S$,  corresponding to the BSR model, this difficulty can be overcome. Details on how to compute efficiently $\hmu_{\uY}(x)$ in this specific case can be found in \cite[Section 3.3 and Appendix F]{troiani2022optimal}, adapting previous work by \cite{pennington2017nonlinear}. The explicit formula for the MMSE for the BSR model is then as follows. 
\begin{res}[MMSE for the bilinear sequence regression model, consequence of Result \ref{res1}]\label{prop:mmse_BSR}
    For the bilinear sequence regression model, $\MMSE = 1-q$ where $q$ satisfies \eqref{eqres1}. The spectral density $\hmu_\uY(x)$ to use in \eqref{eqres1} is characterised as follows.
    Define the Stieltjes transform of $\hmu_\uY(x)$ as 
    \begin{equation}
        g_\uY(z) \coloneqq \int dx \, \frac{\hmu_\uY(x)}{z-x} \, . 
    \end{equation} 
    Then, 
    \begin{equation}
        g_\uY(z) = z g_{\uY^2}(z^2) \, ,
    \end{equation}
    where $g_{\uY^2}$ is the Stieltjes transform of the asymptotic spectral density of $\uY\uY^T$ (if $d \leq L$) or of $\uY^T \uY$ (if $d > L$).
    Moreover, $g_{\uY^2}(z^2)$ is the root with largest imaginary part of the quartic polynomial $\sum_{a=0}^4 a_k X^k$, where
    \begin{equation}\label{eq.quartic}
        \begin{split}
        &a_0 = -\psi^3 \\
        &a_1 = \psi ( \zeta (\psi - \phi) + \psi ( \eta (\phi - \psi) + \psi z^2 ) ) \\
        &a_2 = - \zeta^2  (\phi - \psi)^2 + \zeta ( \eta (\phi - \psi)^2 + \psi z^2 (2 \phi - \psi) ) - \eta \psi^2 z^2 \phi \\
        &a_3 = - \zeta z^2 \phi ( 2 \zeta \psi - 2 \zeta \phi - 2 \eta \psi + 2 \eta \phi + \psi z^2 ) \\
        &a_4 = \zeta z^4 \phi^2 (\eta - \zeta) \, , 
        \end{split}
    \end{equation}
    where $\phi = \rho/\beta$, $\psi = \rho$, $\eta = (1+\delta) \sqrt{\beta}$ and $\zeta = \sqrt{\beta}$. 
    Finally, the symmetrised singular value density can be recovered as
    \begin{equation}
        \hmu_\uY(x) = \frac{1}{\pi} \lim_{\epsilon \to 0^+}\operatorname{Im} g_\uY(x - i \epsilon) \, .
    \end{equation}
\end{res}
We start by highlighting that in this Result the factorized nature of the prior \eqref{eq.prior} enters explicitly through the parameters $\phi, \psi$, both dependent on the width parameter $\rho$ of the BSR model.
Even more, the full form of the spectral density $\hmu_\uY(x)$ actually depends on the choice of prior \eqref{eq.prior}, in the sense that a different prior with the same width parameter $\rho$, but different overall structure, will have in general a different spectral density.

Notice also that whenever $\rho < 1$, the symmetrized singular value density of $\uS$ has a delta contribution at the origin with mass $(1-\rho)$, while the non-trivial bulks of the distribution (positively and negatively supported) each have mass $\rho/2$. Moreover, while the asymptotic spectral distribution of $\uS$ is singular due to the delta peak, the one of its noisy version $\uY = \uS + \sqrt{\delta} \uZ$ can be shown to possess a smooth density for all values of $\delta > 0$~\cite{biane1997free}.

Fainlly, in the limit of large $\beta \gg 1$ (i.e.\ when $L \gg d$ or $L \ll d$) and of factorised Gaussian priors (see eq.~\eqref{eq.prior}), we are able to simplify Result \ref{res1} significantly, bypassing the computation of the non-trivial limiting spectral density. We present this result in Appendix~\ref{app:large_beta_ext}.

\subsection{Message-passing algorithm}\label{sec:amp}

For general rotationally-invariant priors $P_0$, we are also able to derive an algorithm that, in the high-dimensional limit, achieves the MMSE (unless computationally hard phases arise, which we have not observed in the present problem, see the discussion in the remainder of the section), giving an efficient implementation of the posterior mean, an often intractable problem. This provides the answer to question \ref{q2} posed in the introduction. 
The algorithm we present is a variant of the well-known Generalised Approximate Message Passing (GAMP) algorithm~\cite{rangan2011generalized}, with an additional matrix denoising step, similarly to the algorithm designed in~\cite{troiani2024}.
\begin{res}[GAMP-RIE for rotationally invariant priors]\label{res2}
    Consider the same setting as in Result \ref{res1}.
    Define:
    \begin{itemize}
        \item $g_{\rm out}$ as
        \begin{equation}
            g_\mathrm{out}(y,\omega,V) \coloneqq \frac{1}{V} \frac{
                \int dz (z- \omega) e^{-\frac{(z-\omega)^2}{2V}} P_{\rm out}(y | z)
            }{
                \int dz e^{-\frac{(z-\omega)^2}{2V}} P_{\rm out}(y | z)
            } \, ,
        \end{equation}
        which reduces to $g_\mathrm{out}(y,\omega,V) = (y-\omega)/(\Delta + V)$ for the Gaussian label noise output channel with variance $\Delta$ (i.e.\ $P_{\rm out}(\cdot | h) = \mathcal{N}(h, \Delta)$). 
        \item $f_{\rm RIE}(\cdot, \delta)$ as the BO rectangular matrix Gaussian denoiser~\cite[Result 1, 2]{troiani2022optimal} with noise-to-signal ratio $\delta$, and $\MMSE_{\rm denoising}(\delta)$ the corresponding MMSE. Explicitly, if $R = U \Lambda V$ is the singular value decomposition of a matrix $R$, with $\Lambda = \text{diag}(\lambda_1, \dots, \lambda_t)$ the denoiser acts on each separate singular value as
        \begin{equation}
            f_{\rm RIE}\left( R = U \Lambda V , \delta \right) = U \text{diag}\left( \lambda_i - \frac{2\delta}{\sqrt{\beta}} \left[ \frac{\beta-1}{2\lambda_i} + \dashint dx \frac{\hmu_\uY(x)}{\lambda_i - x} \right] \right)_{i=1,\dots,t} V \, ,
        \end{equation}
        with $\hmu_\uY$ as defined in Result \ref{res1}, and
        \begin{equation}
            \MMSE_{\rm denoising}(\delta) = \delta - \frac{\delta^2}{\sqrt{\beta}} \left[
                \frac{(\beta - 1)^2}{\beta} \dashint dx \frac{\hmu_\uY(x)}{x^2}
                + \frac{4 \pi^2}{3 \beta} \dashint dx \hmu_\uY(x)^3
            \right] \, ,
        \end{equation}
        is the associated BO mean square error. The dashed integral is regularized as specified in \cite[Appendix A]{troiani2022optimal}.
    \end{itemize}
    Then, Algorithm \ref{alg-main} achieves at convergence an overlap $q = \frac{1}{dL} \EE_{S^*} \EE_{\caD | S^*} \Tr(\hS^T(\caD) \hS(\caD))$ satisfying \eqref{eqres1} in the high-dimensional limit. 
    For the specific case of the BSR model, we can analytically compute the spectral density $\hmu_\uY$, as detailed in Result~\ref{prop:mmse_BSR}.
\end{res}

Notice that Algorithm \ref{alg-main} depends explicitly on the width parameter $\rho$ and on the full factorized form of the prior of the BSR model through the spectral density $\hmu_\uY$, as detailed in Result~\ref{prop:mmse_BSR}.

\begin{algorithm}[t]
\SetAlgoLined
\KwResult{An estimator $\hat{S}_{\rm AMP}$}
\textbf{Input: } 
Dataset $\{(X^{\mu}, y^{\mu})\}_{\mu=1}^n$\;
\emph{Initialize} 
$S_{t=0} \sim P_0$, set $\uS_{t=0} = S_{t=0} / \sqrt[4]{dL}$ and $c_{t=0} = 1,\omega_{t=0} = 1 \times [1, \dots, 1]^T \in \bbR^N,V_{t=0} = 1$ \;
\emph{Rescale} 
$\displaystyle \tilde{X} = X / \sqrt[4]{Ld} $ \; 

\While{not converging}{
$\bullet$ 
\emph{Estimation of the variance and mean of $\Tr(\uS_{t}^T \tilde{X}^\mu)$}\;
$\displaystyle V_{t} = c_t$ 
\hspace{0.2cm}\textrm{ and }\hspace{0.2cm}  
$\displaystyle \omega^\mu_{t} = \Tr(\uS_{t}^T \tilde{X}^\mu) - g_\mathrm{out}(y^\mu,\omega^\mu_{t-1},V_{t-1}) V_t$  \;

$\bullet$ 
\emph{Variance and mean of $\uS$ estimated from the channel observations}\;
$\displaystyle A_t = \frac{\alpha}{n} \sum_{\mu=1}^n g_\mathrm{out}(y^\mu,\omega^\mu_t,V_t)^2$  
\hspace{0.2cm}\textrm{ and }\hspace{0.2cm} 
$\displaystyle R_t = \uS_t + \frac{1}{A_t \, \sqrt{dL}} \sum_{\mu=1}^n g_\mathrm{out}(y^\mu,\omega^\mu_t,V_t) \tilde{X}^\mu$ \;

$\bullet$ \emph{Update of the estimation of $S^\star$ with the prior information}\;
$\displaystyle \uS_{t+1} = f_{\rm RIE}\left(R_t, \frac{1}{A_t}\right)$
\hspace{0.3cm} \hspace{0.2cm}\textrm{ and }\hspace{0.2cm}  \hspace{0.3cm}
$\displaystyle c_{t+1} = \MMSE_{\rm denoising}\left(\frac{1}{A_t}\right) $\;
$t = t + 1$\;
}
\emph{Rescale} 
$\displaystyle \hat{S}_{\rm AMP} = \sqrt[4]{Ld} \, \uS_t$  .  
\caption{GAMP-RIE for the BSR model with extensive width.\label{alg-main}}

\end{algorithm}

Let us now justify the last claim of Result~\ref{res2}, concerning the performance achieved by Algorithm~\ref{alg-main}.
It builds on the connection between approximate-message-passing algorithms and replica theory, which is a well-established result in the theory of generalised linear models~\cite{zdeborova2016statistical}. This connection stems from the fact that AMP algorithms can be tracked in high-dimension by an iterative update equation for the order parameters, such as the overlap $q$ and its conjugate parameter $\hq$, called state evolution \cite{donoho2009message}.
One can show that the state evolution equations for Algorithm \ref{alg-main} are
\begin{equation}\label{eq:se_gamp}
    \begin{split}
    \hq_t &= 
            \frac{\alpha}{q_t} \int Dz \, dy \, \frac{\left(\del_z I_{\rm out}(z, y; q_t)\right)^2}{I_{\rm out}(z, y; q_t)}
            \, , 
            \\
        q_{t+1} &= 1 - \frac{1}{\hq_t} 
            + \frac{2}{\beta^{3/2} \hq^{2}_t}
            \int dx \, \hmu_{\uY_t}(x) \left[
                \frac{(\beta-1)^2}{2x^2} 
                + \frac{2\pi^2}{3} \hmu_{\uY_t}(x) ^2
                \right] 
            = 1 - \MMSE_{\rm denoising}\left(\frac{1}{\hq_t}\right) 
                \, , 
    \end{split}
\end{equation}
where $\uY_t = \uS + \sqrt{1/\hq_t} \, \uZ$, the last equality is justified in \eqref{eqMMSE} and 
\begin{equation}
    q^t = \Tr(\uS_*^T \uS_t) \, .
\end{equation}
This is
just a particular iterative scheme for \eqref{eqres1}, justifying our claim that Algorithm \ref{alg-main} satisfies \eqref{eqres1} at convergence.

To derive the state evolution equation~\eqref{eq:se_gamp}, one can follow directly \cite[Sections 6.3 and 6.4]{zdeborova2016statistical} (see also \cite{troiani2024} for a symmetric version of the same GAMP-RIE algorithm). 
One considers the relaxed-BP algorithm \cite[Algorithm 1]{zdeborova2016statistical} with the substitution $f_a \to f_{\rm RIE}$ and $f_v \to \MMSE_{\rm denoising}$ -- notice that in \cite{zdeborova2016statistical} these functions are applied coordinate-wise on $R_t$, while here they are applied directly to the full matrix, as in \cite{zdeborova2016statistical} they consider factorised distributions $P_0$. Then, one can follow independently \cite[Section 6.3.2]{zdeborova2016statistical} to derive the GAMP-RIE algorithm (our Algorithm \ref{alg-main}) as an asymptotic approximation of the r-BP algorithm, and \cite[Section 6.4.1]{zdeborova2016statistical} to derive the state evolution equations for the GAMP-RIE algorithm from the r-BP algorithm.
Both derivations can be followed step-by-step with the mentioned substitutions.
The first of the state evolution equations is then found directly. For the second state evolution equation, one notices that
\begin{equation}
    R_t \overset{d}{=} \uS_* + \frac{1}{A_t} \uZ
\end{equation}
in distribution, where $\uS_*$ is the ground-truth and $\uZ$ is an i.i.d. Gaussian noise (both normalised to have $\caO(1)$ singular values).
Thus, $\uS_{t+1}$ is the BO estimate of $\uS_*$, and the associated MSE satisfies $c_{t+1} = \MMSE_{\rm denoising}(1/A_t) = 1 - q^{t+1}$, where $q^{t+1}$ is the overlap between the iterate $\hS_{t+1}$ and the ground-truth $\uS_*$, leading to the second, non-trivial, state evolution equation.
A more general treatment of state evolution for GAMP with non-separable denoisers is given in \cite{berthier2020state, gerbelot2023graph}.
 
We stress that, depending on the choice of prior and output channel, Algorithm \ref{alg-main} may not achieve the BO performance when initialised at zero overlap with the ground-truth. This is usually referred to as a computational-to-statistical gap~\cite{donoho2009message, rangan2011generalized, zdeborova2016statistical}, denoting a region where we have $\MMSE < 1$ (i.e. the BO estimator retrieves some information about the ground truth), while the AMP algorithm is stuck at a larger, possibly trivial $\MMSE$.
This can happen if among the solutions of \eqref{eqres1} there are multiple local maximisers of the associated free entropy \eqref{eq:freeentropy}. In that case, GAMP-RIE will find the local maximiser with smallest value $q$, while the BO performance will be given by the global maximiser. A gap arises if the local maximiser with smallest value $q$ is not the global maximiser.

For our case study, i.e. the BSR model with Gaussian label noise, we do not observe any such gap. In Appendix \ref{app.additional} we verify that the free entropy \eqref{eq:freeentropy} has a unique local maximum, and that our solver of \eqref{eqres1} finds that maximum, for a selection of values of $\beta, \rho, \alpha$. 
While not an analytical justification, our observations do not provide any hint to the existence of such a computational-to-statistical gap, allowing us to conjecture that no hard phase is present for this specific choice of prior. Other rotationally-invariant priors may still however exhibit computational gaps, but we leave such a study for future work.

We provide numerical experiments on the GAMP-RIE algorithm in Section \ref{sec:phenomenology}, Figure \ref{fig:amp}.
Notice that to improve the convergence of the algorithm we used a damped version, where the iteration for the variance and mean of $\Tr(\uS_{t}^T \tilde{X}^\mu)$, i.e. $V_t$ and $\omega_t$, becomes (e.g. for $V$) $V_t = (1-\gamma) c_t + \gamma V_{t-1}$
for some damping factor $0 < \gamma < 1$.
We also needed to fine tune the scale of the initialisation $c_{t=0} = \zeta,\omega_{t=0} = \zeta \times [1, \dots, 1]^T \in \bbR^N,V_{t=0} = \zeta$ to $\zeta \approx 20$ to obtain satisfactory results. 
We discuss how we tuned $\gamma, \zeta$ in Appendix \ref{app.additional}.

Notice that Algorithm \ref{alg-main} provides the Bayes optimal versions to the class of matrix denoisers considered in \cite{romanov_near-optimal_2018}, in which the authors approximate the RIE denoiser with different sub-optimal denoisers.

\subsection{MMSE in the low-width case}
\label{sec:low_rank}

We recall here the results of \cite{schulke16} for the MMSE in the low-width case with our notations.
In Appendix \ref{app:replicas} we re-derive the result of \cite{schulke16} in the more general case of correlated low-width priors.
In Appendix \ref{app:large_beta_int} we also derive the large $\beta$ limit of this result.

\begin{prevres}[MMSE for low-width BSR model \cite{schulke16}]\label{prevreslow}
    Consider the factorised Gaussian prior on $S$ \eqref{eq.prior} in the low-width high-dimensional limit, i.e. $d\to\infty$ with $\beta = \max(d,L) / \min(d,L)$ fixed and $r = \mathcal{O}(1)$. Define the sample ratio as 
    \begin{equation}
    \bar{\alpha} = \frac{n}{r (d+L)} = \frac{\beta}{\rho(1+\beta)} \alpha \, , \label{eq:two_alphas}
\end{equation} 
    Then, $\MMSE = 1-q$,  where $q$ is a solution to the non-linear system of equations
    \begin{equation}
        \begin{split}
            q &= g_1 g_2 \, , \\    
            g_1 &= \frac{(\beta+1)^2 \hq^2-\beta}{(\beta+1) \hq (\beta \hq+\hq+1)} \, ,
            \\
            g_2 &= \frac{(\beta+1)^2 \hq^2-\beta}{(\beta+1) \hq (\beta \hq+\hq+\beta)} \, ,
            \\
            \hq &= 
            \frac{\bar\alpha}{q} \int Dz \, dy \, \frac{\left(\del_z I_{\rm out}(z, y; q)\right)^2}{I_{\rm out}(z, y; q)} 
            \, ,
        \end{split}
    \end{equation}
    with the same $I_{\rm out}$ as in Result \ref{res1}.
\end{prevres}
Notice importantly that the result depends on the number of samples $n$ and the width $r$ only through the ratio $\bar{\alpha}$. Consequently, the dependence of the MMSE on the width $r$ is very simple. Compared to this, the MMSE in the extensive width regime depends on the widths in a richer way.

Notice that Result \ref{res1} (for the extensive-width case) and Previous Result~\ref{prevreslow} (for finite width) apply to different scalings for the number of samples $n$. In particular one can see that non-zero overlap in the low-width case happens on a sample scale $n = \mathcal{O}(rd)$, much smaller than the scale $n = \mathcal{O}(d^2)$ in the rotationally-invariant, extensive-width case. 

Previous Result \ref{prevreslow} is derived in the strictly intensive width regime $r = \mathcal{O}(1)$. The extension to all sub-extensive widths $r \ll d$ may be technically non-trivial, see \cite{barbier2024information, barbierMultiscale, pourkamali2024matrix} for related discussion in another model, but it turns out that Previous Result \ref{prevreslow} can be recovered as the limit $\rho \to 0$ (recall $r = \rho \min(d, L)$) of Result \ref{res1} (with an appropriately rescaled sample ratio), see Figure \ref{fig:varyrho}. 

We also note here that optimal algorithms based on message-passing for low-width priors have also been derived and discussed in \cite{schulke16}. 

Let us remark that our Result \ref{prop:mmse_BSR} completes the analysis of all scaling regimes for the BSR model. It turns out that only three non-trivial scaling regimes are present as a function of the width scaling, i.e. $r \ll \caO(d)$, $r = \caO(d)$ and $r \gg \caO(d)$.
Indeed, Figure \ref{fig:varybeta} that in the extensive width regime $r = \caO(d)$, when $r/d \to 0$, the performance reduces to that of the $r \ll \caO(d)$ regime. On the other hand, we show in Appendix \ref{app:large_rho_ext} that in the extensive width regime $r = \caO(d)$, when $r/d \to +\infty$,the performance reduces to that of linear regression, which is the Bayes optimal estimator for i.i.d. Gaussian priors (to which the factorized BSR prior converges in the large width limit).

\section{Consequences of main results}\label{sec:cons}

In this section we study the Bayes Optimal test error, i.e.\ the minimum theoretically-achievable test error, for the BSR model and Gaussian label noise, as a function of the parameters $\beta = \max(d,L) / \min(d,L)$ (the sequence length to embedding dimension ratio), $\rho = r / \min(d,L)$ (the width ratio, note that a lower $\rho$ corresponds to a more structured prior), $\alpha = n / (dL)$ (the number of samples ratio) and $\Delta$ (the label noise). We identify the corresponding phase transitions in the performance, thus answering question \ref{q3} from the introduction.

We will also compare the BO estimator to some other baseline algorithms. Importantly, in Section \ref{sec:regression}, we compare with the linear regression on the vectorized sequence of tokens, thus quantifying the advantage of an estimator specializing in sequences of tokens over basic linear regression, which answers question \ref{q4} from the introduction. 

Again, notice that the factorized nature of the prior \eqref{eq.prior} enters explicitly in all our results, typically through the dependence on the  width parameter $\rho$ of the BSR model.

\subsection{General phenomenology of the Bayes-optimal performance}\label{sec:phenomenology}

Figure \ref{fig:varyrho} and \ref{fig:varybeta} show the BO test error as a function of the sample complexity $\alpha$, for several values of $\beta$ and width $\rho$, and in the noiseless setting $\Delta = 0$. We observe that lower values of $\rho$ and larger values of $\beta$ (i.e. larger structure in the prior) lead to lower test error. We also observe that the BO estimator can achieve zero test error at a finite value $\alpha_{\rm BO} < 1$, defining a threshold called the \textit{strong recovery threshold} that we discuss in detail in Section \ref{sec:strong}.

We provide analogous data for $\Delta > 0$ in Appendix \ref{app.additional}. The overall phenomenology is similar, with the important difference that we observe no strong recovery at finite $\alpha$, and the test error in the noisy case to be a continuous and differentiable function of the parameters. 

In Figure \ref{fig:varyrho}, right column, we highlight the convergence of the extensive-width test error to the low-width result for $\rho \to 0$, after appropriately rescaling the sample ratio $\alpha$ to $\bar{\alpha} = \frac{\beta}{\rho(1+\beta)} \alpha$.
We observe quantitative differences from the low-width result for width ratios as low as $0.05$, stressing the fact that the extensive width analysis is relevant in finite-size applications, where $\rho$ may be small, but not strictly vanishing. 

Figure \ref{fig:amp} shows numerical experiments on GAMP-RIE with $\max(d,L) = 100$ for $\beta = 1,2$, $\rho = 0.2, 1$ and $\Delta = 0, 0.1$, comparing it with the MMSE obtained by solving \eqref{eqres1}. We observe a very nice agreement already at these moderate system sizes.

\begin{figure*}

    \includegraphics{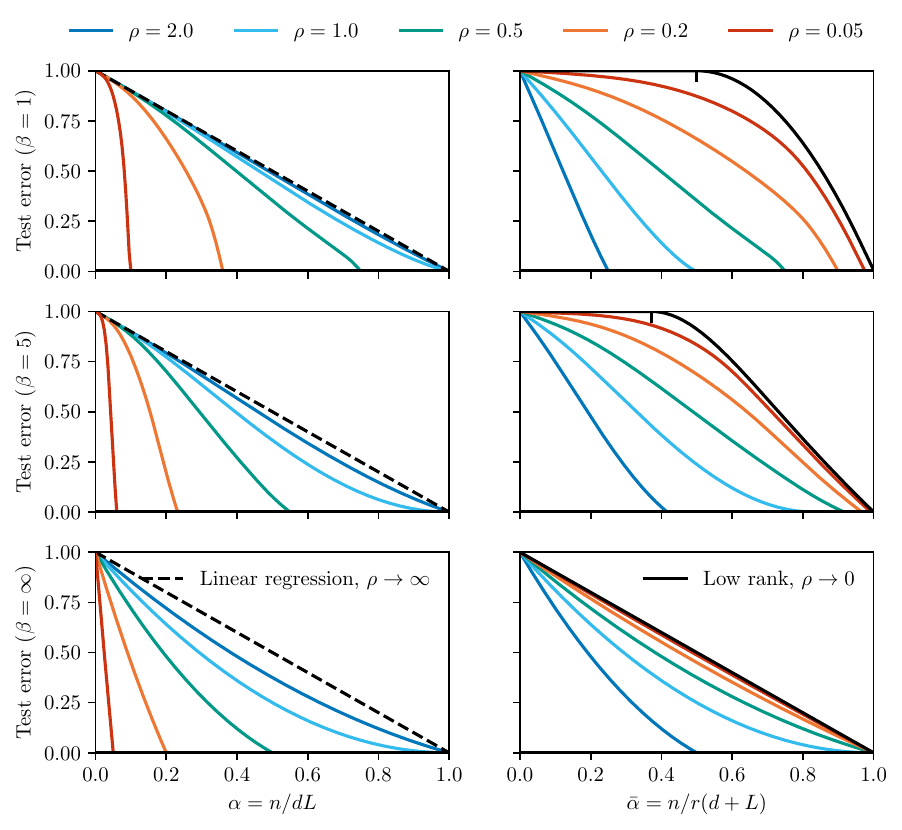}
    
    \caption{
    Bayes Optimal test error for the BSR model with noiseless output channel ($\Delta = 0$) as a function of the sample ratio $\alpha = n/(dL)$ (left column) and of the low-width sample ratio $\bar{\alpha} = n / [r(d+L)]$ (right column). We plot a different value of the aspect ratio $\beta = \max(d,L) / \min(d,L) = 1, 5, +\infty$ for each row from top to bottom, and in each panel compare several values of the width ratio $\rho = r / \min(d,L) = 0.05, 0.1, 0.2, 0.5, 1, 2$ (colored solid lines). In the left column, we also plot for comparison the performance of optimally-regularised linear regression (in this case, $\lambda \to 0^+$) on the vectorized data (it does not depend on $\rho$ and $\beta$) in the black dashed line, which corresponds also to the BO error for $\rho \to \infty$.
    We observe that the BO test error is always better than the linear regression test error, and that it gets better and better as $\rho$ decreases: the more structure in the distribution of the signal, i.e. the lower the width, the better one can estimate it. We also observe that the BO test error vanishes at a finite value of $\alpha$, the so-called strong recovery threshold, and that this threshold is smaller than one for $\rho < 1$. In this regime, there are values of $\alpha$ for which the BO estimator achieves zero test error, while the linear regression estimator has a non-zero test error.
    The middle and bottom panel show the same overall phenomenology as $\beta$ increases from 1 to infinity.
    The right column shows the same curves as a function of the low-width sample ratio $\bar{\alpha} = n / [r(d+L)]$, comparing with the already known low-width BO test error (solid black line) \cite{schulke16}. We observe a clear convergence to the low-width error curve as $\rho \to 0$, but we highlight that, for e.g. at $\beta = 1$, the test error of the BO estimator is still quantitatively better than its low-width counterpart already at $\rho = 0.05$.
    Notice also that for $\rho \to 0$ the BO estimator has a weak recovery threshold at $\bar{\alpha}_{\rm weak} = (1+\Delta)\sqrt{\beta}/(1+\beta)$, i.e. below it has the same performance as the zero estimator ${\hat{S}}_{\rm zero}(\caD) = 0$. As soon as $\rho > 0$, the weak recovery threshold disappears, allowing for better-than-trivial performance at all values of $\bar{\alpha}$. The weak recovery threshold is marked by a vertical black marker: notice that for $\beta \to \infty$ the weak recovery threshold is at zero.
    The Bayes-optimal curves are plotted using Result \ref{prop:mmse_BSR} and \eqref{eqres1} for extensive width and Previous result \ref{prevreslow} for intensive width. Linear regression is plotted using Previous result \ref{prevres-regr}.
    }
    \label{fig:varyrho}
\end{figure*}

\begin{figure*}

    \includegraphics[width=\columnwidth]{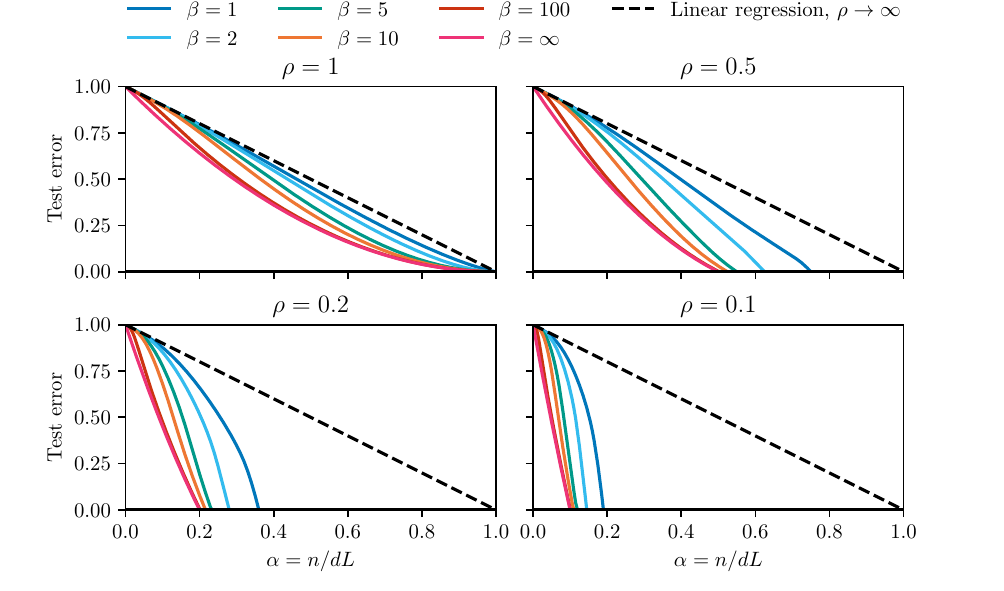}
   
    \caption{
    Bayes Optimal test error for the BSR model with noiseless output channel ($\Delta = 0$) as a function of the sample ratio $\alpha = n/(dL)$. We plot a different value of the
    width ratio $\rho = r / \min(d,L) = 0.05, 0.5, 1, 2$ in each panel, and several values of the aspect ratio $\beta = \max(d,L) / \min(d,L) = 1, 5, +\infty$ (colored solid lines) in all panels. The black dashed line is the performance of optimally-regularised linear regression on the vectorized data. Again we observe that the more structured signals (larger $\beta$ and smaller $\rho$), the better the achieved test errors.
    The Bayes-optimal curves are plotted using Result \ref{prop:mmse_BSR} and \eqref{eqres1}. Linear regression is plotted using Previous result \ref{prevres-regr}.}
    \label{fig:varybeta}
\end{figure*}

\begin{figure*}

    \includegraphics{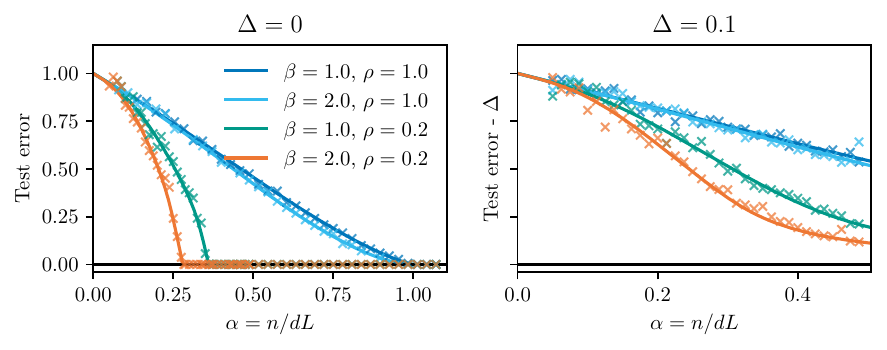}
    
    \caption{
    Comparison between the BO test error and the test error of GAMP-RIE (Algorithm~\ref{alg-main}) for two choices of the aspect ratio $\beta = \max(d,L) / \min(d,L) = 1, 2$ and the width ratio $\rho = r / \min(d,L) = 0.2, 1$ both in the noiseless $\Delta = 0$ (left) and noisy $\Delta  = 0.1$ (right) case. Solid lines are the theoretical prediction from \eqref{eqres1}. The crosses represent numerical experiments for the test error measured after iterating GAMP-RIE until convergence, on instances of size $\min(d,L) = 100$, with initialisation from the prior distribution. Each point is a run over a single realisation of the data and ground truth. 
    The Bayes-optimal curves are plotted using Result \ref{prop:mmse_BSR} and \eqref{eqres1}. The performance of GAMP-RIE is given by \eqref{eq:est} applied to the output of Algorithm \ref{alg-main}.
    \label{fig:amp}
    }
    
\end{figure*}

Result \ref{prop:mmse_BSR}, combined with Previous Result \ref{prevreslow}, provide the full picture for the Bayes-optimal test error in the BSR model, completely solving Question~\ref{q1}. Additionally, Figure \ref{fig:varyrho} and \ref{fig:varybeta} showcase the phenomenology for the noiseless observation channel. 
We have also answered positively to Question \ref{q2}, as the GAMP-RIE algorithm efficiently achieves the BO error in the high-dimensional limit.

\subsection{Strong and weak recovery thresholds}\label{sec:strong}

In the noiseless output channel  
we can provide an explicit characterisation of the strong recovery threshold, i.e. the value of $\alpha_{\rm BO}$ such that for all $\alpha > \alpha_{\rm BO}$, zero test error is achieved.
\begin{res}[BO strong recovery threshold]
\label{res:strong}
    Consider the same setting as in Result \ref{res1}, and specify it to the BSR model \eqref{eq.prior} and noiseless output channel.
    Then, in the high dimensional limit, the strong recovery threshold satisfies
    \begin{equation}\label{strong}
        \alpha_{\rm BO} = 
        \begin{cases}
           \frac{\rho}{\beta} \left( 1 + \beta - \rho  \right) & 0 < \rho < 1 \, , \\ 
           1 & \rho \geq 1 \, .\\
        \end{cases}
    \end{equation}
\end{res}
The derivation of the threshold is performed in Appendix \ref{app:strong_ext}, and involves expanding Result~\ref{res1} in the limit $q \to 1^-$ and $\hq \to +\infty$.
We recall that for non-zero label noise $\Delta$, the strong recovery threshold is at infinity.

For comparison, the strong recovery threshold in the low-width limit equals $\lim_{\rho \to 0} \alpha_{\rm BO} = 0$, 
and in the more appropriate low-width sample scaling
\begin{equation}
    \lim_{\rho \to 0}
    \bar{\alpha}_{\rm BO} 
    = \lim_{\rho \to 0}
    \frac{\beta}{\rho (1+\beta)} \alpha_{\rm BO}
    = 1 \, ,
\end{equation}
which can be derived either by taking the limit of \eqref{strong}, or independently by taking Previous Result \ref{prevreslow}, and solving the corresponding equations, as we do in Appendix \ref{app:strong_int}.

We remark that the strong recovery threshold can be guessed (but not properly justified) also through a counting argument where we compare the number of observations with the number of degrees of freedom (see also \cite{romanov_near-optimal_2018}). 
The spectrum of $S^*$ accounts for a number $\mathcal{O}(d)$ of degrees of freedom. The singular vectors of $S^\star$ are a set of $r$ orthonormal vectors in dimension $d$, and one in dimension $L$. 
It is known that the set of $1 \leq r \leq d$ orthonormal vectors in dimension $d$, as a manifold (called the Stiefel manifold), has dimension \cite{helmke2012optimization}
\begin{equation}
    \text{dim}(r, d) = dr - \frac{r(r+1)}{2} \, .
\end{equation}
Thus, by a dimensional argument the number of samples needed to learn such bases for $r \leq \min(d, L)$ (i.e. $\rho \leq 1$) should equal
\begin{equation}
    n = dr - \frac{r(r+1)}{2} + Lr - \frac{r(r+1)}{2} + \mathcal{O}(d) = (d+L) r - r(r+1) + \mathcal{O}(d) \sim \frac{\rho}{\beta} (1+\beta - \rho) \, dL \, .
\end{equation}
This counting argument recovers the analytically derived threshold \eqref{strong}, and hints to the fact that this threshold will be universal to a larger subset of rotationally invariant priors with rank constraint, not limited to the BSR model for which our derivation of \eqref{strong} holds.

We now turn to the weak recovery threshold. 
 Recall that $\alpha_{\rm weak}$ is the largest $\alpha$ such that the performance of the BO estimator is the same as the performance of randomly sampling the prior. 
 In the extensive-width case the weak recovery threshold $\alpha_{\rm weak}$ is trivial.
 In other words, $\alpha_{\rm weak} = 0$ for $\rho > 0$, and for any $\alpha > 0$ non-trivial recovery is achieved. Instead, in the low-width case, a non-trivial weak recovery threshold arises (also for positive label noise $\Delta$) at
\begin{equation}
    \bar{\alpha}_{\rm weak}  =  (1+\Delta) \frac{\sqrt{\beta}}{1+\beta} \, .
\end{equation}
We derive this threshold in Appendix \ref{app:strong_int}.

This section gives a clear answer to Question \ref{q3}. For noiseless output channels, the BO test error has a sharp threshold, corresponding to a second-order phase transition, between a region of positive error (at a small number of samples) and a region of zero error (at a large number of samples). Result \ref{res:strong} pinpoints the sample-complexity $\alpha$ of the transition analytically. This, together with previous results for the low-width case, provides a full picture of the transition in the noiseless BSR model. 

Notice that our result for the BO strong recovery threshold agrees with lower bounds discussed in \cite[Appendix 4.3]{romanov_near-optimal_2018} and originally shown in \cite{donoho2014minimax} in a related context of matrix denoising. This, along with Algorithm \ref{alg-main}, settles the question posed by \cite{romanov_near-optimal_2018} whether an AMP algorithm can reach the naive dimensional lower bound for the strong recovery threshold. We answer positively, and provide an AMP which not only achieves optimal recovery threshold, but that is also optimal at all sampling ratios $\alpha$.

\subsection{Comparison with linear regression on the vectorized data}\label{sec:regression}

As a crucial baseline motivating this work, we consider here the performance of linear regression performed on the vectorized input data. In the context of learning sequences of tokens, this amounts to flattening the data matrix $X_{ij} \in \bbR^{L \times d}$ into an $Ld$-dimensional vector, thus losing the semantic separation between token space and embedding space. 
The performance of such a procedure is quantified below. 

Comparing this baseline to the Bayes-optimal performance of the BSR model quantifies the gain one can get when performing learning using a specialized sequence model as opposed to vectorizing the data and using fully connected neural networks (of which linear regression is the simplest example), effectively discarding some prior information.

\begin{prevres}[Performance of ridge regression for Gaussian output channels and arbitrary priors]\label{prevres-regr}
    Consider the ridge regression estimator
    \begin{equation}
        \hS_{\rm ridge}(\caD) = \argmin_S \left[ \frac{1}{2} \sum_{\mu} \left(y_\mu - (Ld)^{-1/2} \Tr(S^TX^\mu) \right)^2 + \frac{\lambda}{2} \Tr(S^T S) \right] \, ,
    \end{equation}
    and a dataset $\caD$ generated by \eqref{eq.model} with Gaussian label output channel with variance $\Delta$, and arbitrary prior $P_0$ normalised such that $Q_* = 1$. Define the sample ratio as $\alpha = n / (dL)$.
    Then, the optimal value of the regularisation is $\lambda_{\rm opt} = \Delta$, and the mean square estimation error of the optimally-regularised ridge regression estimator equals (in the large-dimensional limit)
    \begin{equation}
        \MSE^{\rm ridge}_{\alpha, \Delta} = 
        \frac{1+\alpha+\Delta-\sqrt{(\alpha+\Delta+1)^2-4 \alpha}}{2} \, ,
    \end{equation}
    which reduces to $\MSE^{\rm ridge}_{\alpha, \Delta= 0} = \max(1 - \alpha, 0)$ in the noiseless case.
\end{prevres}
The analysis of empirical risk minimisers (ERMs) for convex losses, and in particular for ridge regression, is standard, see \cite{loureiro2021learning} for example for a very generic derivation. We only point out here that the prior, arbitrarily complicated and with extensive width, enters this performance only through its second moment $Q_* = 1$ due to the choice of $\ell_2$ regularisation. This can also be seen directly by the explicit solution of the ridge regression problem, which notably depends only on the second-order statistics of the data and labels.

Figure \ref{fig:varyrho} shows the BO test error as a function of $\alpha$ for several values of $\beta$ and $\rho$, and compares with the performance of linear regression (which is independent on $\rho$ and $\beta$ in the scaling we chose).
We see that in all cases, larger values of $\beta$ and smaller $\rho$ lead to more significant gains in using the prior-aware BO estimator compared with the simple ridge estimator, both in noiseless ($\Delta = 0$) and noisy ($\Delta > 0$, see Appendix \ref{app.additional}) cases. 

Notice that in the noiseless case we see that $\MSE^{\rm ridge}_{\alpha, \Delta= 0} = 0$ at $\alpha = 1$, recovering the trivial fact that an invertible linear system of $p$ equations in $p$ unknowns has a unique solution. More generally, the strong recovery threshold for ridge regression with noiseless data is given by $\alpha_{\rm ridge} = 1$.
Whenever $\rho < 1$, there exists a full range of sample ratios $\alpha_{\rm BO} < \alpha < \alpha_{\rm ridge}$ where the BO estimator achieves zero test error, while the ridge estimator does not.

As $\rho\to \infty$, $S$ 
converges to a matrix with i.i.d. standard Gaussian entries: the problem is effectively vectorised, as there remain no correlations between the token and embedding dimensions $L$ and $d$. 
In this limit, we expect that optimally-regularised ridge regression will achieve the BO performance, as the loss/regularisation choice matches the distribution on $S$ and the generative process for the labels. This is indeed the case, as we show in Appendix \ref{app:large_rho_ext} by explicitly computing the large-$\rho$ limit of \eqref{eqres1} for the BSR model.  

This answers Question \ref{q4} from the introduction: vectorizing data and learning it with linear regression is suboptimal for any finite $\rho$. For $0<\rho<1$, the suboptimality is particularly striking, as there exists a full region of sample ratio $\alpha_{\rm BO} < \alpha < 1$ in which linear regression has non-zero test error, while the BO estimator achieves zero error.

\subsection{Comparison with a minimal nuclear norm estimator}

\begin{figure*}
    \includegraphics{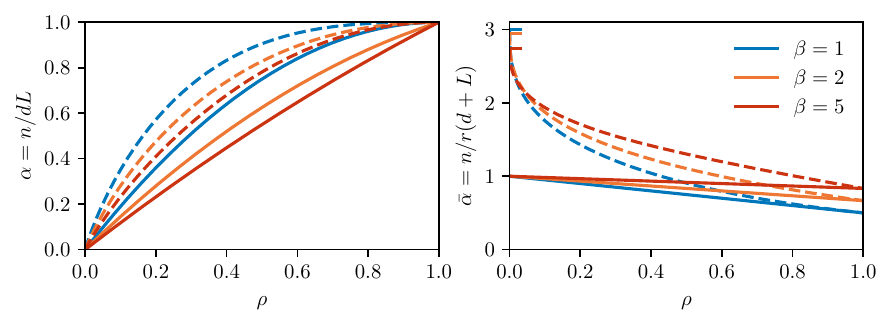}
   
    \caption{
        Comparison between the BO strong recovery threshold (Result \ref{res:strong}, solid lines) and the MNNE strong recovery threshold (Previous Result \ref{mnne_res}, dashed lines) for $\beta = \max(d,L) / \min(d,L) = 1, 2, 10$ as a function of the width ratio $\rho = r / \min(d,L)$.
        In the left panel we plot the strong recovery thresholds in the scaling $\alpha = n/(dL)$, natural in the extensive-width case $\rho >0$.
        In the right panel we plot the same data in the low-width sample scaling $\bar\alpha = n / [r(d+L)]$, highlighting the strong suboptimality of MNNE at low ranks/widths.
        The colored markers on the vertical axis highlight the finite $\rho \to 0$ limit of the strong recovery threshold of MNNE, as given in  \eqref{eq-mnne-strong-low}.
    }
    \label{fig:mnne_strong}
\end{figure*}

As discussed in the introduction, previous works explored algorithms based on nuclear norm minimization to solve the matrix sensing/denoising problem  \cite{recht2010guaranteed,Donoho_2013,donoho2014minimax}. It is instructive to compare the performance of this algorithm to the optimal estimator. 
The \textit{minimal nuclear norm estimator} (MNNE) is defined as
\begin{equation}
    S_{\rm MNNE} \coloneqq \argmin ||S||_{\rm nuc} = \argmin \Tr( \sqrt{S S^T} )
    \quad\text{such that}\quad
    y^\mu = \frac{1}{\sqrt{Ld}} \sum_{a, i=1}^{L, d} X^\mu_{ai} S_{ia} \, ,
\end{equation}
where we recall that the nuclear norm is just the sum of the singular values of a given matrix. This algorithm is the convex relaxation of the minimum rank estimator, where one seeks a matrix with minimal rank fitting the dataset. It has been observed \cite{Donoho_2013,donoho2014minimax} that the MNNE can achieve zero estimation error, provided that the ground-truth matrix $S^*$ has constrained rank (i.e. $0 < \rho < 1$), and that the number of samples $n$ is large enough. The authors are also able to characterise the corresponding strong recovery threshold $\alpha_{\rm MNNE}$, and provide an explicit asymptotic value, which we report here for completeness.

\begin{prevres}[Strong recovery threshold for the MNNE \cite{Donoho_2013,donoho2014minimax}]\label{mnne_res}
    Consider the Mar\u{c}enko--Pastur distribution defined by
    \begin{equation}
    \label{mp:eq} p_\gamma(t) \coloneqq \frac{1}{2\pi\gamma t} \sqrt{(
    \gamma_{+} - t) (t-\gamma_{-}}) \cdot\mathbf{1}_{[\gamma_-,\gamma_+]}(t),
    \end{equation}
    where $\gamma_{\pm}= ( 1\pm\sqrt{\gamma}  )^2$, 
    and define its complementary incomplete moments as
    \begin{equation}
    \label{MP:eq} P_{\gamma}(x;k) \coloneqq \int_x^{\gamma_+}
    t^{k} p_\gamma(t) \,dt.
    \end{equation}
    Let
    \begin{equation}
            \mathbf{M}(\Lambda;\rho,\beta) 
            \coloneqq \rho(1 + \beta - \rho) + (\beta-\rho)
            \biggl[ \rho\Lambda^2 +(1-\rho) \biggl( P_\gamma\bigl( \Lambda^2 ; 1
                \bigr) - 2\Lambda P_\gamma\biggl( \Lambda^2 ;
                \frac{1}{2}\biggr) + \Lambda^2 P_\gamma\bigl(
                \Lambda^2 ; 0\bigr) \biggr) 
            \biggr] \, ,
    \end{equation}
    for $0<\rho<1$ and $\beta \geq 1$, with
    \begin{equation}
        \gamma = \frac{1-\rho}{\beta - \rho} \, .
    \end{equation}
    Then the strong recovery threshold of the MNNE satisfies
    \begin{equation}
        \alpha_{\rm MNNE} = \min_{ 0 \leq \Lambda \leq \gamma_+(\rho,\beta)} \mathbf{M}(\Lambda;\rho,\tilde{\beta})  \, .
    \end{equation}
    The minimum can be computed numerically by solving the zero-derivative condition $d\mathbf{M} / d\Lambda = 0$ on the interval $(0, \gamma_+)$, i.e. 
    \begin{equation}
    P_\gamma\biggl(\Lambda^2;\frac{1}{2}
    \biggr) - \Lambda\cdot P_\gamma\bigl(\Lambda^2;0\bigr) =
    \frac{\Lambda\rho}{
    1-\rho} \, .
    \end{equation}
    This can be done by a bisection algorithm.
\end{prevres}
In Figure \ref{fig:mnne_strong}, we plot the theoretical prediction of $\alpha_{\rm MNNE}$ from \cite{Donoho_2013, donoho2014minimax} and compare it with the BO threshold $\alpha_{\rm BO}$ that we derived in Section \ref{sec:strong} for different values of $\beta$. We observe that for all values of $0 < \rho < 1$ and $\beta \geq 1$ the two thresholds are different, and in particular $\alpha_{\rm BO} < \alpha_{\rm MNNE}$. 
This highlights an intrinsic suboptimality of the MNNE, which for an extended range of values $\alpha_{\rm BO} < \alpha < \alpha_{\rm MNNE}$ fails to achieve the BO performance.

In the low-rank regime $\rho \to 0$, \cite{Donoho_2013} provide the following asymptotic value for the strong recovery threshold of the MNNE
\begin{equation}\label{eq-mnne-strong-low}
    \bar\alpha_{\rm MNNE} 
    =
    \lim_{\rho \to 0} 
    \frac{\beta}{\rho (1+\beta)} \alpha_{\rm MNNE} 
    = 
    2 \left(1 + \frac{\sqrt{\beta}}{1+\beta}\right) \, .
\end{equation}
while for the BO strong recovery threshold we have
$\lim_{\rho \to 0} \bar\alpha_{\rm BO} = 1$.
We thus see that also in the low-rank limit (with the appropriately rescaled sample ratio) the MNNE recovery threshold remains suboptimal. This is akin to what happens in the compressed sensing problem when we compare the Bayes-optimal performance to the performance of the convex relaxation via $L_1$ regularization~\cite{krzakala2012statistical}.
In Appendix \ref{app.additional}, we provide numerical experiments comparing the performance of the MNNE estimator with the BO test error, with GAMP-RIE, and with the prediction for the strong recovery threshold (Previous Result \ref{mnne_res}).

\section{Behaviour of gradient descent}
\label{sec:GD}

Arguably, the most interesting algorithm to study in the context of the BSR model is \textit{gradient descent} (GD) since its variants are the driving horse of state-of-the-art applications of machine learning. We will consider here the BSR model with Gaussian additive noise channel and assume the width parameter $r$ is known. For this case the most natural choice of loss function is 
\begin{equation} \label{eq:loss}
    \mathcal{L}(U, V) = \frac{1}{4}\sum_{\mu = 1}^n\left(y^\mu - \frac{1}{\sqrt{Ldr}} \sum_{a, i=1}^{L, d} X^\mu_{ai} \sum_{j=1}^r U_{ij}V_{ja}\right)^2 \, ,
\end{equation}
where $U \in \mathbb{R}^{d \times r}$ and $V\in \mathbb{R}^{r \times L}$. The loss is then minimized  over the factors $U$ and $V$ using the following gradient descent iterations
\begin{equation}\label{eq.iter}
    U^{t+1} = U^t - \eta \nabla_{U} \mathcal{L}(U^t, V^t)
    \mathand
    V^{t+1} = V^t - \eta \nabla_{V} \mathcal{L}(U^t, V^t) \, .
\end{equation}
with $\eta > 0$ being the learning rate. Unlike in linear regression, the loss \eqref{eq:loss} is non-convex, and thus keeping particular attention to the initialisation, learning rate and the stopping criterion is required to properly understand the properties of the GD estimator. In general non-convex settings the generalization performance of the GD algorithms is mostly a widely open question that is actively studied.  

In this section, we initiate the understanding the performance of the GD in the BSR model, and investigate how the choice of initialization and learning rate influence the performance of the algorithm, thus providing some answers to \ref{q5} from the introduction. We argue that the BSR model is a simple yet very interesting model to further the understanding of the broad set of questions behind the functioning of the GD algorithm. 
Without exhaustively mapping the possible choices of initialization, learning rate and stopping time, we identify two remarkable properties and discuss them further below:
\begin{itemize} 
   \item {\bf GD can reach the Bayes-optimal performance:} For the noiseless BSR model we find that for well-chosen learning rate and factors initialized 
   in the prior distribution, 
   GD behaves \textit{as if} it 
   was sampling uniformly the space of global minimizers, and an averaged version of GD (defined in \eqref{eq:averaged_GD}) reaches the Bayes-optimal generalization error. We stress that there is no \textit{a priori} reason for GD being able to sample the minimizers, this observations is thus very surprising. When noise is present, the behaviour is more complex, and GD does not seem to sample the minimizers anymore.
   Numerical evidence is given in Fig. \ref{fig:GD-BO}. 
   Note that similar properties were observed for a related model in \cite{troiani2024}. 
   \item {\bf Implicit regularization of GD, but not with respect to the minimum nuclear norm:} We find that many choices of the learning rate and initialization lead to a generalization performance that is better than the one of a randomly-chosen global minimizer. An interesting existing line of work proposed that in some settings the implicit regularization may be related to the nuclear norm \cite{gunasekar2017implicit}. We thus ask whether this would be the case in  the BSR model. Our numerical investigation suggests that in the BSR model, even with small learning rate and small amplitude initialization, GD does not minimize the nuclear norm. This is evidenced in Fig.~\ref{fig:GD-Nuclear}. 
\end{itemize}
More work is needed to fully characterize the behaviour of the GD algorithm in the BSR model and this characterization is a prerequisite for further understanding of learning dynamics in more complex sequence models.

\subsection{GD and the Bayes-optimal performance
}\label{sec:GDprior}

\begin{figure*}

    \includegraphics[width=\columnwidth]{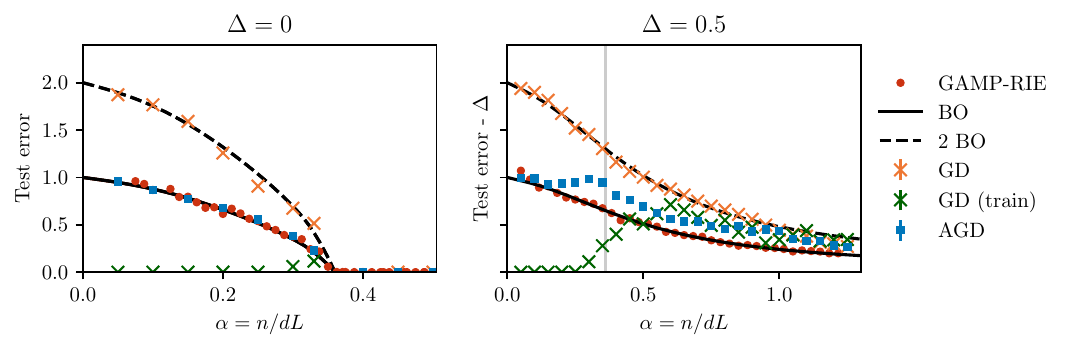}
   
    \caption{
        Comparison between the test error achieved by GD and AGD initialised in the prior and of the BO test error for $\beta = 1$, $\rho = 0.2$ and
        $\Delta = 0,0.5$  (left and right panels respectively). In the noisy case, we depict the test error minus the variance of the noise $\Delta$. 
        Solid lines are the BO test error, dashed lines are twice the BO test error corresponding to the error of the Gibbs sampler. Orange crosses are numerical experiments for the test errors at the end of the run of GD for $d=L = 100$, maximum number of steps $\tau = 50000$, and runs are averaged over 16 instances of the data. 
        Blue squares are numerical experiments for the test errors at the end of the run of AGD (averaged over 32 initial conditions) and they are averaged over 2 instances of the data (8 in the right panel up to $\alpha = 0.6$). 
        The error bars denoting standard error on the mean are negligible.
        In both cases, a fine-tuned value of the learning rate $\eta(\alpha)$ must be used, dependent on the sample ratio $\alpha$. 
        We provide the values used to generate this plot in Appendix \ref{app.additional}. The green crosses mark the value of the training loss at the end of the training for GD. The grey vertical line in the right-hand panel marks where the number of samples equals the number of degrees of freedom.
        Finally, red dots are numerical experiments for GAMP-RIE, with a single random instance of $d= L  = 100$.
        We observe that in the noiseless case $\Delta = 0$ (left), GD achieves a test error compatible with the error of the Gibbs sampler and that AGD achieves a test error compatible with the BO test error.
        Instead, for $\Delta = 0.5$ (right) we observe that AGD does not reach the BO error, and moreover it trivialises (namely, all differently-initialised runs of GD converge to the same estimator) for $\alpha$ large enough, roughly around $\alpha \approx 1$ here.  
        We show qualitatively similar comparisons at $\beta = 2$ in Appendix \ref{app.additional}. 
        The Bayes-optimal curves are plotted using Result \ref{prop:mmse_BSR} and \eqref{eqres1}. The performance of GAMP-RIE, GD and AGD are given by \eqref{eq:gen} applied to the output of the respective algorithm.
    }
    \label{fig:GD-BO}
\end{figure*}

In Fig. \ref{fig:GD-BO} we initialise both $U$ and $V$ as i.i.d.\ Gaussian matrices, with each entry having mean zero and unit variance, i.e.\ from the same distribution as in the BSR model \eqref{eq:teacher}. 
Figure \ref{fig:GD-BO}, left panel, shows that GD run on the BSR model without noise reaches very small training loss if the learning rate $\eta$ is properly tuned, and a test error equal to the one of a uniformly sampled global minimum of the loss  (this is usually called a Gibbs sampler for the posterior distribution, and has test error equal to twice the BO test error as we show in \eqref{eq-gibbs-error}). 
This prompts us to speculate that runs of GD with independent initialization may be close to sampling the space of global minimizers of the loss (as a Gibbs sampler would do).
Notice that GD can only converge to the boundary of the set of global minimizers, and that in high-dimension the uniform measure over such set is plausibly concentrated on the boundary, provided that the set of global minimizers is not pathological.
This observation, plus the numerical observation that the test error matches the Gibbs sampler, justifies our speculation.

The Bayes-optimal estimator in the noiseless case is given by averaging over the global minimizers of the loss. 
Given our hypothesis above that GD samples uniformly the set of global minimizers, we are prompted
to 
average $J$ GD runs to construct a novel estimator.
If our hypothesis is correct, then this averaged estimator should achieve close to BO performance. Thus, we sample $J$ different pairs of initial matrices $(U_j^0, V^0_j)$, iterate GD for $\tau$ iterations obtaining $(U_j^\tau, V^\tau_j)$ to compute the estimator of $S$ as 
\begin{equation}\label{eq:averaged_GD}
    \hat{S}_{\rm GD, avg} = \frac{1}{J}\sum_{j=1}^J \frac{U^\tau_j V^\tau_j}{\sqrt{r}}
\end{equation}
We call this the \textit{averaged GD} (AGD) algorithm.
Figure \ref{fig:GD-BO} shows that this AGD estimator  indeed reaches the Bayes-optimal test error.
This observation holds for different values of $\beta$.

Figure \ref{fig:GD-BO}, right panel, shows experiments in the same setting, but with label noise $\Delta = 0.5$. We observe here a different phenomenology, 
in particular with AGD not reaching the BO test error. We notice also that the training error equals zero before the naive interpolation threshold where the number of samples equals the number of degree of freedom ($\alpha = \alpha_{\rm BO}$ in the BSR model). At the interpolation threshold AGD has a trace of what may be an interpolation peak, which does not appear in the simple GD.
More work is needed to fully understand whether there is a way to tune the parameters of the AGD algorithm to reach the BO error also in the noisy case. 

\subsection{Implicit regularization of GD, comparison to the minimum nuclear norm}\label{sec:GDsmall}

\begin{figure*}
    
    \includegraphics{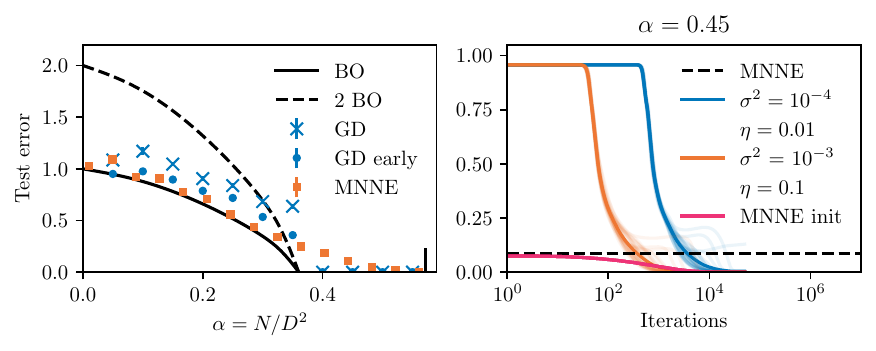}
    
    \caption{
     (Left) Comparison between the test error achieved by GD initialised with small norm (blue crosses, $d=L=50$, maximum iterations $T = 10^4$, learning rate $\eta = 0.2$, initialisation norm $\sigma^2 = 10^{-4}$) and its early stopped version (blue dots) averaged over $16$ instances, the test error of the MNNE (orange squares, $d=L= 50$) and of the BO test error (solid black line, dashed lines is twice the BO error corresponding to the error of the Gibbs sampler) for $\beta = 1$, $\rho = 0.2$ and $\Delta = 0$.  
     We observe that the MNNE performs slightly better than GD with small initialisation up to roughly the BO strong recovery thresholds, while for larger $\alpha$ GD becomes better, and notably has a better strong recovery threshold than MNNE.
     (Right) Comparison between the MNNE and several runs of GD with small initialisation (maximum iterations $\tau = 20000$), all on the same instance of the data and ground-truth with $d =L = 50$, $\rho = 0.2$ and $\Delta = 0$. We perform this comparison at $\alpha = 0.45$, where the left panel suggests that GD will achieve zero error, while MNNE does not. We run GD from 2 different initialisation magnitudes $\sigma^2 = 10^{-3}, 10^{-4}$ (orange and blue curves, thick curves mark the average), and we also run GD from the MNNE initialisation (see main text for a precise definition). We observe that in all cases GD outperforms MNNE at convergence.
     The Bayes-optimal curves are plotted using Result \ref{prop:mmse_BSR} and \eqref{eqres1}. The performance of MNNE and GD are given by \eqref{eq:gen} applied to the output of the respective algorithm.
    }
    \label{fig:GD-Nuclear}
\end{figure*}

The data reported in Fig.~\ref{fig:GD-BO} depend strongly on the choice of the initialization and the learning rate. We will consider initialization where the component of matrices $U^0$ and $V^0$ are still iid Gaussian random variables of zero mean, but this time with variance $\sigma^2$. Fig.~\ref{fig:GD-BO} was for $\sigma^2=1$, but initializing with small $\sigma^2$  is considered more interesting, one reason being that the BO estimator at $\alpha=0$ is simply $0$. 
In Figure \ref{fig:GD-Nuclear}, left panel, we show that GD initialised with small norm $\sigma^2 = 10^{-3}$ or $10^{-4}$ reaches a test error which is slightly larger than the BO one, but still significantly smaller than initialising at $\sigma^2 = 1$ (where non-averaged GD reaches the Gibbs sampling error). When GD leads to a better test error than a random global minimizer the machine learning literature often refers to so-called \textit{implicit regularization} of gradient descent \cite{neyshabur2014search,gunasekar2017implicit,li2020towards,arora2019implicit}. What we observe in Fig.~\ref{fig:GD-Nuclear} is a clear sign of implicit regularization. In this case the role of the learning rate is less influential, as long as it is kept small enough (see Figure \ref{fig:learning_rate_compare} in Appendix \ref{app.additional}). We also find that for small initialization early stopping can be advantageous, as also reported in the figure. 
    
It is natural to compare GD with small initialisation to the MNNE, as it was suggested that in some settings (in particular over-parametrized ones), gradient flow with vanishing norm at initialisation has an implicit bias towards minimising the nuclear norm \cite{gunasekar2017implicit}. Later work then questioned this conjecture, and disproved it in a constructed special case \cite{li2020towards}. 
In the BSR model, we find numerically that GD starting with small initialization does not go to the minimizer corresponding to the smallest nuclear norm. Evidence for this is reported in Fig.~\ref{fig:GD-Nuclear}.
In particular, the most striking difference between GD and the minimum nuclear norm estimator (MNNE) is at values of sample complexity $\alpha$ slightly above the BO strong recovery threshold but below the MNNE strong recovery threshold, where we see that GD already reaches strong recovery and the MNNE does not.

In Figure \ref{fig:GD-Nuclear}, right panel, we show a numerical experiment on a single instance of the data and ground truth, comparing the test error during training for GD with small initialisation and small learning rate ($\sigma^2 = 10^{-3}, 10^{-4}$) and GD initialised in the MNNE solution with the MNNE error.
To initialise GD in the MNNE solution, we consider the SVD of the MNNE $S_{\rm MNNE} = U_{\rm MNNE} D_{\rm MNNE} V_{\rm MNNE}$, and we take $U^0 = U_{\rm MNNE} \sqrt{\tilde{D}_{\rm MNNE}}$ and similar for $V$, where $\tilde{D}_{\rm MNNE}$ is the truncated version of $D_{\rm MNNE}$ to the leading $r$ singular values.
We plot this comparison at $\alpha = 0.45$, where we observed GD to achieve zero error and MNNE to not achieve zero error from the averaged comparison in Figure \ref{fig:GD-Nuclear}, left panel.
We clearly see that at single instance GD outperforms the MNNE estimator, and that the MNNE estimator is not even a stable minimum of the GD landscape.

\section{Discussion and future directions}

We introduced the bilinear sequence regression as a basic model for learning from long sequences of high-dimensional tokens. In this paper, we addressed and answered questions \ref{q1}-\ref{q4} posed in the Introduction. Our analysis involved techniques of statistical physics that are, in general, not mathematically rigorous, and the rigorous establishment of our results is a clear avenue for future work. 
Concerning question \ref{q5} about gradient descent, we investigated the behaviour of this algorithm numerically, and it is clear that, already for this rather simple model, there is a very rich behaviour of gradient descent that can be observed. 
Our experiments reveal some of the intriguing properties of GD in the BSR model. A clear avenue for future work is to analyze the behaviour of gradient descent in the large size limit e.g. via dynamical mean field theory \cite{mignacco2020dynamical} and aim to explain our numerical observation theoretically.

The need for a detailed understanding of the properties of the optimizer is also emphasized by a set of experiments that we performed using the toy transformer architecture presented in Section \ref{subsec:BSR_transformer} on the data from the BSR model. We observed that the performance is comparable to what we report for gradient descent is section \ref{sec:GD}. The transformer model seems able to figure out that the attention part of the architecture is not useful and only the skip connection is. However, again the performance depended strongly on the optimizer used, and thus, it seems to us that attempts to quantify the cases where the attention layer is advantageous need to be preceded by a more complete understanding of the gradient descent and its variants. 

Since sequence models are behind the recent progress of artificial intelligence, having a basic model for studying learning from sequences of tokens opens the avenue to address many of the questions underlying these systems. For this, future work will need to generalize the BSR model in several directions. 
\begin{itemize} 
   \item \textbf{Structured input data and more general tasks}: So far we considered Gaussian iid input data $X$ and a task defined by \eqref{eq:teacher}. One should generalize the model to add non-trivial structure in the input $X$ to mimic correlations present e.g. in natural language. A step towards this that is technically achievable may be done in a similar manner as in the hidden manifold model in \cite{goldt2020modeling}, or in the general Gaussian covariance model treated in \cite{loureiro2021learning}. For sequences of token the correlation can be added both between different tokens and among the different embedding dimensions, two cases which are of interest. 
   
   \item \textbf{Learning with other architectures} (in particular those involving attention layers): In this paper, we start with the analysis of the Bayes-optimal performance for data generated by the BSR model. We then consider gradient descent for the model \eqref{eq:student} that matches the BSR model, $d'=r$. 
   The next step would be to study in detail the mismatched case where the learning model uses a different width than the data generative model $d'\neq r$. The landscape of the corresponding loss \eqref{eq:loss} is of interest as well as the behaviour of the GD algorithm. 
   Future work should also study models that include attention layers and clarify how their use is related to the structure in the data/task. 
   \item \textbf{Training algorithms}: In this work, we studied the Bayes-optimal estimation and a related message-passing algorithm. Understanding the behaviour and properties of gradient descent theoretically is more challenging and is left for future work. Gradient descent should also be analyzed in more general, e.g. overparamaterized $d'>r$ settings. Studying the behaviour and properties of other algorithms such as stochastic gradient descent, variants with momentum and adaptive learning rates are also of interest. Investigation of an algorithm that would eventually lead to better learning than the currently existing variant of gradient descent would be immensely important.  
\end{itemize}

Of course, the investigation of the above three directions cannot be done separately because the right architecture will depend on the structure of the data and the task and on the fact that the used training algorithm needs to run efficiently. The three ingredients -- the data/task structure; the architecture/estimator; and the algorithm interplay in ways that need to be understood better. The present paper initiates this study for sequence models and opens a natural program for future research along these directions, both in terms of more realistic models and avenues for developing the methodological toolbox to study learning in high dimensions.  

\section{Ackwnoledgements}
We would like to thank Florent Krzakala for numerous discussions related to this work, Jason Lee for pointing out the connection to the works on implicit regularization of gradient descent and Christian Keup for early discussions.  
We acknowledge funding from the Swiss National Science Foundation grants SNFS SMArtNet (grant number 212049) and TMPFP2-210012.

\newpage
\appendix

\section{Additional plots} \label{app.additional}

In this section we provide additional details on the plots we presented in the main text, as well as additional plots for different values of the parameters.
\begin{itemize}
    \item In Figure \ref{fig:no-hard-phase}, we verify for $\beta = 1, 2$ and $\rho = 0.5, 1$ that the free entropy as a function of $q$ is maximised at a unique non-trivial $q<1$ solution for all values of $\alpha < \alpha_{\rm BO}$. We cannot check all values of alpha. We check a selection, and claim that by regularity of the free entropy no hard phase (i.e. no maximality of the trivial solution of the state evolution equations $q=1$) is expected up to the largest value of alpha we checked, which is numerically very close to the predicted strong recovery threshold for all values of $\beta, \rho$ showed.
    \item Figure \ref{fig:varyrho-noisy} and Figure \ref{fig:varybeta-noisy} are the noisy versions, with $\Delta = 0.1$, of the main text Figure \ref{fig:varyrho} and Figure \ref{fig:varybeta}.
    \item In Figure \ref{fig:amp} left, 
    for $\alpha < \alpha_{\rm BO}$ we used $\zeta = 20$, $\gamma = 0.5$, while
    for $\alpha > \alpha_{\rm BO}$ we used $\zeta = 20$, $\gamma = 0.8$.
    In Figure \ref{fig:amp} right, we used $\zeta = 20$, $\gamma = 0.5$.
    In all cases the tolerance used to determine convergence was $\epsilon = 10^{-6}$ on the MSE between successive iterates.
    Around the transition $\alpha_{\rm BO}$ in the noiseless case, more iterates are needed, and for this reason we used a lower tolerance $\epsilon = 10^{-8}.$
    \item
    In Figure \ref{fig:mnne_numerics}, we plot the MSE of the MNNE as a function of $\alpha$ obtained through finite-size numerical simulations ($\max(d,L) = 20, 50$ using CVXPY \cite{diamond2016cvxpy, agrawal2018rewriting}) for $\beta = 1, 2$ and $\rho = 0.2, 0.5$. We observe that, somewhat surprisingly, for $0 < \alpha \lessapprox \alpha_{\rm BO}$ the performance of the MNNE is quantitatively very close to the BO performance, while for $\alpha_{\rm BO} < \alpha < \alpha_{\rm MNNE}$ the MNNE performs sizably worse than the BO estimator.
    A theoretical prediction of the MSE of the MNNE in the high-dimensional limit is, as far as we know, not readily available.

    \item Figure \ref{fig:GD-BO-app} is the rectangular version $\beta = 2$ of Figure \ref{fig:GD-BO}, and Table \ref{tab:GD_rect} indicates the learning rates used.
    \item Tables \ref{tab:GD_square_right}, \ref{tab:GD_square_left} lists all learning rates used to produce \ref{fig:GD-BO}.
    \item In Figure \ref{fig:learning_rate_compare} we probe the effect of the learning rate on the test error. We can see that if we initialise with small norm we need the learning rate to be small enough for GD to perform well, while if we initialise in the prior we need to carefully tune our parameters. 
\end{itemize}

\begin{table}[h!]

\begin{tabular}{llllllll}
\hline
Sample ratio $\alpha = n / (dL)$ & 0.05 & 0.1  & 0.15       & 0.2        & 0.25       & 0.3 - 0.45  & 0.5 - 0.6 \\ 
Learning Rate $\eta$    & 0.7  & 0.75 & 0.65       & 0.58       & 0.53       & 0.5         & 0.45      \\ \hline
Sample ratio $\alpha = n / (dL)$ & 0.65 & 0.7  & 0.75 - 0.8 & 0.85 - 0.9 & 0.95 - 1.1 & 1.15 - 1.25 &           \\
Learning Rate $\eta$    & 0.4  & 0.35 & 0.3        & 0.25       & 0.2        & 0.15        &          \\ \hline
\end{tabular}
\caption{Learning rate as a function of sample complexity for Figure \ref{fig:GD-BO} right}
\label{tab:GD_square_right}

\vspace{0.5cm}
\begin{tabular}{llllll}
\hline
Sample ratio $\alpha = n / (dL)$  & 0.05 & 0.1 - 0.33 & 0.45 - 0.5 \\
Learning Rate $\eta$  & 0.5  & 0.7          & 0.3    \\ \hline  
\end{tabular}
\caption{Learning rate as a function of sample complexity for Figure \ref{fig:GD-BO} left}
\label{tab:GD_square_left}

\vspace{0.5cm}
\begin{tabular}{lllllll}
\hline
Sample ratio $\alpha = n / (dL)$ & 0.05 & 0.1  & 0.15 & 0.2  & 0.23 & 0.35 \\ 
Learning Rate $\eta$ & 0.85 & 0.75 & 0.7  & 0.65 & 0.65 & 0.25 \\
\hline
\end{tabular}
\caption{Learning rate as a function of sample complexity for Figure \ref{fig:GD-BO-app}}
\label{tab:GD_rect}

\end{table}

\begin{figure*}
    \includegraphics[width=\columnwidth]{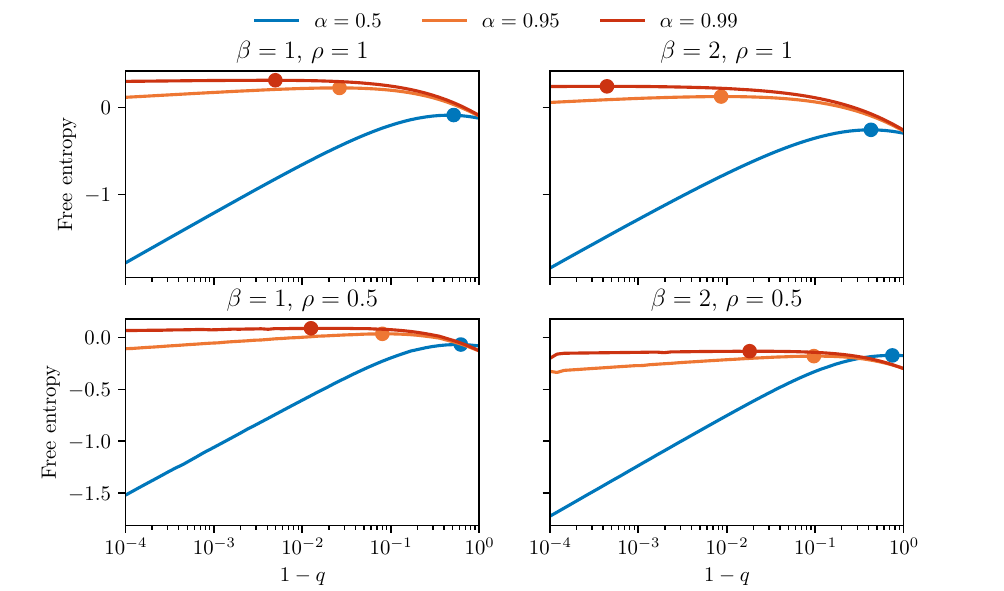}
    \caption{
        Free entropy \eqref{eq:freeentropy} for $\beta = 1, 2$ and $\rho = 0.5, 1$ as a function of $q$. We highlight that the free entropy has only a single maximum in $0<q<1$ in all these cases.
    }
    \label{fig:no-hard-phase}
\end{figure*}
\begin{figure*}
    \includegraphics[width=\columnwidth]{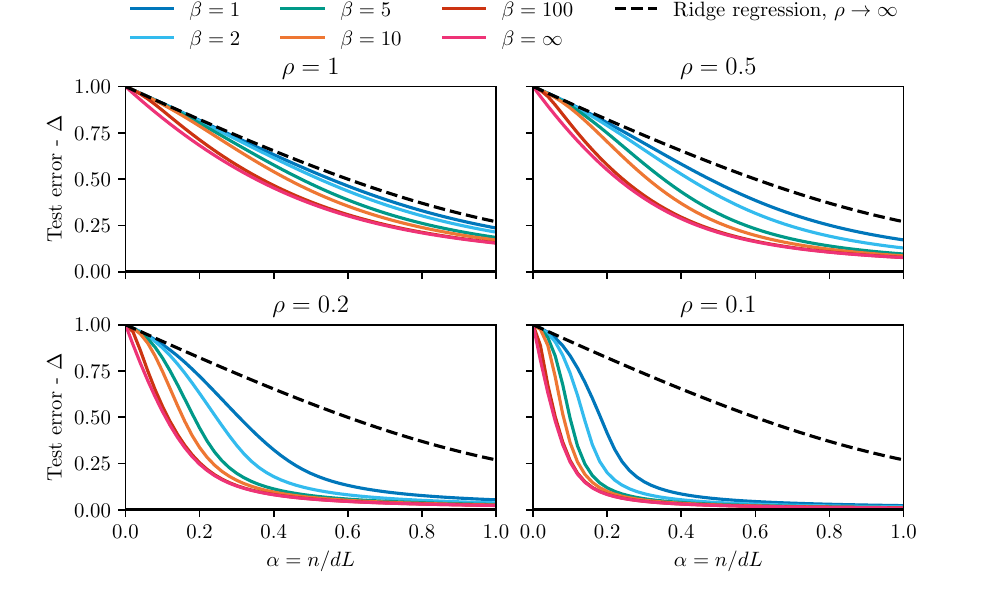}
   
    \caption{
    Same as Figure \ref{fig:varybeta}, only with $\Delta = 0.1$.
    }
    \label{fig:varybeta-noisy}
\end{figure*}

\begin{figure*}
    \includegraphics{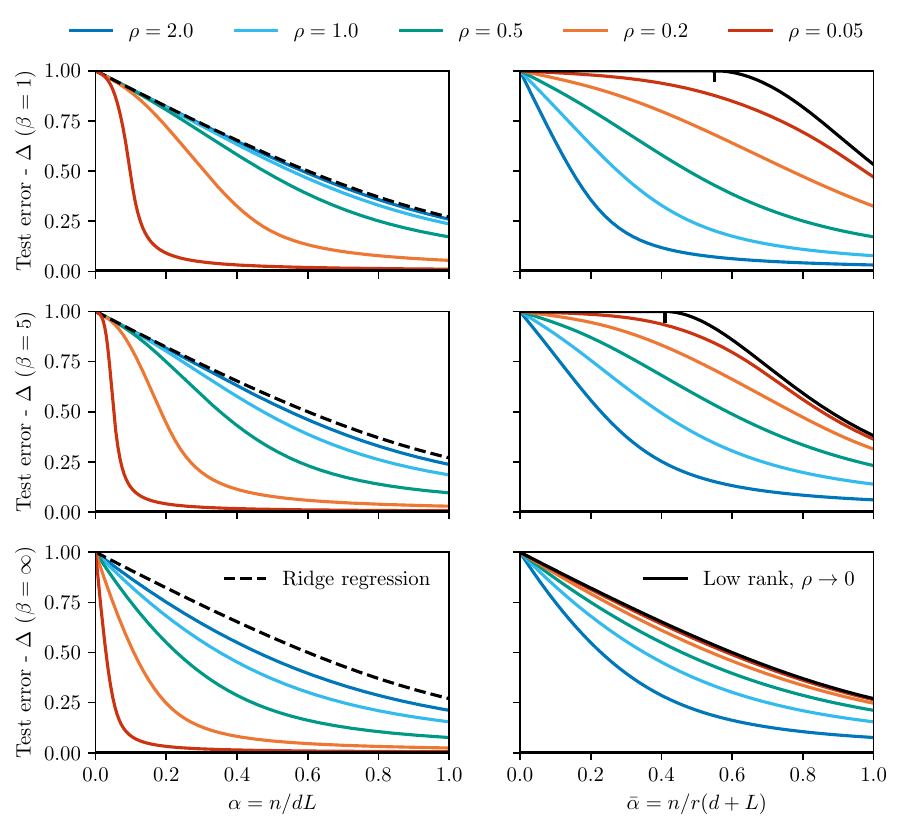}
    
    \caption{
    Same as Figure \ref{fig:varyrho}, only with $\Delta = 0.1$.
    }
    \label{fig:varyrho-noisy}
\end{figure*}

\begin{figure*}
    \includegraphics{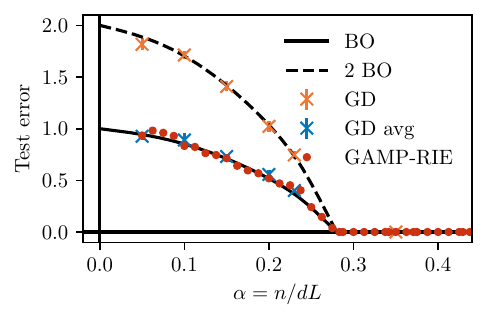}
   
    \caption{
    Same as Figure \ref{fig:GD-BO}, only with $\beta = 2$.
    }
    \label{fig:GD-BO-app}
\end{figure*}

\begin{figure*}
    \includegraphics[scale=0.9]{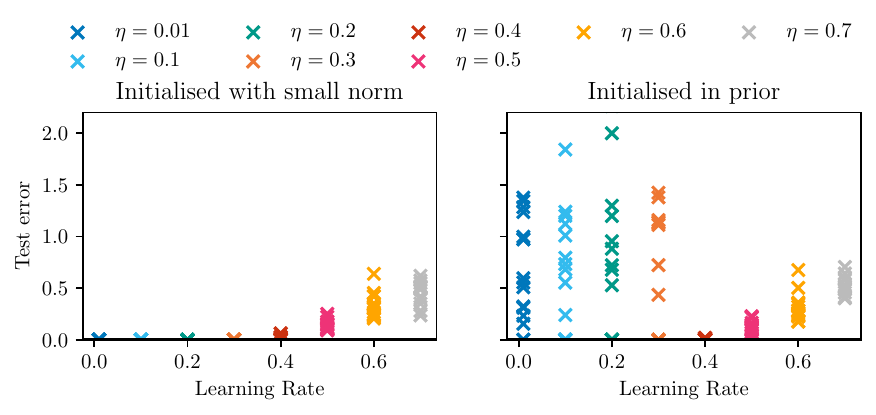}
    \caption{
        Test error as a function of the learning rate for GD initialised with small norm ({\bf left}) or in the prior ({\bf right}). Here $D=50$, $\beta=1$, $\rho = 0.2$, $\alpha = 0.44$. In this regime we expect GAMP-RIE to achieve zero error. We can see how choosing the right learning rate influences the performance of the model rather drastically.
    }
    \label{fig:learning_rate_compare}
\end{figure*}

\begin{figure*}

    \includegraphics[scale=0.9]{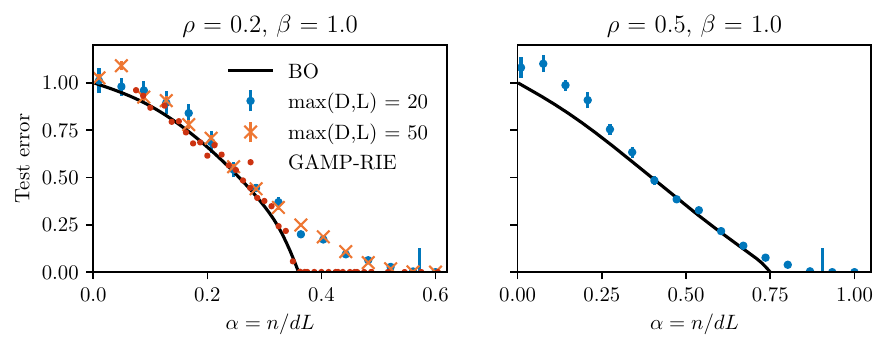}
    \includegraphics[scale=0.9]{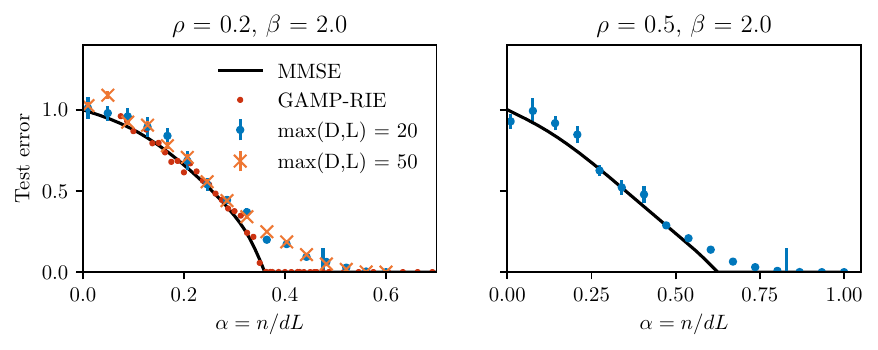}
    \caption{
        Comparison between of the test error of the MNNE estimator  and the BO test error for $\rho = r / \min(d,L) = 0.2, 0.5$ and $\beta = \max(d,L) / \min(d,L) = 1$. Dots are numerical experiments for MNNE at $\max(d,L) = 20,50$  averaged over 16 independent instances of the data and ground truth, and error bars denote the standard error on the mean. Solid lines denote the BO test error, dots/crosses the predicted asympototic strong recovery threshold for the MNNE estimator (Previous Result \ref{mnne_res}), and red dots the performance of AMP (same paramteters as in Figure \ref{fig:amp}.
        We observe two qualitatively different behaviours. 
        For $\alpha \lessapprox \alpha_{\rm BO}$, the MNNE estimator achieves a test error very close to the BO one.
        For $\alpha_{\rm BO} \lessapprox \alpha < \alpha_{\rm MNNE}$ instead, the BO error is precisely zero, while the MNNE error is non-zero.
        Finally, our numerical experiments are compatible with the theoretical prediction for the strong recovery thrshold of MNNE (Previous Result \ref{mnne_res}).
        We show qualitatively similar comparisons at $\beta = 2$ in the second row.
    }
    \label{fig:mnne_numerics}
\end{figure*}

\newpage
\section{Replica computation for the Bayes Optimal case}\label{app.replica}\label{app:replicas}

In this Appendix we derive Result \ref{res1}, and we rederive Previous Result \ref{prevreslow} under more general priors, through the replica method. We study both the intensive and extensive width BSR model using a common framework. We then specialise to the extensive case in \ref{appext}, and to the intensive case in \ref{appint}.

\paragraph{A word on scalings.}
In the following derivation, we consider the sample ratio $\bar{\alpha} = n  (r (d+L))$, where the number of samples scales proportionally to the number of unknown scalars in the ground truth signal. This allows to treat the low width case $r \ll d$ and the extensive width case $r = O(d)$ within a unique framework. 
This choice dictates also the overall scaling for the free entropy (defined below) to be $\caO(r (d+L))$.
In the extensive-width case, one recovers the results in term of the ratio $\alpha = n / (dL)$ through the rescaling
\begin{equation}
    \bar{\alpha} = \frac{n}{r(d+L)} = \frac{\beta}{\rho(1+\beta)} \alpha \, .
\end{equation}
The scaling $\bar{\alpha}$ makes sense only for factorised priors, where $r$ exists. Our computation holds also for non-factorised, but rotationally invariant priors. To obtain the associated results set $\rho = 1$. 

\paragraph{Preliminaries.}
We start by recalling the definition of the posterior distribution and the associated partition function.
The posterior distribution $P(S|\caD)$ is the probability that given an observed dataset $\caD_n = \{ (X^{\mu}, y^{\mu} \}_{\mu=1}^n$, the dataset has been generated from a given set of weights $S$. By the Bayes rule, we have
\begin{equation}
    P(S|\caD_n) 
    = \frac{P(\caD_n | S) P_0(S)}{P(\caD_n)} 
    = \frac{P_0(S)}{P(\caD_n)}\prod_{\mu = 1}^n P_{\rm out} \left( y^\mu \bigg\vert \frac{1}{\sqrt{dL}} \Tr( S^T X^{\mu} ) \right)
\end{equation}
where $P(\caD_n)$ is interpreted as the normalisation factor for the distribution (as it is independent on $\caD_n$), and for this reason is though of as a partition function.
As usual in the statistical mechanics of disordered systems and in its applications in inference, we expect the free entropy $\Phi_\caD = \frac{1}{r(d+L)} \log P(\caD_n)$ to concentrate both w.r.t. the variable $S$ and the quenched disorder $\caD$ in the high dimensional limit.
For this reason, we study the averaged free entropy $\Phi = \EE_{\caD} \Phi_\caD$, and we do so using the replica method.

\paragraph{Replica trick.}
The first step is to study the integer moments of the partition function $\EE_{\caD} P(\caD_n)^u$, from which the averaged free entropy can be recovered (using a carefully chosen analytic continuation in $u$) as
\begin{equation}
    \Phi = \frac{1}{r(d+L)} \EE_{\caD}\log P(\caD_n) = \frac{1}{r(d+L)} \lim_{u \to 0} \frac{\EE_{\caD} P(\caD_n)^u - 1}{u} \, .
\end{equation} 
The averaged replicated partition function is (we call $S_0$ the ground truth $S_*$ to highlight the fact that it behaves identically to other replicas)
\begin{equation}\label{repl-Z-BO}
    \begin{split}
        \EE &P(\caD_n)^u
        = 
        \EE_{X, y, S_0}
        \int \prod_{a=1}^u dS_a \, P_0(S_a)
        \prod_{\mu=1}^n P_{\rm out}\Big(y^\mu | \frac{1}{\sqrt{dL}} \Tr( S_a^T X^{\mu}  )\Big) 
        \\
        &= 
        \EE_{X} \EE_{S_0} \EE_{y | X, S_0} 
        \int \prod_{a=1}^u dS_a \, P_0(S_a)
        \prod_{\mu=1}^n P_{\rm out}\Big(y^\mu | \frac{1}{\sqrt{dL}} \Tr( S_a^T X^{\mu}  )\Big) 
        \\
        &= 
        \EE_{X} 
        \int dS_0 \, P_0(S_0) \prod_{\mu=1}^n dy^\mu
        P_{\rm out}\Big(y^\mu | \frac{1}{\sqrt{dL}} \Tr( X^{\mu} S_0 )\Big)
        \int  \prod_{a=1}^u dS_a \, P_0(S_a)
        \prod_{\mu=1}^n P_{\rm out}\Big(y^\mu | \frac{1}{\sqrt{dL}} \Tr( S_a^T X^{\mu}  )\Big) 
        \\
        &= 
        \int \prod_{a=0}^n \prod_{\mu=1}^n 
        dS_a \, P_0(S_a)
        \EE_{X^{\mu}}
        \int dy^\mu
        P_{\rm out}\Big(y^\mu | \frac{1}{\sqrt{dL}} \Tr( S_a^T X^{\mu}  )\Big)
        \\
        &= 
        \int \prod_{a=0}^n \prod_{\mu=1}^n dS_a \, P_0(S_a) 
        \EE_{X^{\mu}}
        \int dy^\mu \frac{dh^{\mu a} \,d\hh^{\mu a}}{(2\pi)^{Nr}} \,
        P_{\rm out}\Big(y^\mu | h^{\mu a}\Big)
        \exp\Big[ i \hh^{\mu a} \Big(h^{\mu a} - \frac{1}{\sqrt{dL}} \Tr( S_a^T X^{\mu}  )\Big)\Big]
        \\
        &= 
        \int \prod_{a=0}^n \prod_{\mu=1}^n dS_a \, P_0(S_a) 
        \EE_{X^{\mu}}
        \int dy^\mu \, d\hh^{\mu a} \, 
        \hP_{\rm out}\Big(y^\mu , \hh^{\mu a} \Big)
        \exp\Big( - i \frac{1}{\sqrt{dL}} \hh^{\mu a} \Tr( S_a^T X^{\mu}  )\Big)
        \, ,
    \end{split}
\end{equation}
where we defined
\begin{equation}
    \hP_{\rm out}(y, \hh) = 
    \int \frac{dh}{2\pi}  P_{\rm out}\Big(y| h \Big)
    \exp\Big[ i \hh h \Big] \, .
\end{equation}
For the Gaussian noise channel $P_{\rm out}(y|h) = N(y;h, \Delta)$ we have
\begin{equation}
    \begin{split}
        \hP_{\rm out}(y, \hh) 
        = 
        \int \frac{dh}{2\pi \sqrt{2\pi \Delta}} 
        e^{ - \frac{(y-h)^2}{2\Delta} + i \hh h }
        =
        e^{i \hh y} \int \frac{dx}{2\pi \sqrt{2\pi \Delta}} 
        e^{ - \frac{x^2}{2\Delta} - i \hh x }
        =
        \frac{1}{2\pi} e^{i \hh y - \Delta/2 \hh^2} 
    \end{split}
\end{equation}
which reduces to the noiseless case for $\Delta = 0$.
Notice also that by normalisation $\int dy \, P_{\rm out}(y|h) = 1$, implying
\begin{equation}\label{eq:normPhat}
    \int dy \, 
    \hP_{\rm out}(y, \hh) = 
        \int dy \,\frac{dh}{2\pi}  P_{\rm out}\Big(y| h \Big)e^{ i \hh h } 
        = \int \frac{dh}{2\pi} e^{ i \hh h } 
        = \delta(\hh) \, .
\end{equation}

\paragraph{Disorder average.}
We can now perform the average over the data $X^{\mu}$. 
We have
\begin{equation}
    \begin{split}
        \EE_{X^{\mu}} 
        \exp\Big( - i \frac{1}{\sqrt{dL}} \sum_{a=0}^u \hh^{\mu a} \Tr( X^{\mu} S_a )\Big)
        &=
        \EE_{X^{\mu}}
        \exp\Big( - i \frac{1}{\sqrt{dL}} \sum_{i=1}^d \sum_{j=1}^L X^{\mu}_{ij} \sum_{a=0}^u \hh^{\mu a} S_{a, ij} \Big)
        \\
        &=
        \exp\Big( - \frac{1}{2 dL} \sum_{a,b=0}^u \hh^{\mu a}\hh^{\mu b} \sum_{i=1}^d \sum_{j=1}^L S_{a, ij} S_{b, ij} \Big)
        \\
        &=
        \exp\Big( - \frac{1}{2} \sum_{a,b=0}^u \hh^{\mu a}\hh^{\mu b} Q(S_a, S_b) \Big)
        \, ,
    \end{split}
\end{equation}
where we introduced the overlaps
\begin{equation}
    Q(S_a, S_b) = \frac{1}{dL} \sum_{i=1}^d \sum_{j=1}^L S_{a, ij} S_{b, ij} = \frac{1}{dL} \Tr( S_a^T S_b) \, .
\end{equation}
This allows to rewrite the replicated partition function as
\begin{equation}
    \begin{split}
        \EE P(\caD_n)^u
        &= 
        \int dS_a \, dy^\mu \, d\hh^{\mu a} \, \left[ \prod_a P_0(S_a) \prod_{\mu a} \hP_{\rm out}\Big(y^\mu | \hh^{\mu a} \Big)  \right]
        \exp\Big( - \frac{1}{2} \sum_{\mu ab} \hh^{\mu a}\hh^{\mu b} Q(S_a, S_b) \Big)
        \\
        &=
        \int \frac{ dQ_{ab} \, d\hQ_{ab}}{(2\pi)^{u(u+1)/2}} \, dS_a \, dy^\mu \, d\hh^{\mu a} \,
        \left[ \prod_a P_0(S_a) \prod_{\mu a} \hP_{\rm out}\Big(y^\mu | \hh^{\mu a} \Big) \right]
        \\&\qquad\times
        \exp\Big( 
            - \frac{1}{2} \sum_{\mu ab} \hh^{\mu a}\hh^{\mu b} Q_{ab} 
            + i r(d+L) \sum_{a \leq b} Q_{ab} \hQ_{ab}
            - i \frac{r(d+L)}{dL} \sum_{a \leq b} \hQ_{ab} \Tr( S_a^T S_b)
        \Big)
        \\
        &=
        \int \frac{dQ_{ab} \, d\hQ_{ab}}{(2\pi)^{u(u+1)/2}} 
        \exp\Big( 
            i r(d+L) \sum_{a \leq b} Q_{ab} \hQ_{ab}
        \Big)   
        \\&\qquad\times
        \left[
            \int dy \, d\hh^{a}
            \hP_{\rm out}(y | \hh^{a} ) 
            \exp\Big( 
                - \frac{1}{2} \sum_{ab} \hh^{a}\hh^{b} Q_{ab} 
            \Big)
        \right]^n
        \\&\qquad\times
        \left[
            \int dS_a P_0(S_a) 
            \exp\Big( 
                - i \frac{r(d+L)}{dL}\sum_{a \leq b} \hQ_{ab} \Tr( S_a^T S_b)
            \Big)
        \right]
        \, .
    \end{split}
\end{equation}

\paragraph{Replica symmetric ansatz.}
It is well known that in the BO case, the replica symmetric ansatz is correct due to Nishimori's identities~\cite{zdeborova2016statistical}. 
Then, we can take
$Q_{00} = Q_0$, $Q_{0a} = m$, $Q_{aa} = Q$ and $Q_{ab} = q$
and 
$i \hQ_{00} = \hQ_0$, $i \hQ_{0a} = - \hm$, $i \hQ_{aa} = \hQ$ and $i \hQ_{ab} = - \hq$.
Nishimori's identities additionally imply $m=q$ and $Q_0 = Q$, and similarly for the hat variables.
Using these simplifications, we can perform the following rewritings.
The algebraic term becomes
\begin{equation}
    i \sum_{a \leq b} Q_{ab} \hQ_{ab} 
    = (u+1) Q \hQ - \frac{u(u+1)}{2} q \hq 
    \underset{u \to 0}{=} (1+u) Q \hQ - \frac{u}{2} q \hq + O(u^2) 
    \, .
\end{equation}
The output channel term can be treated by a standard decoupling trick involving an Hubbard-Stratonovich transformation. One has first the rewriting
\begin{equation}
    \begin{split}
        \sum_{ab} \hh^{a}\hh^{b} Q_{ab} 
        = Q \sum_a (\hh^a)^2 + q \sum_{a \neq b} \hh^{a}\hh^{b}
        = (Q - q) \sum_a (\hh^a)^2 + q \Big(\sum_{a} \hh^{a}\Big)^2  \, ,
    \end{split}
\end{equation}
then the Hubbard-Stratonovich decoupling
\begin{equation}
    \exp\bigg(- \frac{q}{2} \Big(\sum_{a} \hh^{a}\Big)^2\bigg)
    \propto \int Dz\, \exp( \sqrt{q} z \sum_{a} i \hh^{a} ) \, ,
\end{equation}
where $Dz$ denotes integration against a standard Gaussian measure, from which 
\begin{equation}
   \begin{split}
        \int dy  & \, d\hh^{a}
            \hP_{\rm out}(y | \hh^{a} ) 
            \exp\Big( 
                - \frac{1}{2} \sum_{ab} \hh^{a}\hh^{b} Q_{ab} 
            \Big) =
    \\
    &=
     \int Dz \, dy \, \left[
        \int d\hh
        \hP_{\rm out}(y|\hh) 
        \exp\Big( 
            - \frac{Q - q}{2} \hh^2
            + \sqrt{q} z i \hh
        \Big)
        \right]^{u+1}
    \\&=
    \int Dz \, dy \,  I_{\rm out}(z, y)^{u+1}
    \\&\underset{u \to 0}{=}
    \int Dz \, dy \,  I_{\rm out}(z, y) 
    + u \int Dz \, dy \, I_{\rm out}(z, y)  \log  I_{\rm out}(z, y)  + O(u^2) \, ,
   \end{split}
\end{equation}
where
\begin{equation}
    I_{\rm out}(z, y) = \int d\hh
        \hP_{\rm out}(y|\hh) 
        \exp\Big( 
            - \frac{Q - q}{2} \hh^2
            + \sqrt{q} z i \hh
        \Big) \, .
\end{equation}
A similar procedure can be repeated for the prior term.
We start by
\begin{equation}
\begin{split}
    - i \sum_{a \leq b} \hQ_{ab} \Tr( S_a^T S_b)
    &= 
    - \hQ \sum_a \Tr( S_a^T S_a)
    + \hq \sum_{a<b} \Tr( S_a^T S_b)
    \\
    &= 
    - \left(\hQ + \frac{\hq}{2}\right) \sum_a \Tr( S_a^T S_a)
    + \frac{\hq}{2} \sum_{ab} \Tr( S_a^T S_b)  \, ,
\end{split}
\end{equation}
then
\begin{equation}
    \begin{split}
        \exp\Big( 
            \frac{\hq}{2} \Tr( (\sum_a S_a)^T (\sum_b S_b))
        \Big)
        =
        \int DY \, 
        \exp\Big( 
            - \frac{1}{2} \Tr(Y^T Y) + \sqrt{\hq} \Tr( Y^T (\sum_a S_a))
        \Big) \, ,
    \end{split}
\end{equation}
where $Y$ is a $d \times L$ matrix with standard Gaussian entries.
This gives
\begin{equation}
    \begin{split}
        &\int dS^a P_0(S^a) 
            \exp\Big( 
                - i \frac{r(d+L)}{dL} \sum_{a \leq b} \hQ_{ab} \Tr( S_a^T S_b)
                \Big)
                \\
        &=
        \int DY \, 
        \left[
            \int dS \,P_0(S) 
            \exp\Big( 
                - \frac{r(d+L)}{dL} \left(\hQ + \frac{\hq}{2}\right) \Tr( S^T S)
                + \sqrt{\frac{r(d+L)}{dL}} \sqrt{\hq} \Tr( Y^T S)
            \Big)
        \right]^{u+1}
        \\
        &=
        \int DY \, I_0(Y)^{u+1}
        \\
        &\underset{u \to 0}{=}
        \int DY \, I_0(Y) + u  \int DY \, I_0(Y) \log I_0(Y) + O(u^2) \, ,
    \end{split}
\end{equation}
where
\begin{equation}
    I_0(Y) = \int dS \,P_0(S) 
            \exp\Big( 
                - \frac{r(d+L)}{dL} \left(\hQ + \frac{\hq}{2}\right) \Tr( S^T S)
                + \sqrt{\frac{r(d+L)}{dL}} \sqrt{\hq} \Tr( Y^T S)
            \Big) \, .
\end{equation}

\paragraph{Zero replicas: $Q$ and $\hQ$ equations.}
For $u=0$ we need to recover the trivial result $\EE P(\caD_n)^0 = 1$.
The replicated partition function in that case equals
\begin{equation}
    \begin{split}
        \EE P(\caD_n)^0
        &=
        \int \frac{dQ_{ab} \, d\hQ_{ab}}{(2\pi)^{u(u+1)/2}} 
        \exp\Big( 
            r(d+L) Q \hQ
            +
            n \log \int Dz \, dy \,  I_{\rm out}(z, y) 
            +
            \log \int DY \, I_0(Y)
        \Big)   
    \end{split}
\end{equation}
where (using  \eqref{eq:normPhat})
\begin{equation}
   \begin{split}
        \int Dz \, dy \,  I_{\rm out}(z, y) 
        = 
        \int Dz \, dy \, d\hh \, 
        \hP_{\rm out}(y|\hh) 
        \exp\Big( 
            - \frac{Q - q}{2} \hh^2
            + \sqrt{q} z i \hh
        \Big)
        = 1
   \end{split}
\end{equation}
and
\begin{equation}
    \begin{split}
        \int DY \, I_0(Y) 
        &=
        \int DY \, dS \,P_0(S) 
        \exp\Big( 
            - \frac{r(d+L)}{dL}\left(\hQ + \frac{\hq}{2}\right) \Tr( S^T S)
            + \sqrt{\frac{r(d+L)}{dL}}\sqrt{\hq} \Tr( Y^T S)
        \Big)
        \\
        &=
        \int  dS \,P_0(S) 
        \exp\Big( 
            - \frac{r(d+L)}{dL}\left(\hQ + \frac{\hq}{2}\right) \Tr( S^T S)
            + \frac{r(d+L)}{dL}\frac{\hq}{2} \Tr( S^T S)
        \Big)
        \\
        &=
        \int  dS \,P_0(S) 
        \exp\Big( - \frac{r(d+L)}{dL} \hQ \Tr(S^T S) \Big) \, .
    \end{split}
\end{equation}
Taking a saddle-point approximation on the scale $r(d+L)$ gives two equations for $Q, \hQ$, namely
$\hQ = 0$ and
\begin{equation}
    Q = \frac{1}{r(d+L)} 
        \frac{
            \int dS \, P_0(S) \exp\Big( - \hQ \Tr(S^T S) \Big) \frac{r(d+L)}{dL} \Tr(S^T S)
        }{
            \int dS \, P_0(S) \exp\Big( - \hQ \Tr(S^T S) \Big)
        }
    \underset{\hQ = 0}{=} \frac{1}{dL} \int dS \, P_0(S) \Tr(S^T S)
    = Q_* \, .
\end{equation}
This in turn implies that at the saddle point
\begin{equation}\label{eq:normI0}
    \begin{split}
        \int DY \, I_0(Y) 
        &=
        \int  dS \,P_0(S) 
        = 1 \, .
    \end{split}
\end{equation}

\paragraph{Free entropy.}
At the first non-trivial order in $u \to 0$, the replicated partition function reads (using the above results for $Q, \hQ$, and dropping irrelevant constant factors)
\begin{equation}
    \begin{split}
        \EE P(\caD_n)^u
        &\propto
        \int dq d\hq
        \exp \Bigg[u r(d+L) \Big( 
            - \frac{1}{2} q \hq
            \\&\qquad
            + \bar{\alpha} \log \int DY \, I_0(Y) \log I_0(Y)
            + \frac{1}{r(d+L)} \log \int Dz \, dy \, I_{\rm out}(z, y)  \log  I_{\rm out}(z, y) 
        \Big)   \Bigg]
        \, ,
    \end{split}
\end{equation}
from which we get 
\begin{equation}
    \begin{split}
        \Phi 
        &=
        \extr_{q, \hq}\Big( 
            - \frac{1}{2} q \hq
            + \frac{1}{r(d+L)} \int DY \, I_0(Y) \log I_0(Y)
            + \bar{\alpha} \int Dz \, dy \, I_{\rm out}(z, y)  \log  I_{\rm out}(z, y) 
        \Big)
        \, .
    \end{split}
\end{equation}

\paragraph{Equation w.r.t. $q$.}
Taking the derivative w.r.t. $q$ gives
\begin{equation}
    \begin{split}
        \hq 
        &= 2\bar{\alpha} \del_q \int Dz \, dy \, I_{\rm out}(z, y)  \log  I_{\rm out}(z, y)
        \\
        &= 
        2\bar{\alpha} \int Dz \, dy \,  \left(1+ \log  I_{\rm out}(z, y)\right)\del_q I_{\rm out}(z, y)
    \end{split}
\end{equation}
Now, using~\cite[Eq 19]{schulke16}, we get
\begin{equation}
    \begin{split}
        \del_q I_{\rm out}(z, y)
        &=
        \del_q \int dh \, \caN\left( h ; \sqrt{q} z ; Q_* - q \right) 
        P_{\rm out}\Big(y| h \Big)
        \\
        &=
        \frac{e^{z^2/2}}{2q} \del_z \left[
            e^{-z^2/2} \del_z \int dh \, \caN\left( h ; \sqrt{q} z ; Q_* - q \right) 
            P_{\rm out}\Big(y| h \Big)
        \right]
        \\
        &=
        \frac{e^{z^2/2}}{2q} \del_z \left[
            e^{-z^2/2} \del_z I_{\rm out}(z, y)
        \right]
    \end{split}
\end{equation}
giving
\begin{equation}\label{sp-out}
    \begin{split}
        \hq 
        &= 
        2\bar{\alpha} \int Dz \, dy \,  \left(1+ \log  I_{\rm out}(z, y)\right)\del_q I_{\rm out}(z, y)
        \\
        &= 
        - 2\bar{\alpha} \int Dz \, dy \,  \left(1+ \log  I_{\rm out}(z, y)\right)\frac{e^{z^2/2}}{2q} \del_z \left[
            e^{-z^2/2} \del_z I_{\rm out}(z, y)
        \right]
        \\
        &= 
    - \frac{\bar{\alpha}}{q \sqrt{2\pi}} \int dz \, dy \,  \left(1+ \log  I_{\rm out}(z, y)\right) \del_z \left[
            e^{-z^2/2} \del_z I_{\rm out}(z, y)
        \right]
        \\
        &= 
        \frac{\bar{\alpha}}{q \sqrt{2\pi}} \int dz \, dy \,  \del_z \left(1+ \log  I_{\rm out}(z, y)\right)  \left[
            e^{-z^2/2} \del_z I_{\rm out}(z, y)
        \right]
        \\&= 
        \frac{\bar{\alpha}}{q} \int Dz \, dy \, \frac{\left(\del_z I_{\rm out}(z, y)\right)^2}{I_{\rm out}(z, y)}  
    \end{split}
\end{equation}
which matches~\cite[Eqs (60), (76)]{schulke16} modulo a different definition of the sample ratio $\bar{\alpha}$.
For the Gaussian noise channel we have
\begin{equation}
    \begin{split}
        I_{\rm out}(z, y) 
        &= \int \frac{d\hh}{2\pi}
        \exp\Big( 
            - \frac{\Delta + Q - q}{2} \hh^2
            + (\sqrt{q} z + y) i \hh
        \Big)
        \\
        &= \frac{1}{\sqrt{2\pi (\Delta + Q-q)}}
        \exp\Big( 
            -\frac{(\sqrt{q} z + y)^2}{2(\Delta + Q-q)}
        \Big)
    \end{split}
\end{equation}
from which 
\begin{equation}\label{eq:ap-state-output-gaussian}
    \begin{split}
        \hq 
        &=
        \frac{\bar{\alpha}}{q} \int Dz \, dy \,
        \frac{\left(\del_z I_{\rm out}(z, y)\right)^2}{I_{\rm out}(z, y)}  
        \\
        &=
        \bar{\alpha} \int Dz \, dy \, 
        \frac{\frac{(\sqrt{q} z + y)^2 }{(\Delta + Q-q)^2}}{\sqrt{2\pi (\Delta + Q-q)}}
        \exp\Big( 
            -\frac{(\sqrt{q} z + y)^2}{2(\Delta + Q-q)}
        \Big) 
        \\
        &= 
        \bar{\alpha} \int  dt \, 
        \frac{\frac{t^2 }{(\Delta + Q-q)^2}}{\sqrt{2\pi (\Delta + Q-q)}}
        \exp\Big( 
            -\frac{t^2}{2(\Delta + Q-q)}
        \Big) 
        \\
        &= 
        \frac{\bar{\alpha} }{\Delta + Q_* - q} \, .
    \end{split}
\end{equation}

We also have, for the free entropy term and Gaussian output channel
\begin{equation}
    \begin{split}
        \int Dz \, dy \, I_{\rm out}(z, y)  \log  I_{\rm out}(z, y) 
        &=
        \int Dz \, dy \, \frac{\exp\Big( 
            -\frac{(\sqrt{q} z + y)^2}{2(\Delta + Q-q)}
        \Big) }{\sqrt{2\pi (\Delta + Q-q)}}
         \log  \frac{\exp\Big( 
            -\frac{(\sqrt{q} z + y)^2}{2(\Delta + Q-q)}
        \Big) }{\sqrt{2\pi (\Delta + Q-q)}}
        \\
         &=
        \int Dt \, 
         \log  \frac{\exp\Big( 
            -\frac{t^2}{2}
        \Big) }{\sqrt{2\pi (\Delta + Q-q)}}
        \\
         &=
        \int Dt \, \left( -\frac{t^2}{2} - \frac{1}{2} \log \left(2\pi (\Delta + Q-q)\right) \right)
        \\
         &=
        -\frac{1}{2} - \frac{1}{2} \log \left(2\pi (\Delta + Q-q)\right)
        \\
        &= - \frac{1}{2} \log \left(\Delta + Q-q\right) + \dots
    \end{split}
\end{equation}
where we changed variable $y \to (y + \sqrt{q} z)/\sqrt{\Delta + Q - q} = t$, and kept only the order-parameter-dependent terms in the last passage.

\paragraph{Equation w.r.t. for $\hq$, and a denoising subproblem.}
The non-trivial part of the free entropy, involving $\hq$, is
\begin{equation}
    \begin{split}
        \frac{1}{r(d+L)} \int DY \, I_0(Y) \log I_0(Y) \, ,
    \end{split}
\end{equation}
where (recall that $\beta = L /d$)
\begin{equation}
    I_0(Y) = \int dS \,P_0(S) 
            \exp\Big( 
                - \frac{r(d+L)\hq}{d L} \frac{1}{2} \Tr( S^T S)
                + \sqrt{\frac{r(d+L)\hq}{d L}} \Tr( Y^T S)
            \Big) \, .
\end{equation}
We recognise that this quantity is proportional to the entropy $H(Y) = - \EE_Y \log P(Y)$ of the observation $Y$ in the denoising problem 
\begin{equation}
    Y = \sqrt{\frac{r(d+L)\hq}{dL}} S_* + Z
\end{equation}
with prior $S \sim P_0$ and additive i.i.d. standard Gaussian noise $Z$, by the identification $I_0(Y) = P(Y)$ (notice that we showed in \eqref{eq:normI0} that $I_0$ is properly normalised). Thus, $\hq$ plays the role of a rescaled signal-to-noise ratio.
Moreover, we have that by the I-MMSE theorem the derivative w.r.t. $\hq$ of the denoising observation entropy will be related to the MMSE of the same denoising problem.

Thus, to proceed we need to solve the asymptotics of this denoising free entropy for the prior \eqref{eq.prior}, and in particular of the associated MMSE.
This requires different approaches based on the scaling of $r$.
If $r \ll d$, the computation is already in the literature~\cite{schulke16}, but we reproduce it here for convenience and generalise it to correlated priors. 
If $r = O(d)$, the computation for the denoiser has been carried out in~\cite{troiani2022optimal}, but its application to our problem is novel.

\subsection{Equation w.r.t. \texorpdfstring{$\hq$}{q} in the extensive width case.}
\label{appext}
In the regime where $r = \rho \min(L,d)$ with constant $\rho$, we use the results from~\cite{troiani2022optimal} (which more generally apply to all rotationally invariant priors). Again, in this subsection we assume that $d \leq L$, understanding that the general case is retrieved by substituting $d \to \min(L,d)$.
It is best to rescale the various quantities to match those in~\cite{troiani2022optimal} to easily adapt their results.
We define 
\begin{equation}
    \delta(\hq) = \frac{\beta}{\rho (1+\beta)\hq} \, .
\end{equation}
We want to compute
\begin{equation}
    \begin{split}
        \caI = \frac{1}{\rho (1+\beta) d^2} \int DY \, I_0(Y) \log I_0(Y) \, ,
    \end{split}
\end{equation}
where
\begin{equation}
    I_0(Y) = \int dS \,P_0(S) 
            \exp\Big( 
                - \frac{1}{2\delta} \Tr( S^T S)
                + \sqrt{\frac{1}{\delta}} \Tr( Y^T S)
            \Big) \, .
\end{equation}
We perform the change of variable $Y = \uY \sqrt[4]{Ld} / \sqrt{\delta}$, $S = \uS \sqrt[4]{Ld}$, $S_* = \uS_* \sqrt[4]{Ld}$ and get
\begin{equation}
    \begin{split}
        \caI 
        &= \frac{1}{\rho (1+\beta) d^2} \int d\uY \, d\uS_* \, \uP_0(\uS_*) \, ( 2\pi \delta / \sqrt{Ld} )^{- Ld/2}
            \exp\Big( 
                - \frac{\sqrt{Ld}}{2\delta} \Tr( (\uY - \uS_*)^T (\uY - \uS_*))
            \Big)
        \\&\qquad\qquad\times
        \log 
        \int d\uS \, \uP_0(\uS)
            ( 2\pi \delta / \sqrt{Ld} )^{- Ld/2}
            \exp\Big( 
                - \frac{\sqrt{Ld}}{2\delta} \Tr( (\uY - \uS)^T (\uY - \uS))
            \Big) 
        \\&\qquad+
        \frac{\sqrt{\beta}}{2 \rho (1+\beta) \delta d} 
        \int d\uY \, d\uS_* \, \uP_0(\uS_*) \, N(\uY; \uS_*, \delta / \sqrt{Ld}) 
        \Tr( \uY^T \uY)
        + \frac{1}{\rho (1+\beta) d^2} \frac{Ld}{2} \log \frac{2\pi \delta}{\sqrt{Ld} }
    \end{split}
\end{equation}
where $\uP_0$ is the rescaled prior, still normalised and such that its samples have $\caO(1)$ spectral density.
Notice that after the rescaling we have the denoising problem $\uY = \uS + \sqrt{\delta(\hq)} \uZ$.
The second term can be simplified to
\begin{equation}
    \begin{split}
        \frac{\sqrt{\beta}}{2 \rho (1+\beta) \delta d}&
        \int d\uY \, d\uS_* \, \uP_0(\uS_*) \, N(\uY; \uS_*, \delta / \sqrt{Ld}) 
        \Tr( \uY^T \uY)
        =
        \\
        &=
        \frac{\sqrt{\beta}}{2 \rho (1+\beta) \delta d} 
        \int d\uS_* \, \uP_0(\uS_*) 
        \left(
            \delta \sqrt{Ld}
            + \Tr( \uS_*^T \uS_*)
        \right)
        \\
        &=
        \frac{\beta (\delta 
            + Q_*)}{2 \rho (1+\beta) \delta} 
            \, ,
    \end{split}
\end{equation}
so that
\begin{equation}
    \begin{split}
        \caI 
        &= \frac{\beta (\delta + Q_*)}{2 \rho (1+\beta) \delta} 
            + \frac{\beta}{2 \rho (1+\beta)}  \log \frac{2\pi \delta}{\sqrt{Ld} } 
            \\
            &\qquad+ \frac{1}{\rho (1+\beta) d^2} \int d\uY \, d\uS_* \, \uP_0(\uS_*) \, ( 2\pi \delta / \sqrt[4]{Ld} )^{- Ld/2}
            \exp\Big( 
                - \frac{\sqrt{Ld}}{2\delta} \Tr( (\uY - \uS_*)^T (\uY - \uS_*))
            \Big)
        \\&\qquad\qquad\qquad\qquad\qquad\times
        \log 
        \int d\uS \, \uP_0(\uS)
            \exp\Big( 
                - \frac{\sqrt{Ld}}{2\delta} \Tr( (\uY - \uS)^T (\uY - \uS))
            \Big) 
    \end{split}
\end{equation}
We now recognise that the third term is proportional to~\cite[Eq (15)]{troiani2022optimal}, which is the free entropy of the denoising problem with noise-to-signal ratio $\sqrt{\delta}$. We just need to be careful of a factor $\beta$, as in \cite{troiani2022optimal} the free entropy is normalised by $1/Ld$ and not by $1/d^2$.
Thus, we can use directly~\cite[Eq (18)]{troiani2022optimal} to get (one needs to set $\beta = \beta$)
\begin{equation}
    \begin{split}
        \caI 
        &= \frac{\beta (\delta 
            + Q_*)}{2 \rho (1+\beta) \delta}  
            + \frac{\beta}{2 \rho (1+\beta)}  \log \frac{ 2\pi \delta }{ \sqrt[4]{Ld} }
        + \frac{\beta}{\rho (1+\beta)} 
        \left[
            \text{const}(\beta, \uP_0)
            - \frac{1}{\beta} \Sigma[ \hmu_\uY ]
            - \frac{\beta - 1}{\beta} \Lambda[ \hmu_\uY ]
        \right]
    \end{split}
\end{equation}
where $\hmu_\uY$ is the symmetrised singular value density of a matrix distributed as $\uY$ in the high dimensional limit, and
\begin{equation}
    \Sigma[ \hmu_\uY ] = \dashint dx \, dy \, \hmu_\uY(x) \hmu_\uY(y) \log | x-y |
    \mathand
    \Lambda[ \hmu_\uY ] = \dashint dx \, \hmu_\uY(x) \log | x |
\end{equation}
regularised as in~\cite[Appendix A]{troiani2022optimal}.
Thus, translating this back into a function of $\hq$ we get, dropping all $\hq$ independent terms
\begin{equation}
    \begin{split}
        \caI 
        &= \text{const} 
        +  \frac{Q_*\hq}{2}  
        - \frac{\beta}{2 \rho (1+\beta)}  \log ( \hq ) 
        - \frac{\beta}{\rho (1+\beta)} 
        \left[
            \frac{1}{\beta}\Sigma[ \hmu_{\uS + \sqrt{\delta(\hq)} \uZ} ]
            +\frac{\beta-1}{\beta} \Lambda[\hmu_{\uS + \sqrt{\delta(\hq)} \uZ} ]
        \right] \, .
    \end{split}
\end{equation}
Then, taking the derivative w.r.t. $\hq$ we obtain the second state equation
\begin{equation}
    \begin{split}
        q 
        &= 2 \del_\hq \caI 
        \\
        &= 
        Q_*  
        - \frac{\beta}{\rho (1+\beta) \hq}
        + \frac{2 \beta^2}{\rho^2 (1+\beta)^2 \hq^2} 
        \del_{\delta}
        \left[
            \frac{1}{\beta} \Sigma[ \hmu_{\uS + \sqrt{\delta} \uZ} ]
            +\frac{\beta-1}{\beta} \Lambda[\hmu_{\uS + \sqrt{\delta} \uZ} ]
        \right]_{\delta = \frac{\beta}{\rho (1+\beta)\hq}}
        \\
        &= 
        Q_*  
        - \delta(\hq)
        + \delta(\hq)^2
        \int dx \, \hmu_{\uS + \sqrt{\delta(\hq)} \uZ}(x) \left[
            \frac{(\beta-1)^2}{\beta^{3/2} x^2} 
            + \frac{4\pi^2}{3 \beta^{3/2}} \hmu_{\uS + \sqrt{\delta(\hq)} \uZ}(x)^2
            \right]
        \, ,
    \end{split}
\end{equation}
where the derivative of the spectral term was computed using~\cite[Eqs (35), (70) and (71)]{troiani2022optimal}. 

We also notice that
\begin{equation}\label{eqMMSE}
    \del_\delta 
        \left[
            \frac{1}{\beta} \Sigma[ \hmu_\uY ]
            + \frac{\beta - 1}{\beta} \Lambda[ \hmu_\uY ]
        \right]
    = - \del_\delta \Phi_{\rm denoising}(\delta)
    = \frac{\delta - \MMSE_{\rm denoising}(\delta)}{2 \delta^2} \, ,
\end{equation}
where we used~\cite[Eq (14)]{troiani2022optimal}, so that the state equation can be rewritten as
\begin{equation}\label{eq:ap-state-extensive}
    \begin{split}
        q 
        &= Q_*  
        - \delta(\hq)
        + \delta(\hq)^2 \frac{\delta(\hq) - \MMSE_{\rm denoising}(\delta(\hq))}{\delta(\hq)^2}
        \\
        &= Q_* - \MMSE_{\rm denoising}(\delta(\hq))
        \, .
    \end{split}
\end{equation}

The only non-trivial ingredient needed to solve the state equations is the symmetrised singular value density of $\uY$. 
This can be computed numerically efficiently, as detailed in~\cite[Section 3.3 and Appendix F, setting $R_1 = \beta$, $R_2 = \rho$ and $\Delta = \delta(\hq)$]{troiani2022optimal} for the factorised Gaussian prior. For generic rotationally invariant priors, this spectral density may be difficult to compute accurately.

It is also useful to write the associated free entropy. One gets, discarding all terms that are not dependent on the order parameters (and using $Q = Q_* = 1$)
\begin{equation}
    \Phi \approx 
    - \frac{\bar{\alpha}}{2} \log \left(\Delta + 1 -q\right) 
    +  \frac{\hq ( 1-q)}{2}  
        - \frac{\beta \log ( \hq ) }{2 \rho (1+\beta)}  
        - \frac{\beta}{\rho (1+\beta)} 
        \left[
            \frac{1}{\beta}\Sigma[ \hmu_{\uS + \sqrt{\delta(\hq)} \uZ} ]
            +\frac{\beta-1}{\beta} \Lambda[\hmu_{\uS + \sqrt{\delta(\hq)} \uZ} ]
        \right] \, .
\end{equation}

Notice also that substituting back $\bar\alpha = \frac{\beta}{\rho(1+\beta)} \alpha$ and rescaling $\hq \to \hq \frac{\beta}{\rho(1+\beta)}$ we get
\begin{equation}\label{eq:freeentropy}
    \frac{\rho(1+\beta)}{\beta} \Phi \approx 
    -  \frac{\alpha}{2} \log \left(\Delta + 1 -q\right) 
    +  \frac{\hq ( 1-q)}{2}  
        - \frac{\log ( \hq ) }{2}  
        -  
        \left[
            \frac{1}{\beta}\Sigma[ \hmu_{\uS + \sqrt{1 / \hq} \uZ} ]
            +\frac{\beta-1}{\beta} \Lambda[\hmu_{\uS + \sqrt{1 / \hq} \uZ} ]
        \right] \, .
\end{equation}

In the case of Gaussian label noise, we get $\hq = \alpha / (1-q + \Delta)$ so that (again discaring all terms not dependent on the order parameter $\hq$, the only one left now)
\begin{equation}
    \frac{\rho(1+\beta)}{\beta} \Phi \approx 
    \frac{\alpha - 1}{2} \log (\hq ) 
    -  \frac{\hq \Delta}{2}  
    -  
    \left[
        \frac{1}{\beta}\Sigma[ \hmu_{\uS + \sqrt{1 / \hq} \uZ} ]
        +\frac{\beta-1}{\beta} \Lambda[\hmu_{\uS + \sqrt{1 / \hq} \uZ} ]
    \right] \, .
\end{equation}

\subsection{Equation w.r.t. \texorpdfstring{$\hq$}{q} in the low width case.}
\label{appint}
In the regime where $r \ll d$, we re-derive results from matrix compressed sensing~\cite{schulke16}.
We assume w.l.o.g. $d \leq L$. Not assuming this is equivalent to substituting $d \to \min(L,d)$ everywhere in this subsection.
We want to compute 
\begin{equation}
    \begin{split}
        \caI = \frac{1}{r(d+L)} \int DY \, I_0(Y) \log I_0(Y) \, ,
    \end{split}
\end{equation}
where (recall that $\beta = L /d$)
\begin{equation}
    I_0(Y) = \int dS \,P_0(S) 
            \exp\Big( 
                - \frac{(1+\beta)\hq}{\beta} \frac{r}{d} \frac{1}{2} \Tr( S^T S)
                + \sqrt{\frac{(1+\beta)\hq}{\beta} \frac{r}{d}}  \Tr( Y^T S)
            \Big) \, .
\end{equation}
We consider factorised priors $S = AB^T / \sqrt{r}$ with $A \in \bbR^{L \times r}$, $B \in \bbR^{d \times r}$, where the priors on $A$ and $B$ are row-factorised, but arbitrarily correlated along the width $r$ through distributions $G_A$ and $G_B$. A special case is that of fully factorised prior, already treated in~\cite{schulke16}, among which one finds the i.i.d. Gaussian priors we are interested to in the main text.
To start, we have
\begin{equation}
    \begin{split}
        \mu_2(0) 
        &= \frac{1}{r d L} \EE_{S \sim P_0} \Tr(S^T S) 
        \\
        &= \frac{1}{r d L} \EE_{A, B} \sum_{ijkl} A_{ik}B_{jk}A_{il}B_{jl}
        \\
        &= \frac{1}{r} \sum_{kl} \EE_{A} \left[A_{\cdot k}A_{\cdot l}\right] \EE_B \left[ B_{\cdot k}B_{\cdot l}\right]
        \\
        &= \frac{1}{r} \sum_{kl} \Sigma^A_{kl} \Sigma^B_{kl}
    \end{split}
\end{equation}
where $\Sigma^{A, B}$ are the column covariances of $A$ and $B$ in $\bbR^r$.
To compute $\caI$ and deal with the logarithm, we need to replicate again. 
We replicate $s$ times, so that we need to compute $\caI = \lim_{s\to 0} (\caI_s - 1)/s$ with
\begin{equation}
    \begin{split}
        \caI_s 
        &= \int DY \, I_0(Y)^{s+1} 
        \\
        &= 
        \int \left( \prod_a dS_a P_0(S_a)\right) 
        \exp\left(
            - \frac{(1+\beta)\hq}{\beta} \frac{1}{2} \frac{r}{d} \sum_a \Tr(S^T_a S_a)
        \right)
        \int DY \,
        \exp\left(
            \sqrt{\frac{(1+\beta)\hq}{\beta} \frac{r}{d}} \sum_{ij} Y_{ij} \sum_a S^a_{ij}
        \right)
        \\
        &= 
        \int \left( \prod_a dS_a P_0(S_a)\right) 
        \exp\left(
            \frac{r}{d} \frac{(1+\beta)\hq}{\beta} \sum_{a < b} \Tr(S_a^T S_b)
        \right)
    \end{split}
\end{equation}
Then we use the prior
\begin{equation}
    dS_a P_{0}(S_a) = dA_a \, dB_a \, P_A(A_a) P_B(B_a)
\end{equation}
to get
\begin{equation}
    \Tr(S_a^T S_b) 
    = \sum_{ij} S^a_{ij} S^b_{ij}
    = \frac{1}{r}\sum_{ijkl} A^a_{ik} B^a_{jk} A^b_{il} B^b_{jl}
    = \frac{1}{r}\sum_{kl} \left(\sum_i A^a_{ik}A^b_{il} \right) \left(\sum_j B^a_{jk}B^b_{jl} \right)
\end{equation}
where the terms we highlighted inside the parentheses are summed over an extensive coordinate $i,j$, and have two free indices $k, l$ that run up to $r$, which are intensive quantities.
Thus, we can introduce two overlaps (call them $g$ instead of $q$ to avoid confusion with the overlaps of the original problem)
\begin{equation}
    \begin{split}
        g^A_{ab, kl} &= \frac{1}{d} \sum_{i=1}^d A^a_{ik}A^b_{il}
        \\
        g^B_{ab, kl} &= \frac{1}{L} \sum_{j=1}^L B^a_{jk}B^b_{jl}
    \end{split}
\end{equation}
giving
\begin{equation}
    \frac{r}{d} \Tr(S_a^T S_b) 
    = \beta d \sum_{kl} g^A_{ab, kl}(A) g^B_{ab, kl}(B) \, .
\end{equation}
Thus, we have
\begin{equation}
    \begin{split}
        \caI_s 
        &= 
        \int \left( \prod_a dS_a P_{0}(S_a)\right) 
        \exp\left(
            \frac{r}{d} \frac{(1+\beta)\hq}{\beta} \sum_{a < b} \Tr(S_a^T S_b)
        \right)
        \\
        &= 
        \int \left( \prod_{a<b, kl} dg^A_{ab, kl} \, dg^B_{ab, kl} \, d\hg^A_{ab, kl} \, d\hg^B_{ab, kl} \right)
        \\&\qquad\times
        \exp\left(
            d (1+\beta)\hq \sum_{a < b} \sum_{kl} g^A_{ab, kl} g^B_{ab, kl}
            - i d \sum_{a<b, kl} g^A_{ab, kl} \hg^A_{ab, kl}
            - i d \beta \sum_{a<b, kl} g^B_{ab, kl} \hg^B_{ab, kl}
        \right)
        \\&\qquad\times
        \int \left( \prod_a dA_a \, dB_a \, P_A(A_a) P_B(B_a)\right) 
        \exp\left(
            i \sum_{a<b, kl} \hg^A_{ab, kl} \sum_{i=1}^d A^a_{ik}A^b_{il}
            + i \sum_{a<b, kl} \hg^B_{ab, kl} \sum_{i=1}^L B^a_{ik}B^b_{il}
        \right)
        \, .
    \end{split}
\end{equation}
We take notice that for $s=0$ the argument of the $\exp$ vanishes, and $\caI_{s=0} = 1$ as it should.
Now we can perform the RS ansatz, with no diagonal as that simplified away and using Nishimori's identities to avoid having to single out the zero-th replica, to get
\begin{equation}
    \begin{split}
        \caI_s 
        &= 
        \int \left( \prod_{kl} dg^A_{kl} \, dg^B_{kl} \, d\hg^A_{kl} \, d\hg^B_{kl} \right)
        \\&\qquad\times
        \exp\left(
            d (1+\beta) \hq \frac{s(s+1)}{2} \sum_{kl} g^A_{kl} g^B_{kl}
            - d \frac{s(s+1)}{2} \sum_{kl} g^A_{kl} \hg^A_{kl}
            - d \beta \frac{s(s+1)}{2} \sum_{kl} g^B_{kl} \hg^B_{kl}
        \right)
        \\&\qquad\times
        \int \left( \prod_a dA_a \, dB_a \, P_A(A_a) P_B(B_a)\right) 
        \exp\left(
            \sum_{kl} \hg^A_{kl} \sum_i \sum_{a<b} A^a_{ik}A^b_{il}
            + \sum_{kl} \hg^B_{kl} \sum_i \sum_{a<b} B^a_{ik}B^b_{il}
        \right)\, .
    \end{split}
\end{equation}
Now we use the assumption that the priors on $A, B$ are row-factorised, call the prior on a row $G$ (it is a distribution over $\bbR^{r}$, so intensive), and get
\begin{equation}
    \begin{split}
        \caI_s 
        &= 
        \int \left( \prod_{kl} dg^A_{kl} \, dg^B_{kl} \, d\hg^A_{kl} \, d\hg^B_{kl} \right)
        \\&\qquad\times
        \exp\left(
            d (1+\beta) \hq \frac{s(s+1)}{2} \sum_{kl} g^A_{kl} g^B_{kl}
            - d \frac{s(s+1)}{2} \sum_{kl} g^A_{kl} \hg^A_{kl}
            - d \beta \frac{s(s+1)}{2} \sum_{kl} g^B_{kl} \hg^B_{kl}
        \right)
        \\&\qquad\times
        \left[\int \left( \prod_a da_a \, G_A(a_a) \right) 
        \exp\left(
            \sum_{kl} \hg^A_{kl} \sum_{a<b} a^a_{k}a^b_{l}
        \right)\right]^{d}
        \\&\qquad\times
        \left[\int \left( \prod_a db_a \, G_B(b_a)\right) 
        \exp\left(
            \sum_{kl} \hg^B_{kl} \sum_{a<b} b^a_{k}b^b_{l}
        \right)\right]^{\beta d}\, ,
    \end{split}
\end{equation}
where $a$ and $b$ a rows of the respective matrices.
Finally we need to decouple the replicas.
We use
\begin{equation}
    \begin{split}
        \exp\left(
            \sum_{kl} \hg^B_{kl} \sum_{a<b} b^a_{k}b^b_{l}
        \right)
        &=
        \exp\left(
            \sum_{kl} \hg^B_{kl} \left[ \frac{1}{2} \sum_{a, b}  - \frac{1}{2} \sum_{a}\right] b^a_{k}b^b_{l}
        \right)
        \\
        &=
        \exp\left(
            - \frac{1}{2} \sum_{kl} \hg^B_{kl} \sum_{a} b^a_{k}b^a_{l}
        \right)
        \exp\left(
            \frac{1}{2} \sum_{kl} (\sum_a b^a_{k}) \hg^B_{kl} (\sum_a b^a_{l}) 
        \right)
        \\
        &=
        \exp\left(
            - \frac{1}{2} \sum_{kl} \hg^B_{kl} \sum_{a} b^a_{k}b^a_{l}
        \right)
        \int dz \, \caN(z, 0, \hg^B_{kl})
        \exp\left(
            \sum_k z_k \sum_a b^a_{k}
        \right)
        \\
        &=
        \int dz \, \caN(z, 0, \hg^B_{kl})
        \left[
        \exp\left(
            - \frac{1}{2} \sum_{kl} \hg^B_{kl} b_{k}b_{l}
            + \sum_k z_k b_{k}
        \right)
        \right]^{s+1}\, ,
    \end{split}
\end{equation}
where $\caN(x, \mu, \Sigma)$ to denote the Gaussian density with given mean $\mu$ and covariance $\Sigma$,
so that
\begin{equation}
    \begin{split}
        \caI_s 
        &= 
        \int \left( \prod_{kl} dg^A_{kl} \, dg^B_{kl} \, d\hg^A_{kl} \, d\hg^B_{kl} \right)
        \\&\qquad\times
        \exp\left(
            d (1+\beta) \hq \frac{s(s+1)}{2} \sum_{kl} g^A_{kl} g^B_{kl}
            - d \frac{s(s+1)}{2} \sum_{kl} g^A_{kl} \hg^A_{kl}
            - d \beta \frac{s(s+1)}{2} \sum_{kl} g^B_{kl} \hg^B_{kl}
        \right)
        \\&\qquad\times
        \left[
            \int dz \, \caN(z, 0, \hg^A_{kl})
            \left[
                \int da \, G_A(a)
                \exp\left(
                    - \frac{1}{2} \sum_{kl} \hg^A_{kl} a_{k}a_{l}
                    + \sum_k z_k a_{k}
                \right)
            \right]^{s+1}
        \right]^{d}
        \\&\qquad\times
        \left[
            \int dz \, \caN(z, 0, \hg^B_{kl})
            \left[
                \int db \, G_B(b)
                \exp\left(
                    - \frac{1}{2} \sum_{kl} \hg^B_{kl} b_{k}b_{l}
                    + \sum_k z_k b_{k}
                \right)
            \right]^{s+1}
        \right]^{d \beta}\, .
    \end{split}
\end{equation}
Now we take $s \to 0$ and get
\begin{equation}
    \begin{split}
        &\int dz \, \caN(z, 0, \hg^A_{kl})
        \left[
            \int da \, G_A(a)
            \exp\left(
                - \frac{1}{2} \sum_{kl} \hg^A_{kl} a_{k}a_{l}
                + \sum_k z_k a_{k}
            \right)
        \right]^{s+1}
        \\
        &=
        \int dz \, \caN(z, 0, \hg^A_{kl})
        \left[
            \int da \, G_A(a)
            \exp\left(
                - \frac{1}{2} \sum_{kl} \hg^A_{kl} a_{k}a_{l}
                + \sum_k z_k a_{k}
            \right)
        \right]
        \\&\qquad\times
        \left[
            1 
            + s \log 
                \int da \, G_A(a)
                \exp\left(
                    - \frac{1}{2} \sum_{kl} \hg^A_{kl} a_{k}a_{l}
                    + \sum_k z_k a_{k}
                \right)
            + \dots
        \right]
        \\
        &=
        1 + s 
        \int dz \, \caN(z, 0, \hg^A_{kl})
        \left[
            \int da \, G_A(a)
            \exp\left(
                - \frac{1}{2} \sum_{kl} \hg^A_{kl} a_{k}a_{l}
                + \sum_k z_k a_{k}
            \right)
        \right]
        \\&\qquad\times
        \log 
        \left[
            \int da \, G_A(a)
            \exp\left(
                - \frac{1}{2} \sum_{kl} \hg^A_{kl} a_{k}a_{l}
                + \sum_k z_k a_{k}
            \right)
        \right]
        + \dots
        \\
        &=
        1 + s \Phi_A(\hg^A) + \dots\, ,
    \end{split}
\end{equation}
where $\Phi_A(\hg^A)$ is related to an $r$-dimensional vector denoising problem with Gaussian noise and prior $G_A$.
More importantly, it is just an $r$-dimensional integral and $r \ll d$.
Thus, we get 
\begin{equation}
    \begin{split}
        \caI_s 
        &= 
        \int \left( \prod_{kl} dg^A_{kl} \, dg^B_{kl} \, d\hg^A_{kl} \, d\hg^B_{kl} \right)
        \exp s d \bigg(
            \frac{\hq (1+\beta)}{2} \sum_{kl} g^A_{kl} g^B_{kl}
            - \frac{1}{2} \sum_{kl} g^A_{kl} \hg^A_{kl}
            - \frac{\beta}{2} \sum_{kl} g^B_{kl} \hg^B_{kl}
            \\&\qquad\qquad\qquad\qquad\qquad\qquad\qquad\qquad\qquad
            + \Phi_A(\hg^A)
            + \beta \Phi_B(\hg^B) + \dots)
        \bigg)\, ,
    \end{split}
\end{equation}
from which 
\begin{equation}
   \begin{split}
        \caI 
    &= \frac{1}{r (1+\beta)} \extr_{g^A, g^B, \hg^A, \hg^B} \Bigg[
    \frac{\hq (1+\beta)}{2} \sum_{kl} g^A_{kl} g^B_{kl}
    - \frac{1}{2} \sum_{kl} g^A_{kl} \hg^A_{kl}
    - \frac{\beta}{2} \sum_{kl} g^B_{kl} \hg^B_{kl}
    \\&\qquad\qquad\qquad\qquad\qquad\quad
    + \Phi_{G^A}(\hg^A_{kl})
    + \beta \Phi_{G^B}(\hg^B_{kl})
    \Bigg]\, ,
   \end{split}
\end{equation}
where for any (symmetric, p.s.d.) $r \times r$ matrix $\hg$ we defined 
\begin{equation}
    \begin{split}
        \Phi_{G}(\hg) 
        &= \int dz \, \caN(z, 0, \hg)
    \left[
        \int da \, G(a)
        \exp\left(
            - \frac{1}{2} \sum_{kl} \hg_{kl} a_{k}a_{l}
            + \sum_k z_k a_{k}
        \right)
    \right]
    \\&\qquad\times
    \log 
    \left[
        \int da \, G(a)
        \exp\left(
            - \frac{1}{2} \sum_{kl} \hg_{kl} a_{k}a_{l}
            + \sum_k z_k a_{k}
        \right)
    \right] \, .
    \end{split}
\end{equation}
The extremization conditions for $\caI$ lead to the equations
\begin{equation}
    \begin{split}
        \hg^A_{kl} &= \hq (1+\beta) g^B_{kl} \, , \\ 
        \hg^B_{kl} &= \hq \frac{1+\beta}{\beta} g^A_{kl} \, ,  \\
        g^A_{kl} &= 2 \del_{\hg^A_{kl}}\Phi_A(\hg^A) \, , \\
        g^B_{kl} &= 2 \del_{\hg^B_{kl}}\Phi_B(\hg^B) \, .
    \end{split}
\end{equation}
At the extremiser of $\caI$, the state equation for the overlap of the original problem reads
\begin{equation}
    q = \frac{1}{r} \sum_{kl} g^A_{kl} g^B_{kl} \, .
\end{equation}

In the special case of factorised priors,~\cite{schulke16} showed that one can take the order parameters $g$ and $\hg$ to be diagonal, and with all diagonal elements equal. 
In that case, one can show by explicit computation that
\begin{equation}
    \begin{split}
        \Phi(\hg) 
        &= 
        r
        \int Dt \, da_0 \, G(a_0)
            \log 
            \left[
                \int da \, G(a)
                \exp\left(
                    - \frac{1}{2} \hg a^2 + (\sqrt{\hg} t + \hg a_0) a
                \right)
            \right]\, ,
            \\
        2\del_\hg \Phi_{G}(\hg) 
        &=
        \frac{r}{\sqrt{\hg}} 
        \int dz \, 
        \frac{            
            \left(\int da \, G(a) \, a \, \caN(a, z / \sqrt{\hg}, 1 / \hg)\right)^2
        }{            
            \int da \, G(a) \, \caN(a, z / \sqrt{\hg}, 1 / \hg)
        } \, ,
    \end{split}
\end{equation}
where now all $g$ and $\hg$ are scalars, and
which leads to~\cite[Eqs. (76-78)]{schulke16}.
Notice that in the extremisation conditions an additional factor $r$ comes out due to the sums for $k,l$ in $\caI$ trivialising.
Finally, for Gaussian priors one can solve all integrals in closed form, obtaining the equations 
\begin{equation}
    \begin{split}
        q &= g^A g^B \, \\
        \hg^A_{kl} &= \hq (1+\beta) g^B_{kl} \, , \\ 
        \hg^B_{kl} &= \hq \frac{1+\beta}{\beta} g^A_{kl} \, ,  \\
        g^A_{kl} &= \frac{\hg^A}{1 + \hg^A} \, , \\
        g^B_{kl} &= \frac{\hg^B}{1 + \hg^B} \, ,
    \end{split}
\end{equation}
which can be solved explicitly to 
\begin{equation}\label{eq:ap-state-intensive}
    \begin{split}
       g^A &= \frac{(\beta+1)^2 \hq^2-\beta}{(\beta+1) \hq (\beta \hq+\hq+1)} \, ,
        \\
        g^B &= \frac{(\beta+1)^2 \hq^2-\beta}{(\beta+1) \hq (\beta \hq+\hq+\beta)} \, ,
        \\
        q &= g^A g^B \, .
    \end{split}
\end{equation}
Notice that $q$ does not depend on $r$ in this case, so that all low-rank problems have the same MMSE. 
This is due to fact that the priors over the factors are i.i.d..

\section{Consequences of the main results}

In this appendix we derive various consequences of Result \ref{res1} and Previous Result \ref{prevreslow}.
In particular:
\begin{itemize}
    \item In Appendix \ref{app:strong_int} we derive the weak and strong recovery thresholds for the BSR model in the intensive width regime, as presented in Section \ref{sec:strong}.
    \item In Appendix \ref{app:strong_ext} we derive the strong recovery threshold for the BSR model in the extensive width regime, as presented in Result \ref{res:strong}.
    \item In Appendix \ref{app:large_beta_int} we derive the large $\beta$ limit of the BO error for the BSR model in the intensive width regime.
    \item In Appendix \ref{app:large_beta_ext} we derive the large $\beta$ limit of the BO error for the BSR model in the extensive width regime.
    \item In Appendix \ref{app:large_rho_ext} we derive the large $\rho$ limit of the BO error for the BSR model, and show that it equals the error of optimally-regularised ridge regression.
\end{itemize}

\subsection{Weak and strong recovery thresholds in the intensive width BSR model without noise}\label{app:strong_int}

The state equations are \eqref{eq:ap-state-output-gaussian} and \eqref{eq:ap-state-intensive}, i.e.
\begin{equation}
    \begin{split}
        g^A &= \frac{(\beta+1)^2 \hq^2-\beta}{(\beta+1) \hq (\beta \hq+\hq+1)} \, ,
        \\
        g^B &= \frac{(\beta+1)^2 \hq^2-\beta}{(\beta+1) \hq (\beta \hq+\hq+\beta)} \, ,
        \\
        q &= g^A g^B \, , 
        \\
        \hq &= \frac{\bar{\alpha} }{1 - q} \, .
    \end{split}
\end{equation}
The strong recovery threshold $\bar{\alpha}_{\rm BO}$ is such that 
\begin{equation}\label{eq:ap-strong-rec}
    \bar{\alpha}_{\rm BO} 
    = \lim_{\hq \to \infty} \hq (1-q(\hq)) 
    = 1 \, ,
\end{equation}
as can be seen by explicitly computing the limit.
In the main text scaling $n = \alpha d L$ this translates to
\begin{equation}
    \alpha_{\rm BO} = \frac{\rho(1+\beta)}{\beta} = 0 \, ,
\end{equation}
as for $r \ll d$ then $\rho \to 0$. 
This highlights the importance of the scaling $r(d+L)$ to study both intensive and extensive width in a common scaling.

The weak recovery threshold, i.e. the value $\bar{\alpha}_{\rm weak}$ at which $q=0$ can be found by imposing $q=0$, which gives either $g^A = 0$ or $g^B = 0$ implying $(\beta + 1)^2 \hq^2 - \beta = 0$. Combined with the equation for $\hq$ this gives
\begin{equation}
    \bar{\alpha}_{\rm weak} = (1+\Delta) \hq = (1+\Delta) \frac{\sqrt{\beta}}{1 + \beta} \, .
\end{equation}
Notice that the weak threshold is non-trivial also in the noiseless case.
 
\subsection{Strong recovery threshold in the extensive width BSR model without noise}\label{app:strong_ext}

We have $\MMSE = 1 - q$ with the state equations \eqref{eq:ap-state-output-gaussian} and \eqref{eq:ap-state-extensive}, i.e.
\begin{equation}
\begin{split}
    \delta &= \frac{\beta}{\rho (1+\beta)} \frac{1-q}{\bar{\alpha}}\, ,
    \\
    q &= 1
        - \delta
        + \delta^2
        \int dx \, \hmu_{\uS + \sqrt{\delta} \uZ}(x) \left[
            \frac{(\beta-1)^2}{\beta^{3/2} x^2} 
            + \frac{4\pi^2}{3 \beta^{3/2}} \hmu_{\uS + \sqrt{\delta} \uZ}(x)^2
            \right]\, .
\end{split}
\end{equation}
Strong recovery happens for $\bar{\alpha}_{\rm BO}$ such that $\MMSE = 0$, i.e. for $q \to 1$ and $\delta \to 0$.
Rearranging the equations, we have
\begin{equation}
    \bar{\alpha}_{\rm BO}
    = \lim_{\delta \to 0} \frac{\beta}{\rho (1+\beta)} \frac{1-q(\delta)}{\delta} 
    = \frac{\beta}{\rho (1+\beta)} - \frac{\beta}{\rho (1+\beta)} \lim_{\delta \to 0} \delta
        \int dx \, \hmu_{\uS + \sqrt{\delta} \uZ}(x) \left[
            \frac{(\beta-1)^2}{\beta^{3/2} x^2} 
            + \frac{4\pi^2}{3 \beta^{3/2}} \hmu_{\uS + \sqrt{\delta} \uZ}(x)^2
            \right]\, .
\end{equation}
Thus, we need to study the second equation in the limit of $\delta \to 0$, and specifically look whether the integral terms develop divergencies at $\delta \to 0$.

Notice that, assuming that the last limit-integral term is finite as we will verify later, $q=1, \delta = 0$ is a solution of these equations for all values of $\bar\alpha$.
This is kind of a trivial solution, and we expect that it is not the only solution for low enough values of $\bar\alpha$. Thus, to find the strong recovery threshold, which can be seen as the bifurcation point at which the non-trivial, low-$\bar\alpha$ solution of the equations merges with the trivial one, we assume that $q<1$ and $\delta > 0$, and take the limit $q \to 1^-$ and $\delta \to 0^+$, effectively moving along the non-trivial solution towards the bifurcation point.

Intuitively, for $\delta \to 0$, the spectral density $\hmu_{\uS + \sqrt{\delta} \uZ}(x)$ will be composed either by a single bulk (if the width parameter does not constrain the rank of $\uS$, i.e.  if $\rho \geq \min(1, \beta)$) or by two bulks (if the width parameter constrains the rank of $\uS$, i.e. $0 < \rho < \min(1, \beta)$), one gapped away from zero and the other close to zero. 
This is due to the fact that the spectrum of $\uS$ is either composed by a bulk gapped away from zero or diverging at zero, or by a bulk gapped away from zero and an additional delta accounting for the rank deficiency. Gaussian noise with vanishingly small variance alters this picture only perturbatively.

Then, if no rank deficiency is present ($0 < \rho < \min(1, \beta)$), and assuming that ungapped bulks are not problematic, the integrals will have no divergence and 
\begin{equation}
    \bar{\alpha}_c = \frac{\beta}{\rho (1+\beta)} \, .
\end{equation}
Otherwise, the first integral will develop a divergence due to the interplay of the noised delta peak of the spectrum of $\uS$ and the $1/x^2$ factor in the integral, leading to a non-trivial $\bar{\alpha}_c$.

To follow this intuition, we perform the change of variable $x = \sqrt{\delta} z$ in the integrals, so that
\begin{equation}
    \bar{\alpha}_{\rm BO} 
    = \frac{\beta}{\rho (1+\beta)} - \frac{\beta}{\rho (1+\beta)} \lim_{\delta \to 0}
        \int dz \, \sqrt{\delta} \hmu_{\uS + \sqrt{\delta} \uZ}(\sqrt{\delta} z) \left[
            \frac{(\beta-1)^2}{\beta^{3/2} z^2} 
            + \frac{4\pi^2}{3 \beta^{3/2}} \left(\sqrt{\delta} \hmu_{\uS + \sqrt{\delta} \uZ}(\sqrt{\delta} z) \right)^2
            \right]\, .
\end{equation}
The scaling $\sqrt{\delta}$ can be guessed as it allows to identify an expression which depends on $\delta$ only through the rescaled density $\sqrt{\delta} \hmu_{\uS + \sqrt{\delta} \uZ}(\sqrt{\delta} z)$, which we now study.

As shown in~\cite{pennington2017nonlinear} (but we follow the notations and definitions of \cite{troiani2022optimal}) the Stieltjes transform of $\uY = \uS + \sqrt{\delta} \uZ$ satisfies
\begin{equation}
    g_\uY(x) = z g_{\uY\uY^T}(x^2)    
\end{equation}
where $g_{\uY\uY^T}(x^2)$ is a root of the polynomial $p(G) = \sum_{a=0}^4 a_k(x^2, \delta) G^k$.
We plug in the equation $x = \sqrt{\delta} z$, take the scaling ansatz $G = H / \delta$, and expand everything at leading order for $\delta \to 0$, obtaining (after simplifying an overall factor of $\delta$)
\begin{equation}
    H(z^2) = \frac{\sqrt{\beta z^4-2 \sqrt{\beta} z^2 (\beta-2 \rho+1)+(\beta-1)^2}+\sqrt{\beta} z^2-\beta+1}{2 z^2}
\end{equation}
(the other solutions have either the wrong sign in front of the square root, or no square root).
Here we defined $\rho = \rho / \min(1,\beta)$ for simplicity.
Now, recall that $\mu$ is the discontinuity at branch cuts over the real axes of $H$. Thus, only the square root term will contribute, and only when its argument is negative.
The roots of the argument of the square roots are
\begin{equation}
    z_{\pm} = \frac{\pm 2 \sqrt{(\rho-1) (\rho-\beta)}+\beta-2 \rho+1}{\sqrt{\beta}}
\end{equation} 
giving the distribution
\begin{equation}
    f(z) = \frac{1}{\pi} \lim_{\epsilon \to 0^+} \Im \left[ (x - i \epsilon) H((x - i \epsilon)^2) \right] = \frac{\sqrt{\beta (z_+ - z^2)(z^2 - z_-)}}{2 \pi z}
\end{equation}
We can check using Mathematica that
\begin{equation}
    \begin{split}
        2 \int_{z_-}^{z_+} dz \, f(z) &= (1-\rho) \\
        2 \int_{z_-}^{z_+} dz \, f(z)^3 &= \frac{3 \sqrt{\beta} (1 - \rho)^2}{4 \pi ^2} \\
        2 \int_{z_-}^{z_+} dz \, f(z) z^{-2} &= \frac{\sqrt{\beta} (1-\rho)}{\beta-1} \\
    \end{split}
\end{equation}
The normalisation is correct, as we are in a scaling limit where only the noisy version of the $(1-\rho) \delta_0$ contribution to the non-noisy measure.
Notice that all this derivation holds only for $0 < \rho < 1$, as $0 < z_- < z_+$ holds only in this case.
For $\rho \geq 1$, the square root never develops a branch cut, so that the function $f(z)$ is identically zero. 

Thus, we get for $\rho \geq 1$
\begin{equation}
    \bar{\alpha}_{\rm BO} = \frac{\beta}{\rho (1+\beta)} \, ,
\end{equation}
as expected, 
and for $0 < \rho < 1$
\begin{equation}
    \bar{\alpha}_{\rm BO} 
    = \frac{\beta}{\rho (1+\beta)} - \frac{\beta}{\rho (1+\beta)}
        \frac{(1 - \rho)(\beta- \rho)}{\beta} 
    = \frac{\beta}{\rho (1+\beta)} - \frac{\left(\rho  - 1\right)\left(\rho - \beta\right)}{\rho (1+\beta)}
    = 1 - \frac{\rho}{1+\beta}
    \, .
\end{equation}
For $\rho \to 0$, the threshold reduces to $\bar{\alpha}_c = 1$, as found explicitly in the low-width case \eqref{eq:ap-strong-rec}.

In the main text scaling $n = \alpha \, d L$ this translates to
\begin{equation}
    \alpha_{\rm BO} = 
    \begin{cases}
       \frac{\rho}{\beta} \left( 1 + \beta - \rho  \right) & 0 < \rho < 1 \, , \\ 
       1 & \rho > 1 \, .\\
    \end{cases}
\end{equation}

\subsection{Large \texorpdfstring{$\beta$}{beta} limit for the intensive width BSR model}\label{app:large_beta_int}

The state equations are \eqref{eq:ap-state-output-gaussian} and \eqref{eq:ap-state-intensive}, i.e.
\begin{equation}
    \begin{split}
        g^A &= \frac{(\beta+1)^2 \hq^2-\beta}{(\beta+1) \hq (\beta \hq+\hq+1)} \, ,
        \\
        g^B &= \frac{(\beta+1)^2 \hq^2-\beta}{(\beta+1) \hq (\beta \hq+\hq+\beta)} \, ,
        \\
        q &= g^A g^B \, , 
        \\
        \hq &= \frac{\bar{\alpha} }{1 - q + \Delta} \, .
    \end{split}
\end{equation}

For large $\beta$, assuming that $\hq$ remains finite, we have
\begin{equation}
    \begin{split}
        g^A &\sim \frac{\beta^2 \hq^2}{\beta \hq (\beta \hq)} = 1 \, ,
        \\
        g^B &\sim \frac{\beta^2 \hq^2}{\beta \hq (\beta \hq +\beta)} = \frac{\hq}{1+\hq} \, ,
        \\
        q &\sim \frac{\hq}{1+\hq} \, , 
        \\
        \hq &= \frac{\bar{\alpha} }{1 - q + \Delta} \, .
    \end{split}
\end{equation}
The equations can be solved explicitly to 
\begin{equation}
    q = \frac{1}{2} \left(1 + \Delta + \bar{\alpha} - \sqrt{(\bar{\alpha}-1)^2+2 (\bar{\alpha}+1) \Delta+\Delta^2}\right) \, ,
\end{equation}
which reduces to 
\begin{equation}
    q = \min(1, \bar{\alpha}) \, ,
\end{equation}
in the noiseless case $\Delta = 0$.

\subsection{Large \texorpdfstring{$\beta$}{beta} limit for the extensive width BSR model}\label{app:large_beta_ext}

We have the state equations \eqref{eq:ap-state-output-gaussian} and \eqref{eq:ap-state-extensive}, i.e.
\begin{equation}
\begin{split}
    \delta &= \frac{1-q + \Delta}{\alpha}\, ,
    \\
    q &= 1
        - \delta
        + \delta^2
        \int dx \, \hmu_{\uS + \sqrt{\delta} \uZ}(x) \left[
            \frac{(\beta-1)^2}{\beta^{3/2} x^2} 
            + \frac{4\pi^2}{3 \beta^{3/2}} \hmu_{\uS + \sqrt{\delta} \uZ}(x)^2
            \right]\, ,
\end{split}
\end{equation}
where we rescaled $\alpha$ and $q = 1/\delta$ to match the scaling $\alpha = n / (dL)$ of the main text w.r.t. Appendix \ref{app.replica}.

We will show below that 
\begin{equation}
    f(w) = \lim_{\beta \to \infty} \sqrt[4]{\beta} \hmu_{\uS + \sqrt{\delta} \uZ}\left(\sqrt[4]{\beta} w\right)
\end{equation}
for a finite and compactly supported function $f(w)$ fully independent on $\beta$. 
This implies that
\begin{equation}
\begin{split}
    q &= 1
        - \delta
        + \delta^2
        \int dx \, \hmu_{\uS + \sqrt{\delta} \uZ}(x) \left[
            \frac{(\beta-1)^2}{\beta^{3/2} x^2} 
            + \frac{4\pi^2}{3 \beta^{3/2}} \hmu_{\uS + \sqrt{\delta} \uZ}(x)^2
            \right]
    \\
    &\sim 1
        - \delta
        + \delta^2
        \int dw \, 
        \sqrt[4]{\beta} \hmu_{\uS + \sqrt{\delta} \uZ}\left(\sqrt[4]{\beta} w\right) 
        \left[
            \frac{1}{w^2} 
            + \frac{4\pi^2}{3 \beta^{2}} 
            \sqrt{\beta} \hmu_{\uS + \sqrt{\delta} \uZ}\left(\sqrt[4]{\beta} w\right)^2
            \right]
    \\
    &\sim 1
        - \delta
        + \delta^2
        \int dw \, 
        f(w)
        \left[
            \frac{1}{w^2} 
            + \frac{4\pi^2}{3 \beta^{2}} 
            f(w)^2
            \right]
            \\
    &\sim 1
        - \delta
        + \delta^2
        \int dw \, 
        \frac{f(w)}{w^2} 
\end{split}
\end{equation}
where we changed variable $x = \sqrt[4]{\beta} w$, and used that
\begin{equation}
    \int dw f(w)^3 < + \infty
\end{equation}
as $f(w)$ is finite and compactly supported.
We now need to compute $f$ and show that it is indeed well behaved.

Recall that the Stieltjes transform of $\hmu_{\uS + \sqrt{\delta} \uZ}$ satisfies \cite{pennington2017nonlinear, troiani2022optimal}
\begin{equation}
    g_{\uY}(z) = z G_{\uY^T \uY}(z^2)
\end{equation}
so that
\begin{equation}
    \sqrt[4]{\beta} g_{\uY}(z) \left( \sqrt[4]{\beta} w \right)
    = w \sqrt{\beta} \, G_{\uY^T \uY}\left( \sqrt{\beta} w^2 \right) \, .
\end{equation}
Now, recall that $G_{\uY^T\uY}(z)$ satisfies a quartic polynomial equation. By rescaling $z = \sqrt[4]{\beta} w$ and $G = H / \sqrt{\beta}$ and considering the leading order in $\beta$. one obtains the simplified quadratic equation
\begin{equation}
    H^2 (\delta  -  w^2) 
    + H  (1 + \rho w^2 -(1 + \delta) \rho) 
    -\rho 
    = 0
\end{equation}
which solves to 
\begin{equation}
    w H(w) = w \frac{\sqrt{\left(-\delta \rho+\rho w^2-\rho+1\right)^2+4 \rho \left(\delta-w^2\right)}+\delta \rho-\rho w^2+\rho-1}{2 \left(\delta-w^2\right)} \, .
\end{equation}
leading by the standard inversion formula to the associated measure
\begin{equation}
    f(w) = \frac{\rho w}{2 \pi (w^2 - \delta)} \sqrt{ \left(b_+ - w^2 \right)  \left(w^2 - b_- \right) } \, ,
\end{equation}
supported on 
\begin{equation}
     b_- < w^2 < b_+ \, ,
\end{equation}
and on the symmetric interval,
where we defined
\begin{equation}
    b_\pm = 1 + \delta + \frac{1}{\rho} \pm \frac{2}{\sqrt{\rho}} \, .
\end{equation}
This confirms that $f$ is finite and compactly supported, as $t \leq b_- < b_+$ for all $\delta$ and $\rho$.
We can also verify that
\begin{equation}
    \begin{split}
        \int dw \, 
        f(w)
        &=
        2 \frac{\rho}{2 \pi} \int_{\sqrt{b_-}}^{\sqrt{b_+}} dw w \frac{\sqrt{ \left(b_+ - w^2 \right)  \left(w^2 - b_- \right) } }{w^2 - \delta} \\
        &=
        \frac{\rho}{2 \pi} \int_{{b_-}}^{{b_+}} dt \frac{\sqrt{ \left(b_+ - t \right)  \left(t - b_- \right) } }{t - \delta} \\
        &=
        \frac{\rho}{2 \pi}
        \begin{cases}
            2 \pi & \mathif \rho < 1 \, , \\
            \frac{2\pi}{\rho} & \quad \text{otherwise} \, ,
        \end{cases}
        \\
        &=
        \min(\rho, 1)
        \, ,
    \end{split}
\end{equation}
where we used $t=w^2$. This is the correct normalisation for the bulk of the distribution, excluding the rank-deficiency-induced spike in zero of mass $\max(0, 1-\rho)$.

Now, we need to compute
\begin{equation}
    \begin{split}
        \int dw \, 
        \frac{f(w)}{w^2} 
        &=
        2 \frac{\rho}{2 \pi} \int_{\sqrt{b_-}}^{\sqrt{b_+}} dw w \frac{\sqrt{ \left(b_+ - w^2 \right)  \left(w^2 - b_- \right) } }{w^2 (w^2 - \delta)} \\
        &=
        \frac{\rho}{2 \pi} \int_{{b_-}}^{{b_+}} dt \frac{\sqrt{ \left(b_+ - t \right)  \left(t - b_- \right) } }{t (t - \delta)} \\
        &=
        \frac{\sqrt{1 + \rho \left( \rho - 2 + 2 \delta + (2+\delta) \delta \rho \right)} - \delta \rho + \rho - 1}{2\delta}
        \, ,
    \end{split}
\end{equation}
where we used $t=w^2$.
Thus, we obtain the equations
\begin{equation}
\begin{split}
    \delta &= \frac{1-q + \Delta}{\alpha}\, ,
    \\
    q &= 1 - \delta 
    + \delta \frac{\sqrt{1 + \rho \left( \rho - 2 + 2 \delta + (2+\delta) \delta \rho \right)} - \delta \rho + \rho - 1}{2} \, ,
\end{split}
\end{equation}
to be solved for $q$.

\subsection{Large \texorpdfstring{$\rho$}{rho} limit for the extensive width BSR model}\label{app:large_rho_ext}

We have the state equations \eqref{eq:ap-state-output-gaussian} and \eqref{eq:ap-state-extensive}, i.e.
\begin{equation}
\begin{split}
    \delta &= \frac{1-q + \Delta}{\alpha}\, ,
    \\
    q &= 1
        - \delta
        + \delta^2
        \int dx \, \hmu_{\uS + \sqrt{\delta} \uZ}(x) \left[
            \frac{(\beta-1)^2}{\beta^{3/2} x^2} 
            + \frac{4\pi^2}{3 \beta^{3/2}} \hmu_{\uS + \sqrt{\delta} \uZ}(x)^2
            \right]\, ,
\end{split}
\end{equation}
where we rescaled $\alpha$ and $q = 1/\delta$ to match the scaling $\alpha = n / (dL)$ of the main text w.r.t. Appendix \ref{app.replica}.

We want to compute the $\rho \to \infty$ limit of the equations.
Recall that the Stieltjes transform of $\hmu_{\uS + \sqrt{\delta} \uZ}$ satisfies \cite{pennington2017nonlinear, troiani2022optimal}
\begin{equation}
    g_{\uY}(z) = z G_{\uY^T \uY}(z^2)
\end{equation}
and that $G_{\uY^T\uY}(z)$ satisfies a quartic polynomial equation. One obtains the simplified quadratic equation in the large $\rho$ limit
\begin{equation}
    G^2 \frac{z^2 (1+\delta)}{\sqrt{\beta}}
    + G \left(z^2 - \frac{(1+\delta)(\beta - 1)}{\sqrt{\beta}} \right) 
    -1
    = 0
\end{equation}
which solves to 
\begin{equation}
    z G(z^2) = z
    \frac{\sqrt{\beta} \left(\sqrt{-\frac{2 (\beta+1) (\delta+1) z^2}{\sqrt{\beta}}+\frac{(\beta-1)^2 (\delta+1)^2}{\beta}+z^4}+z^2\right)-\beta (\delta+1)+\delta+1}{2 (\delta+1) z^2}
    \, .
\end{equation}
leading by the standard inversion formula to the associated measure
\begin{equation}
    f(z) = \frac{\sqrt{\beta}}{1+\delta} \frac{\sqrt{ \left(b_+ - z^2 \right)  \left(z^2 - b_- \right) }}{2\pi z} \, ,
\end{equation}
supported on 
\begin{equation}
     b_- < z^2 < b_+ \, ,
\end{equation}
and on the symmetric interval,
where we defined
\begin{equation}
    b_\pm = (\sqrt{\beta} \pm 1)^2 \frac{1+\delta}{\sqrt{\beta}} \, .
\end{equation}

We can verify that
\begin{equation}
    \begin{split}
        2 \int_{b_-}^{b_+} dz \, 
        f(z)
        &=
        1
        \, ,
    \end{split}
\end{equation}
This is the correct normalisation for the bulk of the distribution.
We also have
\begin{equation}
    \begin{split}
        2 \int_{b_-}^{b_+} dz \, 
        f(z) \frac{1}{z^2}
        &=
        \frac{\sqrt\beta}{(\beta-1) (\delta+1)}
        \, ,
    \end{split}
\end{equation}
and
\begin{equation}
    \begin{split}
        2 \int_{b_-}^{b_+} dz \, 
        f(z)^3
        &=
        \frac{3 \sqrt{\beta}}{4 \pi ^2 (1+\delta)}
        \, ,
    \end{split}
\end{equation}
from which one gets the equations
\begin{equation}
\begin{split}
    \delta &= \frac{1-q + \Delta}{\alpha}\, ,
    \\
    q &= \frac{1}{1+\delta}\, ,
\end{split}
\end{equation}
whose solution gives
\begin{equation}
    \frac{1+\alpha+\Delta-\sqrt{(\alpha+\Delta+1)^2-4 \alpha}}{2} \, .
\end{equation}
This coincides with the overlap achieved by optimally-regularised ridge regression Previous Result \ref{prevres-regr}.


\begin{thebibliography}{87}%
\makeatletter
\providecommand \@ifxundefined [1]{%
 \@ifx{#1\undefined}
}%
\providecommand \@ifnum [1]{%
 \ifnum #1\expandafter \@firstoftwo
 \else \expandafter \@secondoftwo
 \fi
}%
\providecommand \@ifx [1]{%
 \ifx #1\expandafter \@firstoftwo
 \else \expandafter \@secondoftwo
 \fi
}%
\providecommand \natexlab [1]{#1}%
\providecommand \enquote  [1]{``#1''}%
\providecommand \bibnamefont  [1]{#1}%
\providecommand \bibfnamefont [1]{#1}%
\providecommand \citenamefont [1]{#1}%
\providecommand \href@noop [0]{\@secondoftwo}%
\providecommand \href [0]{\begingroup \@sanitize@url \@href}%
\providecommand \@href[1]{\@@startlink{#1}\@@href}%
\providecommand \@@href[1]{\endgroup#1\@@endlink}%
\providecommand \@sanitize@url [0]{\catcode `\\12\catcode `\$12\catcode `\&12\catcode `\#12\catcode `\^12\catcode `\_12\catcode `\%12\relax}%
\providecommand \@@startlink[1]{}%
\providecommand \@@endlink[0]{}%
\providecommand \url  [0]{\begingroup\@sanitize@url \@url }%
\providecommand \@url [1]{\endgroup\@href {#1}{\urlprefix }}%
\providecommand \urlprefix  [0]{URL }%
\providecommand \Eprint [0]{\href }%
\providecommand \doibase [0]{https://doi.org/}%
\providecommand \selectlanguage [0]{\@gobble}%
\providecommand \bibinfo  [0]{\@secondoftwo}%
\providecommand \bibfield  [0]{\@secondoftwo}%
\providecommand \translation [1]{[#1]}%
\providecommand \BibitemOpen [0]{}%
\providecommand \bibitemStop [0]{}%
\providecommand \bibitemNoStop [0]{.\EOS\space}%
\providecommand \EOS [0]{\spacefactor3000\relax}%
\providecommand \BibitemShut  [1]{\csname bibitem#1\endcsname}%
\let\auto@bib@innerbib\@empty
\bibitem [{\citenamefont {LeCun}\ \emph {et~al.}(2015)\citenamefont {LeCun}, \citenamefont {Bengio},\ and\ \citenamefont {Hinton}}]{lecun2015deep}%
  \BibitemOpen
  \bibfield  {author} {\bibinfo {author} {\bibfnamefont {Y.}~\bibnamefont {LeCun}}, \bibinfo {author} {\bibfnamefont {Y.}~\bibnamefont {Bengio}},\ and\ \bibinfo {author} {\bibfnamefont {G.}~\bibnamefont {Hinton}},\ }\bibfield  {title} {\bibinfo {title} {Deep learning},\ }\href@noop {} {\bibfield  {journal} {\bibinfo  {journal} {nature}\ }\textbf {\bibinfo {volume} {521}},\ \bibinfo {pages} {436} (\bibinfo {year} {2015})}\BibitemShut {NoStop}%
\bibitem [{\citenamefont {Deng}\ \emph {et~al.}(2009)\citenamefont {Deng}, \citenamefont {Dong}, \citenamefont {Socher}, \citenamefont {Li}, \citenamefont {Li},\ and\ \citenamefont {Fei-Fei}}]{deng2009imagenet}%
  \BibitemOpen
  \bibfield  {author} {\bibinfo {author} {\bibfnamefont {J.}~\bibnamefont {Deng}}, \bibinfo {author} {\bibfnamefont {W.}~\bibnamefont {Dong}}, \bibinfo {author} {\bibfnamefont {R.}~\bibnamefont {Socher}}, \bibinfo {author} {\bibfnamefont {L.-J.}\ \bibnamefont {Li}}, \bibinfo {author} {\bibfnamefont {K.}~\bibnamefont {Li}},\ and\ \bibinfo {author} {\bibfnamefont {L.}~\bibnamefont {Fei-Fei}},\ }\bibfield  {title} {\bibinfo {title} {Imagenet: A large-scale hierarchical image database},\ }in\ \href@noop {} {\emph {\bibinfo {booktitle} {2009 IEEE conference on computer vision and pattern recognition}}}\ (\bibinfo {organization} {Ieee},\ \bibinfo {year} {2009})\ pp.\ \bibinfo {pages} {248--255}\BibitemShut {NoStop}%
\bibitem [{\citenamefont {Krizhevsky}\ \emph {et~al.}(2012)\citenamefont {Krizhevsky}, \citenamefont {Sutskever},\ and\ \citenamefont {Hinton}}]{krizhevsky2012imagenet}%
  \BibitemOpen
  \bibfield  {author} {\bibinfo {author} {\bibfnamefont {A.}~\bibnamefont {Krizhevsky}}, \bibinfo {author} {\bibfnamefont {I.}~\bibnamefont {Sutskever}},\ and\ \bibinfo {author} {\bibfnamefont {G.~E.}\ \bibnamefont {Hinton}},\ }\bibfield  {title} {\bibinfo {title} {Imagenet classification with deep convolutional neural networks},\ }\bibfield  {booktitle} {\emph {\bibinfo {booktitle} {Advances in Neural Information Processing Systems}},\ }\href@noop {} {\ \textbf {\bibinfo {volume} {25}} (\bibinfo {year} {2012})}\BibitemShut {NoStop}%
\bibitem [{\citenamefont {Silver}\ \emph {et~al.}(2016)\citenamefont {Silver}, \citenamefont {Huang}, \citenamefont {Maddison}, \citenamefont {Guez}, \citenamefont {Sifre}, \citenamefont {Van Den~Driessche}, \citenamefont {Schrittwieser}, \citenamefont {Antonoglou}, \citenamefont {Panneershelvam}, \citenamefont {Lanctot} \emph {et~al.}}]{silver2016mastering}%
  \BibitemOpen
  \bibfield  {author} {\bibinfo {author} {\bibfnamefont {D.}~\bibnamefont {Silver}}, \bibinfo {author} {\bibfnamefont {A.}~\bibnamefont {Huang}}, \bibinfo {author} {\bibfnamefont {C.~J.}\ \bibnamefont {Maddison}}, \bibinfo {author} {\bibfnamefont {A.}~\bibnamefont {Guez}}, \bibinfo {author} {\bibfnamefont {L.}~\bibnamefont {Sifre}}, \bibinfo {author} {\bibfnamefont {G.}~\bibnamefont {Van Den~Driessche}}, \bibinfo {author} {\bibfnamefont {J.}~\bibnamefont {Schrittwieser}}, \bibinfo {author} {\bibfnamefont {I.}~\bibnamefont {Antonoglou}}, \bibinfo {author} {\bibfnamefont {V.}~\bibnamefont {Panneershelvam}}, \bibinfo {author} {\bibfnamefont {M.}~\bibnamefont {Lanctot}}, \emph {et~al.},\ }\bibfield  {title} {\bibinfo {title} {Mastering the game of go with deep neural networks and tree search},\ }\href@noop {} {\bibfield  {journal} {\bibinfo  {journal} {nature}\ }\textbf {\bibinfo {volume} {529}},\ \bibinfo {pages} {484} (\bibinfo {year} {2016})}\BibitemShut {NoStop}%
\bibitem [{\citenamefont {Zdeborov{\'a}}(2020)}]{zdeborova2020understanding}%
  \BibitemOpen
  \bibfield  {author} {\bibinfo {author} {\bibfnamefont {L.}~\bibnamefont {Zdeborov{\'a}}},\ }\bibfield  {title} {\bibinfo {title} {Understanding deep learning is also a job for physicists},\ }\href@noop {} {\bibfield  {journal} {\bibinfo  {journal} {Nature Physics}\ }\textbf {\bibinfo {volume} {16}},\ \bibinfo {pages} {602} (\bibinfo {year} {2020})}\BibitemShut {NoStop}%
\bibitem [{\citenamefont {Hopfield}(1982)}]{hopfield1982neural}%
  \BibitemOpen
  \bibfield  {author} {\bibinfo {author} {\bibfnamefont {J.~J.}\ \bibnamefont {Hopfield}},\ }\bibfield  {title} {\bibinfo {title} {Neural networks and physical systems with emergent collective computational abilities.},\ }\href@noop {} {\bibfield  {journal} {\bibinfo  {journal} {Proceedings of the national academy of sciences}\ }\textbf {\bibinfo {volume} {79}},\ \bibinfo {pages} {2554} (\bibinfo {year} {1982})}\BibitemShut {NoStop}%
\bibitem [{\citenamefont {Ackley}\ \emph {et~al.}(1985)\citenamefont {Ackley}, \citenamefont {Hinton},\ and\ \citenamefont {Sejnowski}}]{ackley1985learning}%
  \BibitemOpen
  \bibfield  {author} {\bibinfo {author} {\bibfnamefont {D.~H.}\ \bibnamefont {Ackley}}, \bibinfo {author} {\bibfnamefont {G.~E.}\ \bibnamefont {Hinton}},\ and\ \bibinfo {author} {\bibfnamefont {T.~J.}\ \bibnamefont {Sejnowski}},\ }\bibfield  {title} {\bibinfo {title} {A learning algorithm for boltzmann machines},\ }\href@noop {} {\bibfield  {journal} {\bibinfo  {journal} {Cognitive science}\ }\textbf {\bibinfo {volume} {9}},\ \bibinfo {pages} {147} (\bibinfo {year} {1985})}\BibitemShut {NoStop}%
\bibitem [{\citenamefont {Gardner}\ and\ \citenamefont {Derrida}(1988)}]{gardner1988optimal}%
  \BibitemOpen
  \bibfield  {author} {\bibinfo {author} {\bibfnamefont {E.}~\bibnamefont {Gardner}}\ and\ \bibinfo {author} {\bibfnamefont {B.}~\bibnamefont {Derrida}},\ }\bibfield  {title} {\bibinfo {title} {Optimal storage properties of neural network models},\ }\href@noop {} {\bibfield  {journal} {\bibinfo  {journal} {Journal of Physics A: Mathematical and general}\ }\textbf {\bibinfo {volume} {21}},\ \bibinfo {pages} {271} (\bibinfo {year} {1988})}\BibitemShut {NoStop}%
\bibitem [{\citenamefont {Gardner}\ and\ \citenamefont {Derrida}(1989)}]{gardner1989three}%
  \BibitemOpen
  \bibfield  {author} {\bibinfo {author} {\bibfnamefont {E.}~\bibnamefont {Gardner}}\ and\ \bibinfo {author} {\bibfnamefont {B.}~\bibnamefont {Derrida}},\ }\bibfield  {title} {\bibinfo {title} {Three unfinished works on the optimal storage capacity of networks},\ }\href@noop {} {\bibfield  {journal} {\bibinfo  {journal} {Journal of Physics A: Mathematical and General}\ }\textbf {\bibinfo {volume} {22}},\ \bibinfo {pages} {1983} (\bibinfo {year} {1989})}\BibitemShut {NoStop}%
\bibitem [{\citenamefont {Seung}\ \emph {et~al.}(1992)\citenamefont {Seung}, \citenamefont {Sompolinsky},\ and\ \citenamefont {Tishby}}]{seung1992statistical}%
  \BibitemOpen
  \bibfield  {author} {\bibinfo {author} {\bibfnamefont {H.~S.}\ \bibnamefont {Seung}}, \bibinfo {author} {\bibfnamefont {H.}~\bibnamefont {Sompolinsky}},\ and\ \bibinfo {author} {\bibfnamefont {N.}~\bibnamefont {Tishby}},\ }\bibfield  {title} {\bibinfo {title} {Statistical mechanics of learning from examples},\ }\href@noop {} {\bibfield  {journal} {\bibinfo  {journal} {Physical review A}\ }\textbf {\bibinfo {volume} {45}},\ \bibinfo {pages} {6056} (\bibinfo {year} {1992})}\BibitemShut {NoStop}%
\bibitem [{\citenamefont {Saxe}\ \emph {et~al.}(2014)\citenamefont {Saxe}, \citenamefont {McClelland},\ and\ \citenamefont {Ganguli}}]{saxe2013exact}%
  \BibitemOpen
  \bibfield  {author} {\bibinfo {author} {\bibfnamefont {A.~M.}\ \bibnamefont {Saxe}}, \bibinfo {author} {\bibfnamefont {J.~L.}\ \bibnamefont {McClelland}},\ and\ \bibinfo {author} {\bibfnamefont {S.}~\bibnamefont {Ganguli}},\ }\bibfield  {title} {\bibinfo {title} {Exact solutions to the nonlinear dynamics of learning in deep linear neural networks},\ }\href@noop {} {\bibfield  {journal} {\bibinfo  {journal} {International Conference on Learning Representations}\ } (\bibinfo {year} {2014})}\BibitemShut {NoStop}%
\bibitem [{\citenamefont {Baldassi}\ \emph {et~al.}(2015)\citenamefont {Baldassi}, \citenamefont {Ingrosso}, \citenamefont {Lucibello}, \citenamefont {Saglietti},\ and\ \citenamefont {Zecchina}}]{baldassi2015subdominant}%
  \BibitemOpen
  \bibfield  {author} {\bibinfo {author} {\bibfnamefont {C.}~\bibnamefont {Baldassi}}, \bibinfo {author} {\bibfnamefont {A.}~\bibnamefont {Ingrosso}}, \bibinfo {author} {\bibfnamefont {C.}~\bibnamefont {Lucibello}}, \bibinfo {author} {\bibfnamefont {L.}~\bibnamefont {Saglietti}},\ and\ \bibinfo {author} {\bibfnamefont {R.}~\bibnamefont {Zecchina}},\ }\bibfield  {title} {\bibinfo {title} {Subdominant dense clusters allow for simple learning and high computational performance in neural networks with discrete synapses},\ }\href@noop {} {\bibfield  {journal} {\bibinfo  {journal} {Physical review letters}\ }\textbf {\bibinfo {volume} {115}},\ \bibinfo {pages} {128101} (\bibinfo {year} {2015})}\BibitemShut {NoStop}%
\bibitem [{\citenamefont {Baldassi}\ \emph {et~al.}(2016)\citenamefont {Baldassi}, \citenamefont {Borgs}, \citenamefont {Chayes}, \citenamefont {Ingrosso}, \citenamefont {Lucibello}, \citenamefont {Saglietti},\ and\ \citenamefont {Zecchina}}]{baldassi2016unreasonable}%
  \BibitemOpen
  \bibfield  {author} {\bibinfo {author} {\bibfnamefont {C.}~\bibnamefont {Baldassi}}, \bibinfo {author} {\bibfnamefont {C.}~\bibnamefont {Borgs}}, \bibinfo {author} {\bibfnamefont {J.~T.}\ \bibnamefont {Chayes}}, \bibinfo {author} {\bibfnamefont {A.}~\bibnamefont {Ingrosso}}, \bibinfo {author} {\bibfnamefont {C.}~\bibnamefont {Lucibello}}, \bibinfo {author} {\bibfnamefont {L.}~\bibnamefont {Saglietti}},\ and\ \bibinfo {author} {\bibfnamefont {R.}~\bibnamefont {Zecchina}},\ }\bibfield  {title} {\bibinfo {title} {Unreasonable effectiveness of learning neural networks: From accessible states and robust ensembles to basic algorithmic schemes},\ }\href@noop {} {\bibfield  {journal} {\bibinfo  {journal} {Proceedings of the National Academy of Sciences}\ }\textbf {\bibinfo {volume} {113}},\ \bibinfo {pages} {E7655} (\bibinfo {year} {2016})}\BibitemShut {NoStop}%
\bibitem [{\citenamefont {Barbier}\ \emph {et~al.}(2019)\citenamefont {Barbier}, \citenamefont {Krzakala}, \citenamefont {Macris}, \citenamefont {Miolane},\ and\ \citenamefont {Zdeborov{\'a}}}]{barbier2019optimal}%
  \BibitemOpen
  \bibfield  {author} {\bibinfo {author} {\bibfnamefont {J.}~\bibnamefont {Barbier}}, \bibinfo {author} {\bibfnamefont {F.}~\bibnamefont {Krzakala}}, \bibinfo {author} {\bibfnamefont {N.}~\bibnamefont {Macris}}, \bibinfo {author} {\bibfnamefont {L.}~\bibnamefont {Miolane}},\ and\ \bibinfo {author} {\bibfnamefont {L.}~\bibnamefont {Zdeborov{\'a}}},\ }\bibfield  {title} {\bibinfo {title} {Optimal errors and phase transitions in high-dimensional generalized linear models},\ }\href@noop {} {\bibfield  {journal} {\bibinfo  {journal} {Proceedings of the National Academy of Sciences}\ }\textbf {\bibinfo {volume} {116}},\ \bibinfo {pages} {5451} (\bibinfo {year} {2019})}\BibitemShut {NoStop}%
\bibitem [{\citenamefont {Goldt}\ \emph {et~al.}(2020)\citenamefont {Goldt}, \citenamefont {M{\'e}zard}, \citenamefont {Krzakala},\ and\ \citenamefont {Zdeborov{\'a}}}]{goldt2020modeling}%
  \BibitemOpen
  \bibfield  {author} {\bibinfo {author} {\bibfnamefont {S.}~\bibnamefont {Goldt}}, \bibinfo {author} {\bibfnamefont {M.}~\bibnamefont {M{\'e}zard}}, \bibinfo {author} {\bibfnamefont {F.}~\bibnamefont {Krzakala}},\ and\ \bibinfo {author} {\bibfnamefont {L.}~\bibnamefont {Zdeborov{\'a}}},\ }\bibfield  {title} {\bibinfo {title} {Modeling the influence of data structure on learning in neural networks: The hidden manifold model},\ }\href@noop {} {\bibfield  {journal} {\bibinfo  {journal} {Physical Review X}\ }\textbf {\bibinfo {volume} {10}},\ \bibinfo {pages} {041044} (\bibinfo {year} {2020})}\BibitemShut {NoStop}%
\bibitem [{\citenamefont {Advani}\ and\ \citenamefont {Saxe}(2020)}]{advani2020high}%
  \BibitemOpen
  \bibfield  {author} {\bibinfo {author} {\bibfnamefont {M.~S.}\ \bibnamefont {Advani}}\ and\ \bibinfo {author} {\bibfnamefont {A.~M.}\ \bibnamefont {Saxe}},\ }\bibfield  {title} {\bibinfo {title} {High-dimensional dynamics of generalization error in neural networks},\ }\href@noop {} {\bibfield  {journal} {\bibinfo  {journal} {Neural Networks}\ }\textbf {\bibinfo {volume} {132}},\ \bibinfo {pages} {428} (\bibinfo {year} {2020})}\BibitemShut {NoStop}%
\bibitem [{\citenamefont {Loureiro}\ \emph {et~al.}(2021)\citenamefont {Loureiro}, \citenamefont {Gerbelot}, \citenamefont {Cui}, \citenamefont {Goldt}, \citenamefont {Krzakala}, \citenamefont {Mezard},\ and\ \citenamefont {Zdeborov\'{a}}}]{loureiro2021learning}%
  \BibitemOpen
  \bibfield  {author} {\bibinfo {author} {\bibfnamefont {B.}~\bibnamefont {Loureiro}}, \bibinfo {author} {\bibfnamefont {C.}~\bibnamefont {Gerbelot}}, \bibinfo {author} {\bibfnamefont {H.}~\bibnamefont {Cui}}, \bibinfo {author} {\bibfnamefont {S.}~\bibnamefont {Goldt}}, \bibinfo {author} {\bibfnamefont {F.}~\bibnamefont {Krzakala}}, \bibinfo {author} {\bibfnamefont {M.}~\bibnamefont {Mezard}},\ and\ \bibinfo {author} {\bibfnamefont {L.}~\bibnamefont {Zdeborov\'{a}}},\ }\bibfield  {title} {\bibinfo {title} {Learning curves of generic features maps for realistic datasets with a teacher-student model},\ }\bibfield  {booktitle} {\emph {\bibinfo {booktitle} {Advances in Neural Information Processing Systems}},\ }\href@noop {} {\ \textbf {\bibinfo {volume} {34}},\ \bibinfo {pages} {18137} (\bibinfo {year} {2021})}\BibitemShut {NoStop}%
\bibitem [{\citenamefont {Sorscher}\ \emph {et~al.}(2022)\citenamefont {Sorscher}, \citenamefont {Geirhos}, \citenamefont {Shekhar}, \citenamefont {Ganguli},\ and\ \citenamefont {Morcos}}]{sorscher2022beyond}%
  \BibitemOpen
  \bibfield  {author} {\bibinfo {author} {\bibfnamefont {B.}~\bibnamefont {Sorscher}}, \bibinfo {author} {\bibfnamefont {R.}~\bibnamefont {Geirhos}}, \bibinfo {author} {\bibfnamefont {S.}~\bibnamefont {Shekhar}}, \bibinfo {author} {\bibfnamefont {S.}~\bibnamefont {Ganguli}},\ and\ \bibinfo {author} {\bibfnamefont {A.}~\bibnamefont {Morcos}},\ }\bibfield  {title} {\bibinfo {title} {Beyond neural scaling laws: beating power law scaling via data pruning},\ }\bibfield  {booktitle} {\emph {\bibinfo {booktitle} {Advances in Neural Information Processing Systems}},\ }\href@noop {} {\ \textbf {\bibinfo {volume} {35}},\ \bibinfo {pages} {19523} (\bibinfo {year} {2022})}\BibitemShut {NoStop}%
\bibitem [{\citenamefont {Vaswani}\ \emph {et~al.}(2017)\citenamefont {Vaswani}, \citenamefont {Shazeer}, \citenamefont {Parmar}, \citenamefont {Uszkoreit}, \citenamefont {Jones}, \citenamefont {Gomez}, \citenamefont {Kaiser},\ and\ \citenamefont {Polosukhin}}]{vaswani2017attention}%
  \BibitemOpen
  \bibfield  {author} {\bibinfo {author} {\bibfnamefont {A.}~\bibnamefont {Vaswani}}, \bibinfo {author} {\bibfnamefont {N.}~\bibnamefont {Shazeer}}, \bibinfo {author} {\bibfnamefont {N.}~\bibnamefont {Parmar}}, \bibinfo {author} {\bibfnamefont {J.}~\bibnamefont {Uszkoreit}}, \bibinfo {author} {\bibfnamefont {L.}~\bibnamefont {Jones}}, \bibinfo {author} {\bibfnamefont {A.~N.}\ \bibnamefont {Gomez}}, \bibinfo {author} {\bibfnamefont {L.~u.}\ \bibnamefont {Kaiser}},\ and\ \bibinfo {author} {\bibfnamefont {I.}~\bibnamefont {Polosukhin}},\ }\bibfield  {title} {\bibinfo {title} {Attention is all you need},\ }\bibfield  {booktitle} {\emph {\bibinfo {booktitle} {Advances in Neural Information Processing Systems}},\ }\href@noop {} {\ \textbf {\bibinfo {volume} {30}} (\bibinfo {year} {2017})}\BibitemShut {NoStop}%
\bibitem [{\citenamefont {Radford}\ \emph {et~al.}(2019)\citenamefont {Radford}, \citenamefont {Wu}, \citenamefont {Child}, \citenamefont {Luan}, \citenamefont {Amodei}, \citenamefont {Sutskever} \emph {et~al.}}]{radford2019language}%
  \BibitemOpen
  \bibfield  {author} {\bibinfo {author} {\bibfnamefont {A.}~\bibnamefont {Radford}}, \bibinfo {author} {\bibfnamefont {J.}~\bibnamefont {Wu}}, \bibinfo {author} {\bibfnamefont {R.}~\bibnamefont {Child}}, \bibinfo {author} {\bibfnamefont {D.}~\bibnamefont {Luan}}, \bibinfo {author} {\bibfnamefont {D.}~\bibnamefont {Amodei}}, \bibinfo {author} {\bibfnamefont {I.}~\bibnamefont {Sutskever}}, \emph {et~al.},\ }\bibfield  {title} {\bibinfo {title} {Language models are unsupervised multitask learners},\ }\href@noop {} {\bibfield  {journal} {\bibinfo  {journal} {OpenAI blog}\ }\textbf {\bibinfo {volume} {1}},\ \bibinfo {pages} {9} (\bibinfo {year} {2019})}\BibitemShut {NoStop}%
\bibitem [{\citenamefont {Mann}\ \emph {et~al.}(2020)\citenamefont {Mann}, \citenamefont {Ryder}, \citenamefont {Subbiah}, \citenamefont {Kaplan}, \citenamefont {Dhariwal}, \citenamefont {Neelakantan}, \citenamefont {Shyam}, \citenamefont {Sastry}, \citenamefont {Askell}, \citenamefont {Agarwal} \emph {et~al.}}]{mann2020language}%
  \BibitemOpen
  \bibfield  {author} {\bibinfo {author} {\bibfnamefont {B.}~\bibnamefont {Mann}}, \bibinfo {author} {\bibfnamefont {N.}~\bibnamefont {Ryder}}, \bibinfo {author} {\bibfnamefont {M.}~\bibnamefont {Subbiah}}, \bibinfo {author} {\bibfnamefont {J.}~\bibnamefont {Kaplan}}, \bibinfo {author} {\bibfnamefont {P.}~\bibnamefont {Dhariwal}}, \bibinfo {author} {\bibfnamefont {A.}~\bibnamefont {Neelakantan}}, \bibinfo {author} {\bibfnamefont {P.}~\bibnamefont {Shyam}}, \bibinfo {author} {\bibfnamefont {G.}~\bibnamefont {Sastry}}, \bibinfo {author} {\bibfnamefont {A.}~\bibnamefont {Askell}}, \bibinfo {author} {\bibfnamefont {S.}~\bibnamefont {Agarwal}}, \emph {et~al.},\ }\bibfield  {title} {\bibinfo {title} {Language models are few-shot learners},\ }\href@noop {} {\bibfield  {journal} {\bibinfo  {journal} {arXiv preprint arXiv:2005.14165}\ } (\bibinfo {year} {2020})}\BibitemShut {NoStop}%
\bibitem [{\citenamefont {Achiam}\ \emph {et~al.}(2023)\citenamefont {Achiam}, \citenamefont {Adler}, \citenamefont {Agarwal}, \citenamefont {Ahmad}, \citenamefont {Akkaya}, \citenamefont {Aleman}, \citenamefont {Almeida}, \citenamefont {Altenschmidt}, \citenamefont {Altman}, \citenamefont {Anadkat} \emph {et~al.}}]{achiam2023gpt}%
  \BibitemOpen
  \bibfield  {author} {\bibinfo {author} {\bibfnamefont {J.}~\bibnamefont {Achiam}}, \bibinfo {author} {\bibfnamefont {S.}~\bibnamefont {Adler}}, \bibinfo {author} {\bibfnamefont {S.}~\bibnamefont {Agarwal}}, \bibinfo {author} {\bibfnamefont {L.}~\bibnamefont {Ahmad}}, \bibinfo {author} {\bibfnamefont {I.}~\bibnamefont {Akkaya}}, \bibinfo {author} {\bibfnamefont {F.~L.}\ \bibnamefont {Aleman}}, \bibinfo {author} {\bibfnamefont {D.}~\bibnamefont {Almeida}}, \bibinfo {author} {\bibfnamefont {J.}~\bibnamefont {Altenschmidt}}, \bibinfo {author} {\bibfnamefont {S.}~\bibnamefont {Altman}}, \bibinfo {author} {\bibfnamefont {S.}~\bibnamefont {Anadkat}}, \emph {et~al.},\ }\bibfield  {title} {\bibinfo {title} {Gpt-4 technical report},\ }\href@noop {} {\bibfield  {journal} {\bibinfo  {journal} {arXiv preprint arXiv:2303.08774}\ } (\bibinfo {year} {2023})}\BibitemShut {NoStop}%
\bibitem [{\citenamefont {Bubeck}\ \emph {et~al.}(2023)\citenamefont {Bubeck}, \citenamefont {Chandrasekaran}, \citenamefont {Eldan}, \citenamefont {Gehrke}, \citenamefont {Horvitz}, \citenamefont {Kamar}, \citenamefont {Lee}, \citenamefont {Lee}, \citenamefont {Li}, \citenamefont {Lundberg} \emph {et~al.}}]{bubeck2023sparks}%
  \BibitemOpen
  \bibfield  {author} {\bibinfo {author} {\bibfnamefont {S.}~\bibnamefont {Bubeck}}, \bibinfo {author} {\bibfnamefont {V.}~\bibnamefont {Chandrasekaran}}, \bibinfo {author} {\bibfnamefont {R.}~\bibnamefont {Eldan}}, \bibinfo {author} {\bibfnamefont {J.}~\bibnamefont {Gehrke}}, \bibinfo {author} {\bibfnamefont {E.}~\bibnamefont {Horvitz}}, \bibinfo {author} {\bibfnamefont {E.}~\bibnamefont {Kamar}}, \bibinfo {author} {\bibfnamefont {P.}~\bibnamefont {Lee}}, \bibinfo {author} {\bibfnamefont {Y.~T.}\ \bibnamefont {Lee}}, \bibinfo {author} {\bibfnamefont {Y.}~\bibnamefont {Li}}, \bibinfo {author} {\bibfnamefont {S.}~\bibnamefont {Lundberg}}, \emph {et~al.},\ }\bibfield  {title} {\bibinfo {title} {Sparks of artificial general intelligence: Early experiments with gpt-4},\ }\href@noop {} {\bibfield  {journal} {\bibinfo  {journal} {arXiv preprint arXiv:2303.12712}\ } (\bibinfo {year} {2023})}\BibitemShut {NoStop}%
\bibitem [{\citenamefont {Wei}\ \emph {et~al.}(2022)\citenamefont {Wei}, \citenamefont {Tay}, \citenamefont {Bommasani}, \citenamefont {Raffel}, \citenamefont {Zoph}, \citenamefont {Borgeaud}, \citenamefont {Yogatama}, \citenamefont {Bosma}, \citenamefont {Zhou}, \citenamefont {Metzler}, \citenamefont {Chi}, \citenamefont {Hashimoto}, \citenamefont {Vinyals}, \citenamefont {Liang}, \citenamefont {Dean},\ and\ \citenamefont {Fedus}}]{wei2022emergent}%
  \BibitemOpen
  \bibfield  {author} {\bibinfo {author} {\bibfnamefont {J.}~\bibnamefont {Wei}}, \bibinfo {author} {\bibfnamefont {Y.}~\bibnamefont {Tay}}, \bibinfo {author} {\bibfnamefont {R.}~\bibnamefont {Bommasani}}, \bibinfo {author} {\bibfnamefont {C.}~\bibnamefont {Raffel}}, \bibinfo {author} {\bibfnamefont {B.}~\bibnamefont {Zoph}}, \bibinfo {author} {\bibfnamefont {S.}~\bibnamefont {Borgeaud}}, \bibinfo {author} {\bibfnamefont {D.}~\bibnamefont {Yogatama}}, \bibinfo {author} {\bibfnamefont {M.}~\bibnamefont {Bosma}}, \bibinfo {author} {\bibfnamefont {D.}~\bibnamefont {Zhou}}, \bibinfo {author} {\bibfnamefont {D.}~\bibnamefont {Metzler}}, \bibinfo {author} {\bibfnamefont {E.~H.}\ \bibnamefont {Chi}}, \bibinfo {author} {\bibfnamefont {T.}~\bibnamefont {Hashimoto}}, \bibinfo {author} {\bibfnamefont {O.}~\bibnamefont {Vinyals}}, \bibinfo {author} {\bibfnamefont {P.}~\bibnamefont {Liang}}, \bibinfo {author} {\bibfnamefont {J.}~\bibnamefont {Dean}},\ and\ \bibinfo {author} {\bibfnamefont {W.}~\bibnamefont {Fedus}},\
  }\bibfield  {title} {\bibinfo {title} {Emergent abilities of large language models},\ }\href@noop {} {\bibfield  {journal} {\bibinfo  {journal} {Transactions on Machine Learning Research}\ } (\bibinfo {year} {2022})}\BibitemShut {NoStop}%
\bibitem [{\citenamefont {Kaplan}\ \emph {et~al.}(2020)\citenamefont {Kaplan}, \citenamefont {McCandlish}, \citenamefont {Henighan}, \citenamefont {Brown}, \citenamefont {Chess}, \citenamefont {Child}, \citenamefont {Gray}, \citenamefont {Radford}, \citenamefont {Wu},\ and\ \citenamefont {Amodei}}]{kaplan2020scaling}%
  \BibitemOpen
  \bibfield  {author} {\bibinfo {author} {\bibfnamefont {J.}~\bibnamefont {Kaplan}}, \bibinfo {author} {\bibfnamefont {S.}~\bibnamefont {McCandlish}}, \bibinfo {author} {\bibfnamefont {T.}~\bibnamefont {Henighan}}, \bibinfo {author} {\bibfnamefont {T.~B.}\ \bibnamefont {Brown}}, \bibinfo {author} {\bibfnamefont {B.}~\bibnamefont {Chess}}, \bibinfo {author} {\bibfnamefont {R.}~\bibnamefont {Child}}, \bibinfo {author} {\bibfnamefont {S.}~\bibnamefont {Gray}}, \bibinfo {author} {\bibfnamefont {A.}~\bibnamefont {Radford}}, \bibinfo {author} {\bibfnamefont {J.}~\bibnamefont {Wu}},\ and\ \bibinfo {author} {\bibfnamefont {D.}~\bibnamefont {Amodei}},\ }\bibfield  {title} {\bibinfo {title} {Scaling laws for neural language models},\ }\href@noop {} {\bibfield  {journal} {\bibinfo  {journal} {arXiv preprint arXiv:2001.08361}\ } (\bibinfo {year} {2020})}\BibitemShut {NoStop}%
\bibitem [{\citenamefont {Brown}\ \emph {et~al.}(2020)\citenamefont {Brown}, \citenamefont {Mann}, \citenamefont {Ryder}, \citenamefont {Subbiah}, \citenamefont {Kaplan}, \citenamefont {Dhariwal}, \citenamefont {Neelakantan}, \citenamefont {Shyam}, \citenamefont {Sastry}, \citenamefont {Askell} \emph {et~al.}}]{brown2020language}%
  \BibitemOpen
  \bibfield  {author} {\bibinfo {author} {\bibfnamefont {T.}~\bibnamefont {Brown}}, \bibinfo {author} {\bibfnamefont {B.}~\bibnamefont {Mann}}, \bibinfo {author} {\bibfnamefont {N.}~\bibnamefont {Ryder}}, \bibinfo {author} {\bibfnamefont {M.}~\bibnamefont {Subbiah}}, \bibinfo {author} {\bibfnamefont {J.~D.}\ \bibnamefont {Kaplan}}, \bibinfo {author} {\bibfnamefont {P.}~\bibnamefont {Dhariwal}}, \bibinfo {author} {\bibfnamefont {A.}~\bibnamefont {Neelakantan}}, \bibinfo {author} {\bibfnamefont {P.}~\bibnamefont {Shyam}}, \bibinfo {author} {\bibfnamefont {G.}~\bibnamefont {Sastry}}, \bibinfo {author} {\bibfnamefont {A.}~\bibnamefont {Askell}}, \emph {et~al.},\ }\bibfield  {title} {\bibinfo {title} {Language models are few-shot learners},\ }\href@noop {} {\bibfield  {journal} {\bibinfo  {journal} {Advances in neural information processing systems}\ }\textbf {\bibinfo {volume} {33}},\ \bibinfo {pages} {1877} (\bibinfo {year} {2020})}\BibitemShut {NoStop}%
\bibitem [{\citenamefont {Cybenko}(1989)}]{cybenko1989approximation}%
  \BibitemOpen
  \bibfield  {author} {\bibinfo {author} {\bibfnamefont {G.}~\bibnamefont {Cybenko}},\ }\bibfield  {title} {\bibinfo {title} {Approximation by superpositions of a sigmoidal function},\ }\href@noop {} {\bibfield  {journal} {\bibinfo  {journal} {Mathematics of Control, Signals and Systems}\ }\textbf {\bibinfo {volume} {2}},\ \bibinfo {pages} {303} (\bibinfo {year} {1989})}\BibitemShut {NoStop}%
\bibitem [{\citenamefont {Hornik}\ \emph {et~al.}(1989)\citenamefont {Hornik}, \citenamefont {Stinchcombe},\ and\ \citenamefont {White}}]{hornik1989multilayer}%
  \BibitemOpen
  \bibfield  {author} {\bibinfo {author} {\bibfnamefont {K.}~\bibnamefont {Hornik}}, \bibinfo {author} {\bibfnamefont {M.}~\bibnamefont {Stinchcombe}},\ and\ \bibinfo {author} {\bibfnamefont {H.}~\bibnamefont {White}},\ }\bibfield  {title} {\bibinfo {title} {Multilayer feedforward networks are universal approximators},\ }\href@noop {} {\bibfield  {journal} {\bibinfo  {journal} {Neural networks}\ }\textbf {\bibinfo {volume} {2}},\ \bibinfo {pages} {359} (\bibinfo {year} {1989})}\BibitemShut {NoStop}%
\bibitem [{\citenamefont {Ravent\'{o}s}\ \emph {et~al.}(2023)\citenamefont {Ravent\'{o}s}, \citenamefont {Paul}, \citenamefont {Chen},\ and\ \citenamefont {Ganguli}}]{raventos2024pretraining}%
  \BibitemOpen
  \bibfield  {author} {\bibinfo {author} {\bibfnamefont {A.}~\bibnamefont {Ravent\'{o}s}}, \bibinfo {author} {\bibfnamefont {M.}~\bibnamefont {Paul}}, \bibinfo {author} {\bibfnamefont {F.}~\bibnamefont {Chen}},\ and\ \bibinfo {author} {\bibfnamefont {S.}~\bibnamefont {Ganguli}},\ }\bibfield  {title} {\bibinfo {title} {Pretraining task diversity and the emergence of non-bayesian in-context learning for regression},\ }\bibfield  {booktitle} {\emph {\bibinfo {booktitle} {Advances in Neural Information Processing Systems}},\ }\href@noop {} {\ \textbf {\bibinfo {volume} {36}},\ \bibinfo {pages} {14228} (\bibinfo {year} {2023})}\BibitemShut {NoStop}%
\bibitem [{\citenamefont {Cagnetta}\ and\ \citenamefont {Wyart}(2024)}]{cagnetta2024towards}%
  \BibitemOpen
  \bibfield  {author} {\bibinfo {author} {\bibfnamefont {F.}~\bibnamefont {Cagnetta}}\ and\ \bibinfo {author} {\bibfnamefont {M.}~\bibnamefont {Wyart}},\ }\bibfield  {title} {\bibinfo {title} {Towards a theory of how the structure of language is acquired by deep neural networks},\ }\href@noop {} {\bibfield  {journal} {\bibinfo  {journal} {arXiv preprint arXiv:2406.00048}\ } (\bibinfo {year} {2024})}\BibitemShut {NoStop}%
\bibitem [{\citenamefont {Behrens}\ \emph {et~al.}(2024)\citenamefont {Behrens}, \citenamefont {Biggio},\ and\ \citenamefont {Zdeborov{\'a}}}]{behrens2024understanding}%
  \BibitemOpen
  \bibfield  {author} {\bibinfo {author} {\bibfnamefont {F.}~\bibnamefont {Behrens}}, \bibinfo {author} {\bibfnamefont {L.}~\bibnamefont {Biggio}},\ and\ \bibinfo {author} {\bibfnamefont {L.}~\bibnamefont {Zdeborov{\'a}}},\ }\bibfield  {title} {\bibinfo {title} {Understanding counting in small transformers: The interplay between attention and feed-forward layers},\ }\href@noop {} {\bibfield  {journal} {\bibinfo  {journal} {arXiv preprint arXiv:2407.11542}\ } (\bibinfo {year} {2024})}\BibitemShut {NoStop}%
\bibitem [{\citenamefont {Geshkovski}\ \emph {et~al.}(2023)\citenamefont {Geshkovski}, \citenamefont {Letrouit}, \citenamefont {Polyanskiy},\ and\ \citenamefont {Rigollet}}]{geshkovski2023mathematical}%
  \BibitemOpen
  \bibfield  {author} {\bibinfo {author} {\bibfnamefont {B.}~\bibnamefont {Geshkovski}}, \bibinfo {author} {\bibfnamefont {C.}~\bibnamefont {Letrouit}}, \bibinfo {author} {\bibfnamefont {Y.}~\bibnamefont {Polyanskiy}},\ and\ \bibinfo {author} {\bibfnamefont {P.}~\bibnamefont {Rigollet}},\ }\bibfield  {title} {\bibinfo {title} {A mathematical perspective on transformers},\ }\href@noop {} {\bibfield  {journal} {\bibinfo  {journal} {arXiv preprint arXiv:2312.10794}\ } (\bibinfo {year} {2023})}\BibitemShut {NoStop}%
\bibitem [{\citenamefont {Cowsik}\ \emph {et~al.}(2024)\citenamefont {Cowsik}, \citenamefont {Nebabu}, \citenamefont {Qi},\ and\ \citenamefont {Ganguli}}]{cowsik2024geometric}%
  \BibitemOpen
  \bibfield  {author} {\bibinfo {author} {\bibfnamefont {A.}~\bibnamefont {Cowsik}}, \bibinfo {author} {\bibfnamefont {T.}~\bibnamefont {Nebabu}}, \bibinfo {author} {\bibfnamefont {X.-L.}\ \bibnamefont {Qi}},\ and\ \bibinfo {author} {\bibfnamefont {S.}~\bibnamefont {Ganguli}},\ }\bibfield  {title} {\bibinfo {title} {Geometric dynamics of signal propagation predict trainability of transformers},\ }\href@noop {} {\bibfield  {journal} {\bibinfo  {journal} {arXiv preprint arXiv:2403.02579}\ } (\bibinfo {year} {2024})}\BibitemShut {NoStop}%
\bibitem [{\citenamefont {Rende}\ \emph {et~al.}(2024)\citenamefont {Rende}, \citenamefont {Gerace}, \citenamefont {Laio},\ and\ \citenamefont {Goldt}}]{rende2024mapping}%
  \BibitemOpen
  \bibfield  {author} {\bibinfo {author} {\bibfnamefont {R.}~\bibnamefont {Rende}}, \bibinfo {author} {\bibfnamefont {F.}~\bibnamefont {Gerace}}, \bibinfo {author} {\bibfnamefont {A.}~\bibnamefont {Laio}},\ and\ \bibinfo {author} {\bibfnamefont {S.}~\bibnamefont {Goldt}},\ }\bibfield  {title} {\bibinfo {title} {Mapping of attention mechanisms to a generalized potts model},\ }\href@noop {} {\bibfield  {journal} {\bibinfo  {journal} {Physical Review Research}\ }\textbf {\bibinfo {volume} {6}},\ \bibinfo {pages} {023057} (\bibinfo {year} {2024})}\BibitemShut {NoStop}%
\bibitem [{\citenamefont {Cui}\ \emph {et~al.}(2024)\citenamefont {Cui}, \citenamefont {Behrens}, \citenamefont {Krzakala},\ and\ \citenamefont {Zdeborov{\'a}}}]{cui2024phase}%
  \BibitemOpen
  \bibfield  {author} {\bibinfo {author} {\bibfnamefont {H.}~\bibnamefont {Cui}}, \bibinfo {author} {\bibfnamefont {F.}~\bibnamefont {Behrens}}, \bibinfo {author} {\bibfnamefont {F.}~\bibnamefont {Krzakala}},\ and\ \bibinfo {author} {\bibfnamefont {L.}~\bibnamefont {Zdeborov{\'a}}},\ }\bibfield  {title} {\bibinfo {title} {A phase transition between positional and semantic learning in a solvable model of dot-product attention},\ }\href@noop {} {\bibfield  {journal} {\bibinfo  {journal} {to appear at NeurIPS2024}\ } (\bibinfo {year} {2024})}\BibitemShut {NoStop}%
\bibitem [{\citenamefont {Lu}\ \emph {et~al.}(2024)\citenamefont {Lu}, \citenamefont {Letey}, \citenamefont {Zavatone-Veth}, \citenamefont {Maiti},\ and\ \citenamefont {Pehlevan}}]{lu2024asymptotic}%
  \BibitemOpen
  \bibfield  {author} {\bibinfo {author} {\bibfnamefont {Y.~M.}\ \bibnamefont {Lu}}, \bibinfo {author} {\bibfnamefont {M.~I.}\ \bibnamefont {Letey}}, \bibinfo {author} {\bibfnamefont {J.~A.}\ \bibnamefont {Zavatone-Veth}}, \bibinfo {author} {\bibfnamefont {A.}~\bibnamefont {Maiti}},\ and\ \bibinfo {author} {\bibfnamefont {C.}~\bibnamefont {Pehlevan}},\ }\bibfield  {title} {\bibinfo {title} {Asymptotic theory of in-context learning by linear attention},\ }\href@noop {} {\bibfield  {journal} {\bibinfo  {journal} {arXiv preprint arXiv:2405.11751}\ } (\bibinfo {year} {2024})}\BibitemShut {NoStop}%
\bibitem [{\citenamefont {Dubey}\ \emph {et~al.}(2024)\citenamefont {Dubey}, \citenamefont {Jauhri}, \citenamefont {Pandey}, \citenamefont {Kadian}, \citenamefont {Al-Dahle}, \citenamefont {Letman}, \citenamefont {Mathur}, \citenamefont {Schelten}, \citenamefont {Yang}, \citenamefont {Fan} \emph {et~al.}}]{dubey2024llama}%
  \BibitemOpen
  \bibfield  {author} {\bibinfo {author} {\bibfnamefont {A.}~\bibnamefont {Dubey}}, \bibinfo {author} {\bibfnamefont {A.}~\bibnamefont {Jauhri}}, \bibinfo {author} {\bibfnamefont {A.}~\bibnamefont {Pandey}}, \bibinfo {author} {\bibfnamefont {A.}~\bibnamefont {Kadian}}, \bibinfo {author} {\bibfnamefont {A.}~\bibnamefont {Al-Dahle}}, \bibinfo {author} {\bibfnamefont {A.}~\bibnamefont {Letman}}, \bibinfo {author} {\bibfnamefont {A.}~\bibnamefont {Mathur}}, \bibinfo {author} {\bibfnamefont {A.}~\bibnamefont {Schelten}}, \bibinfo {author} {\bibfnamefont {A.}~\bibnamefont {Yang}}, \bibinfo {author} {\bibfnamefont {A.}~\bibnamefont {Fan}}, \emph {et~al.},\ }\bibfield  {title} {\bibinfo {title} {The llama 3 herd of models},\ }\href@noop {} {\bibfield  {journal} {\bibinfo  {journal} {arXiv preprint arXiv:2407.21783}\ } (\bibinfo {year} {2024})}\BibitemShut {NoStop}%
\bibitem [{\citenamefont {Mignacco}\ \emph {et~al.}(2021)\citenamefont {Mignacco}, \citenamefont {Urbani},\ and\ \citenamefont {Zdeborov{\'a}}}]{mignacco2021stochasticity}%
  \BibitemOpen
  \bibfield  {author} {\bibinfo {author} {\bibfnamefont {F.}~\bibnamefont {Mignacco}}, \bibinfo {author} {\bibfnamefont {P.}~\bibnamefont {Urbani}},\ and\ \bibinfo {author} {\bibfnamefont {L.}~\bibnamefont {Zdeborov{\'a}}},\ }\bibfield  {title} {\bibinfo {title} {Stochasticity helps to navigate rough landscapes: comparing gradient-descent-based algorithms in the phase retrieval problem},\ }\href@noop {} {\bibfield  {journal} {\bibinfo  {journal} {Machine Learning: Science and Technology}\ }\textbf {\bibinfo {volume} {2}},\ \bibinfo {pages} {035029} (\bibinfo {year} {2021})}\BibitemShut {NoStop}%
\bibitem [{\citenamefont {Sarao~Mannelli}\ \emph {et~al.}(2020)\citenamefont {Sarao~Mannelli}, \citenamefont {Biroli}, \citenamefont {Cammarota}, \citenamefont {Krzakala}, \citenamefont {Urbani},\ and\ \citenamefont {Zdeborov\'{a}}}]{sarao2020complex}%
  \BibitemOpen
  \bibfield  {author} {\bibinfo {author} {\bibfnamefont {S.}~\bibnamefont {Sarao~Mannelli}}, \bibinfo {author} {\bibfnamefont {G.}~\bibnamefont {Biroli}}, \bibinfo {author} {\bibfnamefont {C.}~\bibnamefont {Cammarota}}, \bibinfo {author} {\bibfnamefont {F.}~\bibnamefont {Krzakala}}, \bibinfo {author} {\bibfnamefont {P.}~\bibnamefont {Urbani}},\ and\ \bibinfo {author} {\bibfnamefont {L.}~\bibnamefont {Zdeborov\'{a}}},\ }\bibfield  {title} {\bibinfo {title} {Complex dynamics in simple neural networks: Understanding gradient flow in phase retrieval},\ }\bibfield  {booktitle} {\emph {\bibinfo {booktitle} {Advances in Neural Information Processing Systems}},\ }\href@noop {} {\ \textbf {\bibinfo {volume} {33}},\ \bibinfo {pages} {3265} (\bibinfo {year} {2020})}\BibitemShut {NoStop}%
\bibitem [{\citenamefont {Belkin}\ \emph {et~al.}(2019)\citenamefont {Belkin}, \citenamefont {Hsu}, \citenamefont {Ma},\ and\ \citenamefont {Mandal}}]{belkin2019reconciling}%
  \BibitemOpen
  \bibfield  {author} {\bibinfo {author} {\bibfnamefont {M.}~\bibnamefont {Belkin}}, \bibinfo {author} {\bibfnamefont {D.}~\bibnamefont {Hsu}}, \bibinfo {author} {\bibfnamefont {S.}~\bibnamefont {Ma}},\ and\ \bibinfo {author} {\bibfnamefont {S.}~\bibnamefont {Mandal}},\ }\bibfield  {title} {\bibinfo {title} {Reconciling modern machine-learning practice and the classical bias--variance trade-off},\ }\href@noop {} {\bibfield  {journal} {\bibinfo  {journal} {Proceedings of the National Academy of Sciences}\ }\textbf {\bibinfo {volume} {116}},\ \bibinfo {pages} {15849} (\bibinfo {year} {2019})}\BibitemShut {NoStop}%
\bibitem [{\citenamefont {Gerace}\ \emph {et~al.}(2020)\citenamefont {Gerace}, \citenamefont {Loureiro}, \citenamefont {Krzakala}, \citenamefont {M\'ezard},\ and\ \citenamefont {Zdeborov\'a}}]{gerace2020generalisation}%
  \BibitemOpen
  \bibfield  {author} {\bibinfo {author} {\bibfnamefont {F.}~\bibnamefont {Gerace}}, \bibinfo {author} {\bibfnamefont {B.}~\bibnamefont {Loureiro}}, \bibinfo {author} {\bibfnamefont {F.}~\bibnamefont {Krzakala}}, \bibinfo {author} {\bibfnamefont {M.}~\bibnamefont {M\'ezard}},\ and\ \bibinfo {author} {\bibfnamefont {L.}~\bibnamefont {Zdeborov\'a}},\ }\bibfield  {title} {\bibinfo {title} {Generalisation error in learning with random features and the hidden manifold model},\ }in\ \href@noop {} {\emph {\bibinfo {booktitle} {Proceedings of the 37th International Conference on Machine Learning}}},\ \bibinfo {series} {Proceedings of Machine Learning Research}, Vol.\ \bibinfo {volume} {119}\ (\bibinfo  {publisher} {PMLR},\ \bibinfo {year} {2020})\ pp.\ \bibinfo {pages} {3452--3462}\BibitemShut {NoStop}%
\bibitem [{\citenamefont {Tolstikhin}\ \emph {et~al.}(2021)\citenamefont {Tolstikhin}, \citenamefont {Houlsby}, \citenamefont {Kolesnikov}, \citenamefont {Beyer}, \citenamefont {Zhai}, \citenamefont {Unterthiner}, \citenamefont {Yung}, \citenamefont {Steiner}, \citenamefont {Keysers}, \citenamefont {Uszkoreit} \emph {et~al.}}]{tolstikhin2021mlp}%
  \BibitemOpen
  \bibfield  {author} {\bibinfo {author} {\bibfnamefont {I.~O.}\ \bibnamefont {Tolstikhin}}, \bibinfo {author} {\bibfnamefont {N.}~\bibnamefont {Houlsby}}, \bibinfo {author} {\bibfnamefont {A.}~\bibnamefont {Kolesnikov}}, \bibinfo {author} {\bibfnamefont {L.}~\bibnamefont {Beyer}}, \bibinfo {author} {\bibfnamefont {X.}~\bibnamefont {Zhai}}, \bibinfo {author} {\bibfnamefont {T.}~\bibnamefont {Unterthiner}}, \bibinfo {author} {\bibfnamefont {J.}~\bibnamefont {Yung}}, \bibinfo {author} {\bibfnamefont {A.}~\bibnamefont {Steiner}}, \bibinfo {author} {\bibfnamefont {D.}~\bibnamefont {Keysers}}, \bibinfo {author} {\bibfnamefont {J.}~\bibnamefont {Uszkoreit}}, \emph {et~al.},\ }\bibfield  {title} {\bibinfo {title} {Mlp-mixer: An all-mlp architecture for vision},\ }\href@noop {} {\bibfield  {journal} {\bibinfo  {journal} {Advances in neural information processing systems}\ }\textbf {\bibinfo {volume} {34}},\ \bibinfo {pages} {24261} (\bibinfo {year} {2021})}\BibitemShut {NoStop}%
\bibitem [{\citenamefont {Recht}\ \emph {et~al.}(2010)\citenamefont {Recht}, \citenamefont {Fazel},\ and\ \citenamefont {Parrilo}}]{recht2010guaranteed}%
  \BibitemOpen
  \bibfield  {author} {\bibinfo {author} {\bibfnamefont {B.}~\bibnamefont {Recht}}, \bibinfo {author} {\bibfnamefont {M.}~\bibnamefont {Fazel}},\ and\ \bibinfo {author} {\bibfnamefont {P.~A.}\ \bibnamefont {Parrilo}},\ }\bibfield  {title} {\bibinfo {title} {Guaranteed minimum-rank solutions of linear matrix equations via nuclear norm minimization},\ }\href@noop {} {\bibfield  {journal} {\bibinfo  {journal} {SIAM Review}\ }\textbf {\bibinfo {volume} {52}},\ \bibinfo {pages} {471} (\bibinfo {year} {2010})}\BibitemShut {NoStop}%
\bibitem [{\citenamefont {Donoho}\ \emph {et~al.}(2013)\citenamefont {Donoho}, \citenamefont {Gavish},\ and\ \citenamefont {Montanari}}]{Donoho_2013}%
  \BibitemOpen
  \bibfield  {author} {\bibinfo {author} {\bibfnamefont {D.~L.}\ \bibnamefont {Donoho}}, \bibinfo {author} {\bibfnamefont {M.}~\bibnamefont {Gavish}},\ and\ \bibinfo {author} {\bibfnamefont {A.}~\bibnamefont {Montanari}},\ }\bibfield  {title} {\bibinfo {title} {The phase transition of matrix recovery from gaussian measurements matches the minimax mse of matrix denoising},\ }\href@noop {} {\bibfield  {journal} {\bibinfo  {journal} {Proceedings of the National Academy of Sciences}\ }\textbf {\bibinfo {volume} {110}},\ \bibinfo {pages} {8405–8410} (\bibinfo {year} {2013})}\BibitemShut {NoStop}%
\bibitem [{\citenamefont {Sch\"ulke}\ \emph {et~al.}(2016)\citenamefont {Sch\"ulke}, \citenamefont {Schniter},\ and\ \citenamefont {Zdeborov\'a}}]{schulke16}%
  \BibitemOpen
  \bibfield  {author} {\bibinfo {author} {\bibfnamefont {C.}~\bibnamefont {Sch\"ulke}}, \bibinfo {author} {\bibfnamefont {P.}~\bibnamefont {Schniter}},\ and\ \bibinfo {author} {\bibfnamefont {L.}~\bibnamefont {Zdeborov\'a}},\ }\bibfield  {title} {\bibinfo {title} {Phase diagram of matrix compressed sensing},\ }\href@noop {} {\bibfield  {journal} {\bibinfo  {journal} {Phys. Rev. E}\ }\textbf {\bibinfo {volume} {94}},\ \bibinfo {pages} {062136} (\bibinfo {year} {2016})}\BibitemShut {NoStop}%
\bibitem [{\citenamefont {Giraud}(2011)}]{giraud2011low}%
  \BibitemOpen
  \bibfield  {author} {\bibinfo {author} {\bibfnamefont {C.}~\bibnamefont {Giraud}},\ }\bibfield  {title} {\bibinfo {title} {Low rank multivariate regression},\ }\href@noop {} {\bibfield  {journal} {\bibinfo  {journal} {Electronic Journal of Statistics}\ }\textbf {\bibinfo {volume} {5}},\ \bibinfo {pages} {775} (\bibinfo {year} {2011})}\BibitemShut {NoStop}%
\bibitem [{\citenamefont {Hoff}(2015)}]{hoff2015multilinear}%
  \BibitemOpen
  \bibfield  {author} {\bibinfo {author} {\bibfnamefont {P.~D.}\ \bibnamefont {Hoff}},\ }\bibfield  {title} {\bibinfo {title} {Multilinear tensor regression for longitudinal relational data},\ }\href@noop {} {\bibfield  {journal} {\bibinfo  {journal} {The annals of applied statistics}\ }\textbf {\bibinfo {volume} {9}},\ \bibinfo {pages} {1169} (\bibinfo {year} {2015})}\BibitemShut {NoStop}%
\bibitem [{\citenamefont {Chen}\ and\ \citenamefont {Fan}(2023)}]{chen2023statistical}%
  \BibitemOpen
  \bibfield  {author} {\bibinfo {author} {\bibfnamefont {E.~Y.}\ \bibnamefont {Chen}}\ and\ \bibinfo {author} {\bibfnamefont {J.}~\bibnamefont {Fan}},\ }\bibfield  {title} {\bibinfo {title} {Statistical inference for high-dimensional matrix-variate factor models},\ }\href@noop {} {\bibfield  {journal} {\bibinfo  {journal} {Journal of the American Statistical Association}\ }\textbf {\bibinfo {volume} {118}},\ \bibinfo {pages} {1038} (\bibinfo {year} {2023})}\BibitemShut {NoStop}%
\bibitem [{\citenamefont {Gunasekar}\ \emph {et~al.}(2017)\citenamefont {Gunasekar}, \citenamefont {Woodworth}, \citenamefont {Bhojanapalli}, \citenamefont {Neyshabur},\ and\ \citenamefont {Srebro}}]{gunasekar2017implicit}%
  \BibitemOpen
  \bibfield  {author} {\bibinfo {author} {\bibfnamefont {S.}~\bibnamefont {Gunasekar}}, \bibinfo {author} {\bibfnamefont {B.~E.}\ \bibnamefont {Woodworth}}, \bibinfo {author} {\bibfnamefont {S.}~\bibnamefont {Bhojanapalli}}, \bibinfo {author} {\bibfnamefont {B.}~\bibnamefont {Neyshabur}},\ and\ \bibinfo {author} {\bibfnamefont {N.}~\bibnamefont {Srebro}},\ }\bibfield  {title} {\bibinfo {title} {Implicit regularization in matrix factorization},\ }\bibfield  {booktitle} {\emph {\bibinfo {booktitle} {Advances in Neural Information Processing Systems}},\ }\href@noop {} {\ \textbf {\bibinfo {volume} {30}} (\bibinfo {year} {2017})}\BibitemShut {NoStop}%
\bibitem [{\citenamefont {Li}\ \emph {et~al.}(2021)\citenamefont {Li}, \citenamefont {Luo},\ and\ \citenamefont {Lyu}}]{li2020towards}%
  \BibitemOpen
  \bibfield  {author} {\bibinfo {author} {\bibfnamefont {Z.}~\bibnamefont {Li}}, \bibinfo {author} {\bibfnamefont {Y.}~\bibnamefont {Luo}},\ and\ \bibinfo {author} {\bibfnamefont {K.}~\bibnamefont {Lyu}},\ }\bibfield  {title} {\bibinfo {title} {Towards resolving the implicit bias of gradient descent for matrix factorization: Greedy low-rank learning},\ }in\ \href@noop {} {\emph {\bibinfo {booktitle} {International Conference on Learning Representations}}}\ (\bibinfo {year} {2021})\BibitemShut {NoStop}%
\bibitem [{\citenamefont {Schmidt}(2018)}]{schmidt2018statistical}%
  \BibitemOpen
  \bibfield  {author} {\bibinfo {author} {\bibfnamefont {H.~C.}\ \bibnamefont {Schmidt}},\ }\emph {\bibinfo {title} {Statistical physics of sparse and dense models in optimization and inference}},\ \href@noop {} {Ph.D. thesis},\ \bibinfo  {school} {Universit{\'e} Paris Saclay (COmUE)} (\bibinfo {year} {2018})\BibitemShut {NoStop}%
\bibitem [{\citenamefont {Maillard}\ \emph {et~al.}(2022)\citenamefont {Maillard}, \citenamefont {Krzakala}, \citenamefont {M{\'e}zard},\ and\ \citenamefont {Zdeborov{\'a}}}]{maillard2022perturbative}%
  \BibitemOpen
  \bibfield  {author} {\bibinfo {author} {\bibfnamefont {A.}~\bibnamefont {Maillard}}, \bibinfo {author} {\bibfnamefont {F.}~\bibnamefont {Krzakala}}, \bibinfo {author} {\bibfnamefont {M.}~\bibnamefont {M{\'e}zard}},\ and\ \bibinfo {author} {\bibfnamefont {L.}~\bibnamefont {Zdeborov{\'a}}},\ }\bibfield  {title} {\bibinfo {title} {Perturbative construction of mean-field equations in extensive-rank matrix factorization and denoising},\ }\href@noop {} {\bibfield  {journal} {\bibinfo  {journal} {Journal of Statistical Mechanics: Theory and Experiment}\ }\textbf {\bibinfo {volume} {2022}},\ \bibinfo {pages} {083301} (\bibinfo {year} {2022})}\BibitemShut {NoStop}%
\bibitem [{\citenamefont {Barbier}\ and\ \citenamefont {Macris}(2022)}]{barbier2022statistical}%
  \BibitemOpen
  \bibfield  {author} {\bibinfo {author} {\bibfnamefont {J.}~\bibnamefont {Barbier}}\ and\ \bibinfo {author} {\bibfnamefont {N.}~\bibnamefont {Macris}},\ }\bibfield  {title} {\bibinfo {title} {Statistical limits of dictionary learning: random matrix theory and the spectral replica method},\ }\href@noop {} {\bibfield  {journal} {\bibinfo  {journal} {Physical Review E}\ }\textbf {\bibinfo {volume} {106}},\ \bibinfo {pages} {024136} (\bibinfo {year} {2022})}\BibitemShut {NoStop}%
\bibitem [{\citenamefont {Troiani}\ \emph {et~al.}(2022)\citenamefont {Troiani}, \citenamefont {Erba}, \citenamefont {Krzakala}, \citenamefont {Maillard},\ and\ \citenamefont {Zdeborov{\'a}}}]{troiani2022optimal}%
  \BibitemOpen
  \bibfield  {author} {\bibinfo {author} {\bibfnamefont {E.}~\bibnamefont {Troiani}}, \bibinfo {author} {\bibfnamefont {V.}~\bibnamefont {Erba}}, \bibinfo {author} {\bibfnamefont {F.}~\bibnamefont {Krzakala}}, \bibinfo {author} {\bibfnamefont {A.}~\bibnamefont {Maillard}},\ and\ \bibinfo {author} {\bibfnamefont {L.}~\bibnamefont {Zdeborov{\'a}}},\ }\bibfield  {title} {\bibinfo {title} {Optimal denoising of rotationally invariant rectangular matrices},\ }in\ \href@noop {} {\emph {\bibinfo {booktitle} {Mathematical and Scientific Machine Learning}}},\ \bibinfo {series} {Proceedings of Machine Learning Research}, Vol.\ \bibinfo {volume} {190},\ \bibinfo {editor} {edited by\ \bibinfo {editor} {\bibfnamefont {B.}~\bibnamefont {Dong}}, \bibinfo {editor} {\bibfnamefont {Q.}~\bibnamefont {Li}}, \bibinfo {editor} {\bibfnamefont {L.}~\bibnamefont {Wang}},\ and\ \bibinfo {editor} {\bibfnamefont {Z.-Q.~J.}\ \bibnamefont {Xu}}},\ \bibinfo {organization} {PMLR}\ (\bibinfo  {publisher} {PMLR},\ \bibinfo {year} {2022})\ pp.\
  \bibinfo {pages} {97--112}\BibitemShut {NoStop}%
\bibitem [{\citenamefont {Semerjian}(2024)}]{semerjian2024matrix}%
  \BibitemOpen
  \bibfield  {author} {\bibinfo {author} {\bibfnamefont {G.}~\bibnamefont {Semerjian}},\ }\bibfield  {title} {\bibinfo {title} {Matrix denoising: Bayes-optimal estimators via low-degree polynomials},\ }\href@noop {} {\bibfield  {journal} {\bibinfo  {journal} {arXiv preprint arXiv:2402.16719}\ } (\bibinfo {year} {2024})}\BibitemShut {NoStop}%
\bibitem [{\citenamefont {Camilli}\ and\ \citenamefont {M{\'e}zard}(2023)}]{camilli2023matrix}%
  \BibitemOpen
  \bibfield  {author} {\bibinfo {author} {\bibfnamefont {F.}~\bibnamefont {Camilli}}\ and\ \bibinfo {author} {\bibfnamefont {M.}~\bibnamefont {M{\'e}zard}},\ }\bibfield  {title} {\bibinfo {title} {Matrix factorization with neural networks},\ }\href@noop {} {\bibfield  {journal} {\bibinfo  {journal} {Physical Review E}\ }\textbf {\bibinfo {volume} {107}},\ \bibinfo {pages} {064308} (\bibinfo {year} {2023})}\BibitemShut {NoStop}%
\bibitem [{\citenamefont {Pourkamali}\ and\ \citenamefont {Macris}(2023)}]{pourkamali2023rectangular}%
  \BibitemOpen
  \bibfield  {author} {\bibinfo {author} {\bibfnamefont {F.}~\bibnamefont {Pourkamali}}\ and\ \bibinfo {author} {\bibfnamefont {N.}~\bibnamefont {Macris}},\ }\bibfield  {title} {\bibinfo {title} {Rectangular rotational invariant estimator for general additive noise matrices},\ }in\ \href@noop {} {\emph {\bibinfo {booktitle} {2023 IEEE International Symposium on Information Theory (ISIT)}}}\ (\bibinfo {year} {2023})\ pp.\ \bibinfo {pages} {2081--2086}\BibitemShut {NoStop}%
\bibitem [{\citenamefont {Fyodorov}(2019)}]{fyodorov2019spin}%
  \BibitemOpen
  \bibfield  {author} {\bibinfo {author} {\bibfnamefont {Y.~V.}\ \bibnamefont {Fyodorov}},\ }\bibfield  {title} {\bibinfo {title} {A spin glass model for reconstructing nonlinearly encrypted signals corrupted by noise},\ }\href@noop {} {\bibfield  {journal} {\bibinfo  {journal} {Journal of Statistical Physics}\ }\textbf {\bibinfo {volume} {175}},\ \bibinfo {pages} {789} (\bibinfo {year} {2019})}\BibitemShut {NoStop}%
\bibitem [{\citenamefont {Kamali}\ and\ \citenamefont {Urbani}(2023{\natexlab{a}})}]{kamali2023dynamical}%
  \BibitemOpen
  \bibfield  {author} {\bibinfo {author} {\bibfnamefont {P.~J.}\ \bibnamefont {Kamali}}\ and\ \bibinfo {author} {\bibfnamefont {P.}~\bibnamefont {Urbani}},\ }\bibfield  {title} {\bibinfo {title} {Dynamical mean field theory for models of confluent tissues and beyond},\ }\href@noop {} {\bibfield  {journal} {\bibinfo  {journal} {SciPost Physics}\ }\textbf {\bibinfo {volume} {15}},\ \bibinfo {pages} {219} (\bibinfo {year} {2023}{\natexlab{a}})}\BibitemShut {NoStop}%
\bibitem [{\citenamefont {Montanari}\ and\ \citenamefont {Subag}(2023)}]{montanari2023solving}%
  \BibitemOpen
  \bibfield  {author} {\bibinfo {author} {\bibfnamefont {A.}~\bibnamefont {Montanari}}\ and\ \bibinfo {author} {\bibfnamefont {E.}~\bibnamefont {Subag}},\ }\bibfield  {title} {\bibinfo {title} {Solving overparametrized systems of random equations: I. model and algorithms for approximate solutions},\ }\href@noop {} {\bibfield  {journal} {\bibinfo  {journal} {arXiv preprint arXiv:2306.13326}\ } (\bibinfo {year} {2023})}\BibitemShut {NoStop}%
\bibitem [{\citenamefont {Kamali}\ and\ \citenamefont {Urbani}(2023{\natexlab{b}})}]{kamali2023stochastic}%
  \BibitemOpen
  \bibfield  {author} {\bibinfo {author} {\bibfnamefont {P.~J.}\ \bibnamefont {Kamali}}\ and\ \bibinfo {author} {\bibfnamefont {P.}~\bibnamefont {Urbani}},\ }\bibfield  {title} {\bibinfo {title} {Stochastic gradient descent outperforms gradient descent in recovering a high-dimensional signal in a glassy energy landscape},\ }\href@noop {} {\bibfield  {journal} {\bibinfo  {journal} {arXiv preprint arXiv:2309.04788}\ } (\bibinfo {year} {2023}{\natexlab{b}})}\BibitemShut {NoStop}%
\bibitem [{\citenamefont {Maillard}\ \emph {et~al.}(2024)\citenamefont {Maillard}, \citenamefont {Troiani}, \citenamefont {Martin}, \citenamefont {Krzakala},\ and\ \citenamefont {Zdeborov{\'a}}}]{troiani2024}%
  \BibitemOpen
  \bibfield  {author} {\bibinfo {author} {\bibfnamefont {A.}~\bibnamefont {Maillard}}, \bibinfo {author} {\bibfnamefont {E.}~\bibnamefont {Troiani}}, \bibinfo {author} {\bibfnamefont {S.}~\bibnamefont {Martin}}, \bibinfo {author} {\bibfnamefont {F.}~\bibnamefont {Krzakala}},\ and\ \bibinfo {author} {\bibfnamefont {L.}~\bibnamefont {Zdeborov{\'a}}},\ }\bibfield  {title} {\bibinfo {title} {Bayes-optimal learning of an extensive-width neural network from quadratically many samples},\ }\href@noop {} {\bibfield  {journal} {\bibinfo  {journal} {to appear at NeurIPS 2024}\ } (\bibinfo {year} {2024})}\BibitemShut {NoStop}%
\bibitem [{\citenamefont {Bhojanapalli}\ \emph {et~al.}(2016)\citenamefont {Bhojanapalli}, \citenamefont {Neyshabur},\ and\ \citenamefont {Srebro}}]{bhojanapalli2016global}%
  \BibitemOpen
  \bibfield  {author} {\bibinfo {author} {\bibfnamefont {S.}~\bibnamefont {Bhojanapalli}}, \bibinfo {author} {\bibfnamefont {B.}~\bibnamefont {Neyshabur}},\ and\ \bibinfo {author} {\bibfnamefont {N.}~\bibnamefont {Srebro}},\ }\bibfield  {title} {\bibinfo {title} {Global optimality of local search for low rank matrix recovery},\ }\href@noop {} {\bibfield  {journal} {\bibinfo  {journal} {Advances in Neural Information Processing Systems}\ }\textbf {\bibinfo {volume} {29}} (\bibinfo {year} {2016})}\BibitemShut {NoStop}%
\bibitem [{\citenamefont {Ding}\ \emph {et~al.}(2024)\citenamefont {Ding}, \citenamefont {Drusvyatskiy}, \citenamefont {Fazel},\ and\ \citenamefont {Harchaoui}}]{ding2024flat}%
  \BibitemOpen
  \bibfield  {author} {\bibinfo {author} {\bibfnamefont {L.}~\bibnamefont {Ding}}, \bibinfo {author} {\bibfnamefont {D.}~\bibnamefont {Drusvyatskiy}}, \bibinfo {author} {\bibfnamefont {M.}~\bibnamefont {Fazel}},\ and\ \bibinfo {author} {\bibfnamefont {Z.}~\bibnamefont {Harchaoui}},\ }\bibfield  {title} {\bibinfo {title} {Flat minima generalize for low-rank matrix recovery},\ }\href@noop {} {\bibfield  {journal} {\bibinfo  {journal} {Information and Inference: A Journal of the IMA}\ }\textbf {\bibinfo {volume} {13}},\ \bibinfo {pages} {iaae009} (\bibinfo {year} {2024})}\BibitemShut {NoStop}%
\bibitem [{\citenamefont {Okajima}\ and\ \citenamefont {Takahashi}(2024)}]{okajima2024asymptotic}%
  \BibitemOpen
  \bibfield  {author} {\bibinfo {author} {\bibfnamefont {K.}~\bibnamefont {Okajima}}\ and\ \bibinfo {author} {\bibfnamefont {T.}~\bibnamefont {Takahashi}},\ }\bibfield  {title} {\bibinfo {title} {Asymptotic dynamics of alternating minimization for bilinear regression},\ }\href@noop {} {\bibfield  {journal} {\bibinfo  {journal} {arXiv preprint arXiv:2402.04751}\ } (\bibinfo {year} {2024})}\BibitemShut {NoStop}%
\bibitem [{\citenamefont {Hu}\ and\ \citenamefont {Lu}(2022)}]{hu2022universality}%
  \BibitemOpen
  \bibfield  {author} {\bibinfo {author} {\bibfnamefont {H.}~\bibnamefont {Hu}}\ and\ \bibinfo {author} {\bibfnamefont {Y.~M.}\ \bibnamefont {Lu}},\ }\bibfield  {title} {\bibinfo {title} {Universality laws for high-dimensional learning with random features},\ }\href@noop {} {\bibfield  {journal} {\bibinfo  {journal} {IEEE Transactions on Information Theory}\ }\textbf {\bibinfo {volume} {69}},\ \bibinfo {pages} {1932} (\bibinfo {year} {2022})}\BibitemShut {NoStop}%
\bibitem [{\citenamefont {Wang}\ \emph {et~al.}(2024)\citenamefont {Wang}, \citenamefont {Nichani},\ and\ \citenamefont {Lee}}]{wang2023learninghierarchicalpolynomialsthreelayer}%
  \BibitemOpen
  \bibfield  {author} {\bibinfo {author} {\bibfnamefont {Z.}~\bibnamefont {Wang}}, \bibinfo {author} {\bibfnamefont {E.}~\bibnamefont {Nichani}},\ and\ \bibinfo {author} {\bibfnamefont {J.~D.}\ \bibnamefont {Lee}},\ }\bibfield  {title} {\bibinfo {title} {Learning hierarchical polynomials with three-layer neural networks},\ }in\ \href@noop {} {\emph {\bibinfo {booktitle} {The Twelfth International Conference on Learning Representations}}}\ (\bibinfo {year} {2024})\BibitemShut {NoStop}%
\bibitem [{\citenamefont {Cover}\ and\ \citenamefont {Thomas}(1991)}]{cover1991information}%
  \BibitemOpen
  \bibfield  {author} {\bibinfo {author} {\bibfnamefont {T.~M.}\ \bibnamefont {Cover}}\ and\ \bibinfo {author} {\bibfnamefont {J.~A.}\ \bibnamefont {Thomas}},\ }\bibfield  {title} {\bibinfo {title} {Information theory and statistics},\ }\href@noop {} {\bibfield  {journal} {\bibinfo  {journal} {Elements of information theory}\ }\textbf {\bibinfo {volume} {1}},\ \bibinfo {pages} {279} (\bibinfo {year} {1991})}\BibitemShut {NoStop}%
\bibitem [{\citenamefont {Zdeborov{\'a}}\ and\ \citenamefont {Krzakala}(2016)}]{zdeborova2016statistical}%
  \BibitemOpen
  \bibfield  {author} {\bibinfo {author} {\bibfnamefont {L.}~\bibnamefont {Zdeborov{\'a}}}\ and\ \bibinfo {author} {\bibfnamefont {F.}~\bibnamefont {Krzakala}},\ }\bibfield  {title} {\bibinfo {title} {Statistical physics of inference: Thresholds and algorithms},\ }\href@noop {} {\bibfield  {journal} {\bibinfo  {journal} {Advances in Physics}\ }\textbf {\bibinfo {volume} {65}},\ \bibinfo {pages} {453} (\bibinfo {year} {2016})}\BibitemShut {NoStop}%
\bibitem [{\citenamefont {Pennington}\ and\ \citenamefont {Worah}(2017)}]{pennington2017nonlinear}%
  \BibitemOpen
  \bibfield  {author} {\bibinfo {author} {\bibfnamefont {J.}~\bibnamefont {Pennington}}\ and\ \bibinfo {author} {\bibfnamefont {P.}~\bibnamefont {Worah}},\ }\bibfield  {title} {\bibinfo {title} {Nonlinear random matrix theory for deep learning},\ }\bibfield  {booktitle} {\emph {\bibinfo {booktitle} {Advances in Neural Information Processing Systems}},\ }\href@noop {} {\ \textbf {\bibinfo {volume} {30}} (\bibinfo {year} {2017})}\BibitemShut {NoStop}%
\bibitem [{\citenamefont {Biane}(1997)}]{biane1997free}%
  \BibitemOpen
  \bibfield  {author} {\bibinfo {author} {\bibfnamefont {P.}~\bibnamefont {Biane}},\ }\bibfield  {title} {\bibinfo {title} {On the free convolution with a semi-circular distribution},\ }\href@noop {} {\bibfield  {journal} {\bibinfo  {journal} {Indiana University Mathematics Journal}\ }\textbf {\bibinfo {volume} {46}},\ \bibinfo {pages} {705} (\bibinfo {year} {1997})}\BibitemShut {NoStop}%
\bibitem [{\citenamefont {Rangan}(2011)}]{rangan2011generalized}%
  \BibitemOpen
  \bibfield  {author} {\bibinfo {author} {\bibfnamefont {S.}~\bibnamefont {Rangan}},\ }\bibfield  {title} {\bibinfo {title} {Generalized approximate message passing for estimation with random linear mixing},\ }in\ \href@noop {} {\emph {\bibinfo {booktitle} {2011 IEEE International Symposium on Information Theory Proceedings}}}\ (\bibinfo {year} {2011})\ pp.\ \bibinfo {pages} {2168--2172}\BibitemShut {NoStop}%
\bibitem [{\citenamefont {Donoho}\ \emph {et~al.}(2009)\citenamefont {Donoho}, \citenamefont {Maleki},\ and\ \citenamefont {Montanari}}]{donoho2009message}%
  \BibitemOpen
  \bibfield  {author} {\bibinfo {author} {\bibfnamefont {D.~L.}\ \bibnamefont {Donoho}}, \bibinfo {author} {\bibfnamefont {A.}~\bibnamefont {Maleki}},\ and\ \bibinfo {author} {\bibfnamefont {A.}~\bibnamefont {Montanari}},\ }\bibfield  {title} {\bibinfo {title} {Message-passing algorithms for compressed sensing},\ }\href@noop {} {\bibfield  {journal} {\bibinfo  {journal} {Proceedings of the National Academy of Sciences}\ }\textbf {\bibinfo {volume} {106}},\ \bibinfo {pages} {18914} (\bibinfo {year} {2009})}\BibitemShut {NoStop}%
\bibitem [{\citenamefont {Berthier}\ \emph {et~al.}(2020)\citenamefont {Berthier}, \citenamefont {Montanari},\ and\ \citenamefont {Nguyen}}]{berthier2020state}%
  \BibitemOpen
  \bibfield  {author} {\bibinfo {author} {\bibfnamefont {R.}~\bibnamefont {Berthier}}, \bibinfo {author} {\bibfnamefont {A.}~\bibnamefont {Montanari}},\ and\ \bibinfo {author} {\bibfnamefont {P.-M.}\ \bibnamefont {Nguyen}},\ }\bibfield  {title} {\bibinfo {title} {State evolution for approximate message passing with non-separable functions},\ }\href@noop {} {\bibfield  {journal} {\bibinfo  {journal} {Information and Inference: A Journal of the IMA}\ }\textbf {\bibinfo {volume} {9}},\ \bibinfo {pages} {33} (\bibinfo {year} {2020})}\BibitemShut {NoStop}%
\bibitem [{\citenamefont {Gerbelot}\ and\ \citenamefont {Berthier}(2023)}]{gerbelot2023graph}%
  \BibitemOpen
  \bibfield  {author} {\bibinfo {author} {\bibfnamefont {C.}~\bibnamefont {Gerbelot}}\ and\ \bibinfo {author} {\bibfnamefont {R.}~\bibnamefont {Berthier}},\ }\bibfield  {title} {\bibinfo {title} {Graph-based approximate message passing iterations},\ }\href@noop {} {\bibfield  {journal} {\bibinfo  {journal} {Information and Inference: A Journal of the IMA}\ }\textbf {\bibinfo {volume} {12}},\ \bibinfo {pages} {2562} (\bibinfo {year} {2023})}\BibitemShut {NoStop}%
\bibitem [{\citenamefont {Romanov}\ and\ \citenamefont {Gavish}(2018)}]{romanov_near-optimal_2018}%
  \BibitemOpen
  \bibfield  {author} {\bibinfo {author} {\bibfnamefont {E.}~\bibnamefont {Romanov}}\ and\ \bibinfo {author} {\bibfnamefont {M.}~\bibnamefont {Gavish}},\ }\bibfield  {title} {\bibinfo {title} {Near-optimal matrix recovery from random linear measurements},\ }\href@noop {} {\bibfield  {journal} {\bibinfo  {journal} {Proceedings of the National Academy of Sciences}\ }\textbf {\bibinfo {volume} {115}},\ \bibinfo {pages} {7200} (\bibinfo {year} {2018})}\BibitemShut {NoStop}%
\bibitem [{\citenamefont {Barbier}\ \emph {et~al.}(2024{\natexlab{a}})\citenamefont {Barbier}, \citenamefont {Ko},\ and\ \citenamefont {Rahman}}]{barbier2024information}%
  \BibitemOpen
  \bibfield  {author} {\bibinfo {author} {\bibfnamefont {J.}~\bibnamefont {Barbier}}, \bibinfo {author} {\bibfnamefont {J.}~\bibnamefont {Ko}},\ and\ \bibinfo {author} {\bibfnamefont {A.~A.}\ \bibnamefont {Rahman}},\ }\bibfield  {title} {\bibinfo {title} {Information-theoretic limits for sublinear-rank symmetric matrix factorization},\ }in\ \href@noop {} {\emph {\bibinfo {booktitle} {International Zurich Seminar on Information and Communication (IZS 2024). Proceedings}}}\ (\bibinfo {organization} {ETH Z{\"u}rich},\ \bibinfo {year} {2024})\ pp.\ \bibinfo {pages} {16--16}\BibitemShut {NoStop}%
\bibitem [{\citenamefont {Barbier}\ \emph {et~al.}(2024{\natexlab{b}})\citenamefont {Barbier}, \citenamefont {Ko},\ and\ \citenamefont {Rahman}}]{barbierMultiscale}%
  \BibitemOpen
  \bibfield  {author} {\bibinfo {author} {\bibfnamefont {J.}~\bibnamefont {Barbier}}, \bibinfo {author} {\bibfnamefont {J.}~\bibnamefont {Ko}},\ and\ \bibinfo {author} {\bibfnamefont {A.~A.}\ \bibnamefont {Rahman}},\ }\bibfield  {title} {\bibinfo {title} {A multiscale cavity method for sublinear-rank symmetric matrix factorization},\ }\href@noop {} {\bibfield  {journal} {\bibinfo  {journal} {arXiv preprint arXiv:2403.07189}\ } (\bibinfo {year} {2024}{\natexlab{b}})}\BibitemShut {NoStop}%
\bibitem [{\citenamefont {Pourkamali}\ \emph {et~al.}(2024)\citenamefont {Pourkamali}, \citenamefont {Barbier},\ and\ \citenamefont {Macris}}]{pourkamali2024matrix}%
  \BibitemOpen
  \bibfield  {author} {\bibinfo {author} {\bibfnamefont {F.}~\bibnamefont {Pourkamali}}, \bibinfo {author} {\bibfnamefont {J.}~\bibnamefont {Barbier}},\ and\ \bibinfo {author} {\bibfnamefont {N.}~\bibnamefont {Macris}},\ }\bibfield  {title} {\bibinfo {title} {Matrix inference in growing rank regimes},\ }\href@noop {} {\bibfield  {journal} {\bibinfo  {journal} {IEEE Transactions on Information Theory}\ ,\ \bibinfo {pages} {1}} (\bibinfo {year} {2024})}\BibitemShut {NoStop}%
\bibitem [{\citenamefont {Helmke}\ and\ \citenamefont {Moore}(2012)}]{helmke2012optimization}%
  \BibitemOpen
  \bibfield  {author} {\bibinfo {author} {\bibfnamefont {U.}~\bibnamefont {Helmke}}\ and\ \bibinfo {author} {\bibfnamefont {J.~B.}\ \bibnamefont {Moore}},\ }\href@noop {} {\emph {\bibinfo {title} {Optimization and dynamical systems}}}\ (\bibinfo  {publisher} {Springer Science \& Business Media},\ \bibinfo {year} {2012})\BibitemShut {NoStop}%
\bibitem [{\citenamefont {Donoho}\ and\ \citenamefont {Gavish}(2014)}]{donoho2014minimax}%
  \BibitemOpen
  \bibfield  {author} {\bibinfo {author} {\bibfnamefont {D.}~\bibnamefont {Donoho}}\ and\ \bibinfo {author} {\bibfnamefont {M.}~\bibnamefont {Gavish}},\ }\bibfield  {title} {\bibinfo {title} {Minimax risk of matrix denoising by singular value thresholding},\ }\href@noop {} {\bibfield  {journal} {\bibinfo  {journal} {The Annals of Statistics}\ }\textbf {\bibinfo {volume} {42}},\ \bibinfo {pages} {2413 } (\bibinfo {year} {2014})}\BibitemShut {NoStop}%
\bibitem [{\citenamefont {Krzakala}\ \emph {et~al.}(2012)\citenamefont {Krzakala}, \citenamefont {M\'ezard}, \citenamefont {Sausset}, \citenamefont {Sun},\ and\ \citenamefont {Zdeborov\'a}}]{krzakala2012statistical}%
  \BibitemOpen
  \bibfield  {author} {\bibinfo {author} {\bibfnamefont {F.}~\bibnamefont {Krzakala}}, \bibinfo {author} {\bibfnamefont {M.}~\bibnamefont {M\'ezard}}, \bibinfo {author} {\bibfnamefont {F.}~\bibnamefont {Sausset}}, \bibinfo {author} {\bibfnamefont {Y.~F.}\ \bibnamefont {Sun}},\ and\ \bibinfo {author} {\bibfnamefont {L.}~\bibnamefont {Zdeborov\'a}},\ }\bibfield  {title} {\bibinfo {title} {Statistical-physics-based reconstruction in compressed sensing},\ }\href@noop {} {\bibfield  {journal} {\bibinfo  {journal} {Phys. Rev. X}\ }\textbf {\bibinfo {volume} {2}},\ \bibinfo {pages} {021005} (\bibinfo {year} {2012})}\BibitemShut {NoStop}%
\bibitem [{\citenamefont {Neyshabur}\ \emph {et~al.}(2014)\citenamefont {Neyshabur}, \citenamefont {Tomioka},\ and\ \citenamefont {Srebro}}]{neyshabur2014search}%
  \BibitemOpen
  \bibfield  {author} {\bibinfo {author} {\bibfnamefont {B.}~\bibnamefont {Neyshabur}}, \bibinfo {author} {\bibfnamefont {R.}~\bibnamefont {Tomioka}},\ and\ \bibinfo {author} {\bibfnamefont {N.}~\bibnamefont {Srebro}},\ }\bibfield  {title} {\bibinfo {title} {In search of the real inductive bias: On the role of implicit regularization in deep learning},\ }\href@noop {} {\bibfield  {journal} {\bibinfo  {journal} {arXiv preprint arXiv:1412.6614}\ } (\bibinfo {year} {2014})}\BibitemShut {NoStop}%
\bibitem [{\citenamefont {Arora}\ \emph {et~al.}(2019)\citenamefont {Arora}, \citenamefont {Cohen}, \citenamefont {Hu},\ and\ \citenamefont {Luo}}]{arora2019implicit}%
  \BibitemOpen
  \bibfield  {author} {\bibinfo {author} {\bibfnamefont {S.}~\bibnamefont {Arora}}, \bibinfo {author} {\bibfnamefont {N.}~\bibnamefont {Cohen}}, \bibinfo {author} {\bibfnamefont {W.}~\bibnamefont {Hu}},\ and\ \bibinfo {author} {\bibfnamefont {Y.}~\bibnamefont {Luo}},\ }\bibfield  {title} {\bibinfo {title} {Implicit regularization in deep matrix factorization},\ }\bibfield  {booktitle} {\emph {\bibinfo {booktitle} {Advances in Neural Information Processing Systems}},\ }\href@noop {} {\ \textbf {\bibinfo {volume} {32}} (\bibinfo {year} {2019})}\BibitemShut {NoStop}%
\bibitem [{\citenamefont {Mignacco}\ \emph {et~al.}(2020)\citenamefont {Mignacco}, \citenamefont {Krzakala}, \citenamefont {Urbani},\ and\ \citenamefont {Zdeborov\'{a}}}]{mignacco2020dynamical}%
  \BibitemOpen
  \bibfield  {author} {\bibinfo {author} {\bibfnamefont {F.}~\bibnamefont {Mignacco}}, \bibinfo {author} {\bibfnamefont {F.}~\bibnamefont {Krzakala}}, \bibinfo {author} {\bibfnamefont {P.}~\bibnamefont {Urbani}},\ and\ \bibinfo {author} {\bibfnamefont {L.}~\bibnamefont {Zdeborov\'{a}}},\ }\bibfield  {title} {\bibinfo {title} {Dynamical mean-field theory for stochastic gradient descent in gaussian mixture classification},\ }\bibfield  {booktitle} {\emph {\bibinfo {booktitle} {Advances in Neural Information Processing Systems}},\ }\href@noop {} {\ \textbf {\bibinfo {volume} {33}},\ \bibinfo {pages} {9540} (\bibinfo {year} {2020})}\BibitemShut {NoStop}%
\bibitem [{\citenamefont {Diamond}\ and\ \citenamefont {Boyd}(2016)}]{diamond2016cvxpy}%
  \BibitemOpen
  \bibfield  {author} {\bibinfo {author} {\bibfnamefont {S.}~\bibnamefont {Diamond}}\ and\ \bibinfo {author} {\bibfnamefont {S.}~\bibnamefont {Boyd}},\ }\bibfield  {title} {\bibinfo {title} {{CVXPY}: {A} {P}ython-embedded modeling language for convex optimization},\ }\href@noop {} {\bibfield  {journal} {\bibinfo  {journal} {Journal of Machine Learning Research}\ }\textbf {\bibinfo {volume} {17}},\ \bibinfo {pages} {1} (\bibinfo {year} {2016})}\BibitemShut {NoStop}%
\bibitem [{\citenamefont {Akshay~Agrawal}\ and\ \citenamefont {Boyd}(2018)}]{agrawal2018rewriting}%
  \BibitemOpen
  \bibfield  {author} {\bibinfo {author} {\bibfnamefont {S.~D.}\ \bibnamefont {Akshay~Agrawal}, \bibfnamefont {Robin~Verschueren}}\ and\ \bibinfo {author} {\bibfnamefont {S.}~\bibnamefont {Boyd}},\ }\bibfield  {title} {\bibinfo {title} {A rewriting system for convex optimization problems},\ }\href@noop {} {\bibfield  {journal} {\bibinfo  {journal} {Journal of Control and Decision}\ }\textbf {\bibinfo {volume} {5}},\ \bibinfo {pages} {42} (\bibinfo {year} {2018})}\BibitemShut {NoStop}%
\end{thebibliography}
\end{document}